%% file: main.tex
\documentclass[a4paper,10pt]{article}

\usepackage[top=1in, left=0.9in, bottom=1in, right=0.9in]{geometry}

\usepackage{amsmath,amssymb,amsfonts}

\usepackage[utf8]{inputenc}
\usepackage{cite}

\usepackage{graphicx}
\usepackage{xcolor}
\usepackage{bm}

\usepackage{booktabs}

\usepackage[colorlinks=true,linkcolor=blue,citecolor=blue,urlcolor=blue]{hyperref}

\let\counterwithin\relax
\usepackage{chngcntr}

\usepackage{longtable}
\usepackage{multirow}
\usepackage{makecell}
\usepackage{diagbox}
\usepackage{adjustbox}

\usepackage{tikz}
\usepackage[normalem]{ulem}
\usepackage{verbatim}
\usepackage[dvipsnames]{xcolor}
\usepackage{xcolor}
\usepackage{ytableau}
\usepackage[vcentermath]{youngtab}

\newcommand\mydiagbox[2]{\hbox{\tabcolsep=\arraycolsep\diagbox{#1}{#2}}}

\newcommand{\beq}{\begin {equation}}
\newcommand{\eeq}{\end   {equation}}
\newcommand{\bea}{\begin {eqnarray}}
\newcommand{\eea}{\end   {eqnarray}}
\newcommand{\baa}{\begin {array}   }
\newcommand{\eaa}{\end   {array}   }
\newcommand{\bit}{\begin {itemize} }
\newcommand{\eit}{\end   {itemize} }
\newcommand{\be }{\begin {equation}}
\newcommand{\ee }{\end   {equation}}

\begin{document}


\begin{center}


{\Large \textbf  {Massive On-shell Splitting Functions in Spinor-Helicity Formalism}}\\[10mm]

Yi-Ning Wang$^{a, b}$\footnote{wangyining@itp.ac.cn}, Chao Wu$^{a, b}$\footnote{wuch7@itp.ac.cn}, Jiang-Hao Yu$^{a, b, c, d}$\footnote{jhyu@itp.ac.cn}\\[10mm]

\noindent
$^a${\em \small Institute of Theoretical Physics, Chinese Academy of Sciences, Beijing 100190, P. R. China}  \\
$^b${\em \small School of Physical Sciences, University of Chinese Academy of Sciences, Beijing 100049, P.R. China}   \\
$^c${\em \small School of Fundamental Physics and Mathematical Sciences, Hangzhou Institute for Advanced Study, \\ UCAS, Hangzhou 310024, China} \\
$^d${\em \small International Centre for Theoretical Physics Asia-Pacific, Beijing/Hangzhou, China}\\[10mm]

\date{\today}

\end{center}

\begin{abstract}
\noindent

Collinear splitting functions govern parton evolution, parton showers, and resummation at high-energy colliders. While on-shell spinor-helicity methods have successfully yielded massless QCD splitting functions, a complete on-shell construction for massive particles, systematically incorporating finite-mass effects, is less developed. We present an on-shell constructive formalism for massive collinear splitting functions based on Soper-Weinberg collinear spinors, whose transformation properties follow from a light-front Galilean subgroup of the Poincar\'e group. Decomposing massive momenta and spinors with respect to fixed lightlike vectors $n$ and $\bar n$ makes the expansion in the alignment regime $m<p_T\ll p_+$ manifest. The leading-order structures are matched to massless three-point amplitudes, while an additional Higgs momentum along $\bar n$ probes the subleading spinor components and relates them to massless four-point amplitudes. We derive the complete set of leading and subleading massive splitting functions for all Standard Model particles and establish a systematic matching dictionary between massless and massive coupling coefficients at both the three- and four-point levels. Higher-point splitting functions are obtained through the recursive bootstrap relation with a universal substitution rule as a consequence of the Galilean symmetry. This constructive framework extends naturally to effective field theory operators and higher perturbative orders, providing a flexible computational tool for precision collider physics and parton shower development.

\end{abstract}

\newpage

\tableofcontents

\setcounter{footnote}{0}
\counterwithin{equation}{section}

\newpage

\input{sec1_intro}

\input{sec2_particle}

\input{sec3_projection}

\input{sec4_3pt}

\input{sec5_splitting}

\input{sec6_recursive}

\input{sec7_con}

\section*{Acknowledgments}
This work is supported by the National Science Foundation of China under Grants No. 12347105, No. 12375099 and No. 12047503, and the National Key Research and Development Program of China Grant No. 2020YFC2201501, No. 2021YFA0718304.

\input{app_conformal}

\bibliographystyle{JHEP}
\bibliography{reference}

\end{document}

%% file: sec1_intro.tex
\section{Introduction}

Collinear splitting functions are among the most fundamental objects in perturbative quantum field theory. They govern the evolution of parton distribution functions via the DGLAP equations \cite{Altarelli:1977zs,Dokshitzer:1977sg,Gribov:1972ri}, form the kernel of parton shower Monte Carlo programs at the LHC, and underpin resummation calculations for a wide class of infrared-sensitive observables. In the Standard Model, both Quantum Chromodynamics (QCD) and electroweak interactions contribute to collinear evolution.  Finite-mass effects are especially relevant for the top quark, the $W$ and $Z$ bosons, and the Higgs boson, and must be controlled in precision predictions for present and future colliders. The traditional approach to computing splitting functions proceeds by extracting the collinear limit of squared matrix elements computed from Feynman diagrams \cite{Altarelli:1977zs}. Electroweak splitting functions including mass corrections have been computed in the traditional diagrammatic approach~\cite{Ciafaloni:2001mu,Chiu:2009ft,Chen:2016wkt,Bauer:2017bnh,Cuomo:2019siu,Han:2020uid,Nardi:2024tce,Dittmaier:2025htf} and Standard Model (SM) parton distributions at very high energies have been studied in the context of electroweak resummation~\cite{Ciafaloni:1998xg,Fadin:1999bq,Ciafaloni:2000df,Chiu:2007yn,Manohar:2018kfx,Pagani:2021vyk}. While systematic, this method obscures the underlying group-theoretic and on-shell structure of the splitting process.

In recent years, on-shell methods rooted in the spinor-helicity formalism have transformed our understanding of scattering amplitudes \cite{Dixon:2013uaa,Elvang:2015rqa,Cheung:2017pzi,Badger:2023eqz}. For massless particles, this formalism have enabled the efficient calculations of tree-level and loop-level amplitudes~\cite{Dixon:1996wi}, and these ideas have been successfully applied to derive massless QCD splitting functions from on-shell factorized amplitudes alone~\cite{Bern:1994zx,Bern:2004cz,Cohen:2024xuf}.
Pioneered by Refs.~\cite{Kleiss:1985yh,Kleiss:1986qc,Dittmaier:1998nn,Schwinn:2007ee,Boels:2011zz}, a practical massive helicity formalism for perturbative calculations has been developed by introducing reference spinors and arbitrary spin quantization axes. Massive splitting function has been calculated~\cite{Larkoski:2011fd,Kleiss:2020rcg,Brooks:2021kji} using these massive spinor-helicity variables.
The Arkani-Hamed-Huang-Huang (AHH) formalism \cite{Arkani-Hamed:2017jhn} utilizes the little group covariance and enables a systematic construction of amplitudes for particles of arbitrary mass and spin, providing a unified language for massive on-shell amplitudes. 
However, an on-shell constructive calculation of massive splitting functions in the modern spinor-helicity formalism, which makes the group-theoretic structure manifest and unifies the treatment of leading and subleading mass corrections, has remained lacking.

This gap is significant for several reasons. First, the traditional approach does not utilize spacetime symmetry to determine the splitting function, while in on-shell method the little group covariance determines the 3-point amplitudes and thus splitting function, which fully exploit the advantages of scattering amplitudes.
Second, traditional calculation is usually based on the helicity spin amplitudes, whose spin axis is not aligned with the collinear direction for massive particles. Unlike the massless splitting, one must convert these spin directions to the collinear direction $n$, which not only leads to lengthy calculations, but also obscures the transverse Galilean symmetry. 
Third, the traditional approach treats mass effects as additive power corrections to massless splitting functions, obscuring the fact that, from an on-shell perspective, they arise from a systematic expansion organized by little-group and Galilean symmetries. Finally, the diagrammatic extraction of collinear limits becomes increasingly cumbersome for higher-point splitting functions and for effective field theory (EFT) extensions, while sequential and contact higher-multiplicity splittings are more transparent at
the amplitude level.


In this paper, we present a complete on-shell formalism for constructing collinear splitting functions of massive particles using the spinor-helicity method. Our approach is built on three key elements. First, we adopt the Soper-Weinberg (SW) collinear spinors \cite{Weinberg:1966jm,Soper:1971wn}, which arise from a light-front Galilean subgroup of the Poincar\'e group and provide a systematic decomposition of massive momenta and spinors in terms of fixed collinear reference vectors $n$ and $\bar{n}$. Second, we contrast this collinear decomposition with the AHH helicity decomposition and establish that their power counting coincides in the alignment limit $m < p_T  \ll p_+$, where the collinear physics becomes dominant. Third, we employ massless four-point amplitudes with an additional Higgs insertion to capture subleading mass corrections. In this setup, the anti-collinear momentum of the Higgs serves as a probe of the subleading spinor components that are invisible at leading order. This framework is constructive by nature: the elementary three-point massive amplitudes determine the $1\to 2$ splitting functions, and higher-point splitting functions are obtained through recursive constructions from these elementary $1\to 2$ process.

The main results of this paper are as follows. We derive the complete set of leading and subleading massive splitting functions for all SM particles, covering fermions, vector bosons, and scalars. The results are expressed in terms of the kinematic variables $z$ (longitudinal momentum fraction) and $p_T$ (transverse momentum). We establish a systematic matching between massless and massive three-point amplitudes, and extend this matching to the subleading order via four-point massless amplitudes with a Higgs insertion, providing a dictionary that relates massless and massive coupling coefficients. We demonstrate that higher-point splitting functions can be constructed through a universal substitution rule derived from the Galilean symmetry of the collinear subgroup, requiring no case-by-case diagrammatic calculations. The formalism is directly extensible to EFT operators and to higher perturbative orders, offering a versatile tool for collider physics resummation programs.

The remainder of this paper is organized as follows. In section~\ref{sec:particle_state}, we construct the collinear particle states from the 2D Galilean subgroup of the Poincar\'e group, introduce the Soper-Weinberg spinors, and establish the spinor-helicity formalism for massive particles. Section~\ref{sec:collinear_projection} develops the collinear decomposition of massive spinors onto fixed $n$ and $\bar{n}$ reference vectors, contrasts it with the helicity decomposition, and derives the power-counting rules that govern the high-energy expansion. In section~\ref{sec:3pt}, we construct all Standard Model three-point massive amplitudes, analyze their high-energy behavior, resolve unitarity violations in the longitudinal vector sector via the Goldstone equivalence theorem, and establish the matching with massless three- and four-point amplitudes. Section~\ref{sec:splitting} presents the core derivation: the factorization of collinear amplitudes into splitting probabilities, yielding the complete set of leading and subleading massive splitting functions. In section~\ref{sec:recursive}, we develop the recursive bootstrap for higher-point splitting functions and provide explicit worked examples. We conclude in section~\ref{sec:summary} with a summary of our results and a discussion of their implications for parton shower development, electroweak resummation, and future extensions.

%% file: sec2_particle.tex
\section{Light-front and Collinear Particle States}
\label{sec:particle_state}

As irreducible unitary representations of the Poincar\'e group, single-particle states are characterized by their mass $m$ and spin $s$ \cite{Wigner:1939cj}. A particle state $|p,\sigma\rangle$ carries definite four-momentum $p$ and spin projection $\sigma$ with respect to a spin-quantization axis $s_\mu$ orthogonal to $p$, satisfying
\begin{eqnarray}
P^2 |p, s, \sigma\rangle &=& m^2 |p, s, \sigma\rangle, \qquad  \qquad \qquad \ \
    P_\mu |p, s, \sigma\rangle \ =\  p_\mu |p, s, \sigma\rangle, \\
W^2 |p, s, \sigma\rangle &=& -m^2 s (s + 1) |p, s, \sigma\rangle, \quad
s_\mu W^\mu |p, s, \sigma\rangle \ =\  -m\sigma |p, s, \sigma\rangle,
\end{eqnarray}
where $P_\mu$ and $W_\mu$ denote the momentum and the Pauli-Lubanski pseudovector operators, respectively.

There are several possible choices of the spin axis for massive particles. Following Wigner's construction \cite{Wigner:1939cj}, the state $|p,\sigma\rangle$ is obtained by applying a standard boost $L_p$ that maps the reference momentum $k$, conventionally the rest-frame momentum for massive particles, to $p$: $|p,\sigma\rangle\equiv U(L_p)|k,\sigma\rangle$. The choice of $L_p$ is correlated with the choice of the spin axis. In this work we adopt the standard boost that defines \textit{collinear particle states}, i.e., states adapted to kinematics in which all momenta are nearly parallel to a fixed lightlike direction $n^\mu$.

\subsection{2D Galilean State on Light-front}

When a system is highly boosted along a particular direction, such as the $n$-direction with a large $P^+ \equiv n\cdot P$ momentum, the algebra of the Poincar\'e generators contracts to the 2D Galilean algebra in the transverse plane \cite{Weinberg:1966jm,Susskind:1967rg,Chang:1968bh,Bardakci:1968zqb,Kogut:1969xa,Soper:1971wn}.

We adopt the light-front coordinates
\begin{eqnarray}
    x^\pm = \frac{1}{2}(x^0 \pm x^3), \qquad x^i = (x^1, x^2),
\end{eqnarray}
in which the ten Poincar\'e generators take the form
\begin{equation}
\begin{aligned} \label{eq:lightfront_P}
P_+ &=  P_0 + P_3,\qquad& P_- &=  P_0 - P_3,\qquad& &P_i,&  \\
T_1 &=  K_1 - J_2,\qquad& 
T_2 &=  K_2 + J_1,\qquad& &K_3,& \\
F_1 &=  K_1 + J_2,\qquad& 
F_2 &=  K_2 - J_1,\qquad& &J_3.&
\end{aligned}
\end{equation}
Here $K_j$ and $J_j$ with $j=1,2,3$ denote the Lorentz boost and rotation generators, respectively.

In a highly boosted frame along the $n$ direction, $P_+$ is large, while $P_{-} = (P^i P_i + m^2)/2P_+$ becomes suppressed for a system with fixed transverse momentum and mass. Taking the $P_+ \to \infty$ limit while keeping $P_i/P_+$ finite, the Poincar\'e generators organize into the following identification with the 2D Galilean algebra:
\begin{eqnarray}
P_- &\to& \text{Hamiltonian}, \nonumber\\
P_i &\to& \text{momentum in the 2D plane}, \nonumber\\
J_3 &\to& \text{rotation around the } z\text{-axis}, \label{eq:galilean_id}\\
T_i &\to& \text{generator of boosts in the plane}, \nonumber\\
P_+ &\to& \text{mass (central charge)}. \nonumber
\end{eqnarray}
The commutation relations among the surviving generators are
\begin{eqnarray}
    [T_i,P_j]=iP_+\delta_{ij}, \quad [T_i,P_-]=2i P_i,\quad [J_3,T_i]=i\epsilon_{ij}T_j,\quad [J_3,P_i]=i\epsilon_{ij}P_j,
\end{eqnarray}
where $\epsilon_{ij}$ is the Levi-Civita tensor for the two-dimensional transverse space, and all other commutators vanish in the infinite-momentum limit. From these relations we identify $T_i$ as the Galilean boost generators, $P_i$ the Galilean momenta, $P_+$ the Galilean mass (a central charge), and $J_3$ the generator of transverse rotations. Since $P_+$ commutes with the 2D Galilean generators, it has a fixed value within each Galilean representation. We refer to this 2D Galilean group as $\text{Gal}(2)$ in the following.

For convenience, we reorganize the two transverse degrees of freedom into a complex one, so the generators and transverse momentum become
\begin{equation}
\begin{aligned}
T&=T_1+iT_2,& \quad
\bar{T}&=T_1-iT_2 \\
P_T&=P_1+iP_2,& \quad
P_T^*&=P_1-iP_2.
\end{aligned}
\end{equation}
For real momentum, $P_T$ and $P_T^*$ are Hermitian conjugates and are not independent. This means that the momentum can be labeled by the three quantities $(p_+,p_T, p_-)$, without explicitly emphasizing $p_T^*$.

Following Wigner's induced representation construction on the Bargmann group~\footnote{The Bargmann algebra is the central extension of the Galilei algebra, with the nonrelativistic mass appearing as a central charge \cite{Bargmann:1954gh,Levy-Leblond:1963qdx}.}, single-particle states with respect to the invariant subgroup of $P_+$ are classified by the boost $\Lambda$ and the SO(2) little group of rotations $R$ in the transverse plane on the standard momentum. The standard momentum is taken to be the one for the transverse rest particle $(p_+,p_-,p_T)=(\eta_3,V,0)$, where the central charge is $\eta_3$ and the potential energy is taken to be $V$. These two transformations can be written as
\begin{eqnarray}
   \Lambda &=&  \exp(ivT+iv^*\Bar{T}), \\
   R &=& \exp(i\varphi J_3)
\end{eqnarray}
The boost $\Lambda$ acts on the transverse rest particle and yields
   \begin{align}
       \exp(ivT+iv^*\Bar{T})&:\quad (\eta_3,V,0)\to(\eta_3,V+\frac{p_T^2}{\eta_3},p_T). \nonumber
   \end{align}
Here $T$ and $\bar{T}$ are now the Galilean boost generators. This defines a trajectory in the transverse momentum space. We choose a standard point on this trajectory and induce the Galilean representation from its little group. Taking the standard point to be the transverse rest particle, the little group is generated by the $SO(2)$ rotation $R$, which leaves
   \begin{align}
       \exp(i\varphi J_3)&:\quad (\eta_3,V,0)\to(\eta_3,V,0). 
   \end{align}
Here $J_3$ is the little-group generator of $\text{Gal}(2)$, and its eigenvalue denotes the quantum number $s_z$ for the 2D particle state. 

The representation of $\text{Gal}(2)$ can thus be defined by $\{\eta_3,V,s_z\}$, where $\eta_3$ and $V$ are invariant under Galilean transformation. Combining with $p_T$, we can define the general \textit{2D Galilean state}
\begin{equation}
    |\eta_3, V ; p_T,s_z\rangle=\exp(ivT+iv^*\Bar{T})|\eta_3, V;0,s_z\rangle,
\end{equation}
The variable $V \equiv p_- -p_T^2/(\eta_3)$ has the interpretation of the potential energy in the Galilean analogy, and $\eta_3$ serve as twice Galilean mass. Acting  Galilean generators on this Galilean state gives
\begin{equation}
\begin{aligned}
P_+ |\eta_3, V ; p_T,s_z\rangle &=\eta_3 |\eta_3, V ; p_T,s_z\rangle, \\
P_- |\eta_3, V ; p_T,s_z\rangle &=\left(V+\frac{p_T^2}{\eta_3}\right) |\eta_3, V ; p_T,s_z\rangle, \\
J_3 |\eta_3, V ; p_T,s_z\rangle &=s_z |\eta_3, V ; p_T,s_z\rangle.
\end{aligned}
\end{equation}

\subsection{Collinear Soper-Weinberg State}

We now consider particle states under the full Poincar\'e transformation. The full Poincar\'e algebra $\mathfrak{iso}(3,1)$ admits a coset decomposition with respect to the Galilean subgroup. The Galilean subalgebra is $\text{Gal}(2) = \{T_i, P_i, P^\pm, J_3\}$, while the coset generators are $\mathfrak{iso}(3,1)/\text{Gal}(2) = \{J_i, K_3\}$.~\footnote{The coset construction is not unique. Our choice is motivated by the fact that the coset generators have a natural relation to the Galilean rotation $J_3$: the $J_i$ together with $J_3$ form a closed subalgebra, and $K_3$ serves as a natural boost relative to $J_3$.} The latter do not preserve $P^+$ and therefore transform states between different Galilean representations.

The label $\{\eta_3,V,s_z\}$ is invariant under the action of the Galilean subgroup, but transforms nontrivially under the coset generators. The kinematic labels $\eta_3$ and $V$ transform under the coset boost $K_3$ as
\begin{equation}
    e^{i\zeta K_3}|\eta_3,V;0,s_z\rangle = |\eta_3 e^{\zeta}, V e^{-\zeta};0,s_z\rangle,
\end{equation}
which shows that $\eta_3$ and $V$ scale reciprocally. The spin component $s_z$ shifts by $\pm 1$ under the ladder operators $J_{\pm}=J_1 \pm i J_2$, 
\begin{equation}
    J_3\bigl(J_{\pm}|\eta_3,V;0,s_z\rangle\bigr) = (s_z\pm1)\bigl(J_{\pm}|\eta_3,V;0,s_z\rangle\bigr).
\end{equation}

A crucial observation is that the product $\eta_3 V \equiv m^2$ is invariant under the full Poincar\'e group. This product corresponds to composite operator $\hat{\eta}_3 \hat{V}=P_+ P_- - P_T P_T^*$, so this invariance can be verified by noting that $[J_\pm, P_+ P_-] = [J_\pm, P_T P_T^*] = \pm 2 P_3(P_1 \pm i P_2)$, from which it follows that
\begin{equation}
    \hat{\eta}_3\hat{V}\,J_\pm|\eta_3,V;0,s_z\rangle = J_\pm\hat{\eta}_3\hat{V}|\eta_3,V;0,s_z\rangle = m^2\,J_\pm|\eta_3,V;0,s_z\rangle,
\end{equation}
confirming that $m^2$ is a Poincar\'e-invariant. If $\eta_3 V = m^2 > 0$, the particle is massive and $|s_z| \leq s$. For the massless case $m^2 = 0$, the spin projection $s_z$ is fixed (the little group degenerates to the helicity subgroup).

A general standard boost can be constructed in the product form
\begin{equation}
    L_p = \exp(ivT+iv^*\Bar{T})\,\exp(i\zeta K_3),
\end{equation}
where the Galilean boost $\exp(ivT+iv^*\Bar{T})$ belongs to the subgroup and the longitudinal boost $\exp(i\zeta K_3)$ belongs to the coset. Here the order of these factors can be exchanged, with a corresponding redefinition of the parameters $v$ and $\zeta$. Within the Lorentz group, different choices of the standard boost lead to physically equivalent particle states that differ by a little-group rotation.

For collinear processes, we adopt the Soper-Weinberg (SW) states, defined via the 2D Galilean subgroup of the Poincar\'e group \cite{Soper:1971wn}. Specifically, a collinear particle state is constructed through the sequential actions
\begin{eqnarray}
    |(m,m,0),s_z\rangle \xrightarrow{B_z} |(p_+,\tfrac{m^2}{p_+},0),s_z\rangle \xrightarrow{\text{2D boost}} |(p_+,p_-,p_T),s_z\rangle \equiv |p, s_z\rangle.
    \label{eq:collinear_state}
\end{eqnarray}
Here $B_z=\exp(i\zeta K_3)$ denotes a longitudinal boost along the $z$-direction that takes the rest-frame momentum to a state with zero transverse momentum but non-zero $p_+$, and the 2D Galilean boost subsequently generates the transverse momentum $p_T$.

For completeness, the action of the general standard boost $L_p = \exp(i v T + i v^* \bar{T}) \exp(i \zeta K_3)$ on the light-front momentum is
\begin{equation}
\begin{aligned}
p_{+} &\to p_+, \\
p_{-} &\to p_- + 2 v (e^{-i \zeta} p_{T} + e^{i \zeta} p_{T}^*) + v^2 p_+, \\
p_{T} &\to p_{T} + v e^{i \zeta} p_{+}, \\
p_{T}^* &\to p_{T}^* + v e^{-i \zeta} p_{+},
\end{aligned}
\end{equation}
which confirms that  $p_+$ is preserved, as required by the Galilean subgroup structure.

\subsection{Collinear Spin-spinor Formalism}

In four-dimensional spacetime, the momentum of a particle admits a representation as a spinor bilinear \cite{Elvang:2015rqa}
\begin{eqnarray}
p_{\alpha\dot\alpha}=\lambda_{\alpha}^I\tilde{\lambda}_{\dot\alpha I},
\end{eqnarray}
which treats the spinor variables as more fundamental than the four-momentum itself.
Here $\lambda$ belongs to the left-handed $(\frac{1}{2},0)$ Lorentz representation, while $\tilde{\lambda}$ belongs to the right-handed $(0,\frac{1}{2})$ representation. The index $I$ is a little-group index: for massive particles, $I=+,-$ labels the fundamental representation of the $\mathrm{SU}(2)$ little group; for massless particles, the little group reduces to $\mathrm{U}(1)$ and the index is no longer carried explicitly. 

For both massive and massless particles, the spinors obey a little-group covariance condition: under a Lorentz transformation $\Lambda$, they transform as
\begin{eqnarray}
    \Lambda |p^I\rangle=|\Lambda p^J\rangle\, W_J^{\;I}(\Lambda,p),\qquad
    \Lambda |p^I]=|\Lambda p^J]\, W_J^{\;I}(\Lambda,p),
\end{eqnarray}
where $W_J^{\;I}(\Lambda,p)$ is the little-group Wigner rotation matrix. This redundancy under little-group redefinitions,
\begin{eqnarray}
     |p^I\rangle \to |\Lambda p^J\rangle\, R_J^{\;I},\qquad
     |p^I] \to |\Lambda p^J]\, R_J^{\;I},
\end{eqnarray}
implies that different choices of the spinor basis---corresponding to different standard boosts---are physically equivalent. 

We first focus on the left-handed spinor $\lambda^I$ for massive particles. In the rest frame, the spinor takes a natural diagonal form:
\begin{equation}
    (\lambda_{\text{rest}})^I_{a} =
    \begin{pmatrix}
        \sqrt{m} & 0  \\
        0 & \sqrt{m}
    \end{pmatrix},
\end{equation}
where the two columns ($I=+,-$) form an orthonormal basis in spinor space. Starting from this rest-frame spinor, one obtains the spinor for a general momentum $p$ by applying a Lorentz standard boost.

\begin{itemize}

\item \textbf{AHH boost (helicity basis).} The Arkani-Hamed--Huang--Huang (AHH) standard boost \cite{Arkani-Hamed:2017jhn} first boosts along the $z$-direction and then rotates to the desired direction:
\begin{equation}\begin{aligned}
    \mathcal{L}_p = \exp(-i \phi J_{3})\exp(-i \theta J_{2})\exp(i \phi J_{3})\,\exp(i\zeta K_3).
\end{aligned}\end{equation}
Applying this boost to the rest-frame spinor yields the AHH spinor,
\begin{equation}\begin{aligned} \label{eq:AHH_spinor}
    (\lambda_{\text{rest}})^I_{a} \;\xrightarrow{\mathcal{L}_{p}} \; (\lambda_{\text{AHH}})^I_{a}=
    \begin{pmatrix}
        \sqrt{E+p}\,c & -\sqrt{E-p}\,s^*\\
        \sqrt{E+p}\,s &  \sqrt{E-p}\,c
    \end{pmatrix},
\end{aligned}\end{equation}
where the angular variables are
\begin{equation}
    c=\cos \frac{\theta}{2},\qquad s=\sin \frac{\theta}{2}\, e^{i\phi},
\end{equation}
with $\theta$ the polar angle and $\phi$ the azimuthal angle of the momentum direction.

\item \textbf{Collinear boost (SW basis).} In this work we adopt the collinear standard boost---first a longitudinal boost along $z$, followed by a Galilean boost:
\begin{equation}\begin{aligned}
    L_p = \exp(ivT+iv^*\bar{T})\,\exp(i\zeta K_3).
\end{aligned}\end{equation}

Boosting the rest-frame spinor with $L_p$ produces the Soper-Weinberg (SW) spinor \cite{Soper:1971wn},
\begin{equation}\begin{aligned}  \label{eq:SW_spinor}
    (\lambda_{\text{rest}})^I_{a} \;\xrightarrow{L_{p}} \;
    (\lambda_{\text{SW}})^I_{a}=
    \begin{pmatrix}
        \sqrt{p_+} & 0\\[4pt]
        \dfrac{p_T}{\sqrt{p_+}} & \dfrac{m}{\sqrt{p_+}}
    \end{pmatrix}.
\end{aligned}\end{equation}
A distinctive feature of the SW spinor is that one off-diagonal element vanishes identically. As will become clear in the next section, this property is particularly advantageous in collinear kinematics because it manifestly separates the leading and subleading components in the power counting.

\end{itemize}

The two spinor bases are related by a momentum-dependent little-group (Wigner) rotation:
\begin{equation}
    (\lambda_{\text{SW}})_{a}^I = W^{I}_{\;J}\,(\lambda_{\text{AHH}})_{a}^J,
\end{equation}
with the transformation matrix given by
\begin{equation}\begin{aligned}
    W^I_{\;J}
    = \frac{\langle \lambda_{\text{SW}}^I \,\lambda_{\text{AHH},J} \rangle}{m}.
\end{aligned}\end{equation}
Thus, the freedom in choosing the standard boost is entirely absorbed into a little-group rotation, and physical observables are independent of this choice. To make the expression for this little-group transformation explicit, we substitute Eqs.~\eqref{eq:AHH_spinor} and \eqref{eq:SW_spinor}, yielding
\begin{equation} \label{eq:W_matrix}
W^I_{\;J}=\begin{pmatrix}
\sqrt{ \frac{E+p}{p_{+}} }c & \sqrt{ \frac{E-p}{p_{+}} }s^*  \\
-\sqrt{ \frac{E-p}{p_{+}} }s  & \sqrt{ \frac{E+p}{p_{+}} }c
\end{pmatrix},
\end{equation}
where we have used the relation between the two parametrizations to simplify the result, e.g. $p_{+}c+p_{T} s^*  = (E+p) c$. 
This matrix shows that the little-group index $I$ does not carry the same meaning for AHH and SW spinors. In general, a SW spinor with a given index $I$ can be expanded as a combination of AHH spinors $\lambda^{J=1}_{\text{AHH}}$ and $\lambda^{J=2}_{\text{AHH}}$. This mixing vanishes only in the massless limit $m\to 0$, where the off-diagonal terms in $W$ matrix disappear.

The right-handed spinors are obtained by applying the same two Lorentz boosts to the right-handed rest-frame spinor. For real momentum, the left- and right-handed spinors are related by complex conjugation, $\tilde{\lambda}_I = (\lambda^I)^*$, giving
\begin{equation}
(\tilde{\lambda}_{\text{AHH}})_{\dot{a} I}=
    \begin{pmatrix}
        \sqrt{E+p}\,c & -\sqrt{E-p}\,s\\
        \sqrt{E+p}\,s^* & \sqrt{E-p}\,c
    \end{pmatrix},\qquad
(\tilde{\lambda}_{\text{SW}})_{\dot\alpha I}=
    \begin{pmatrix}
        \sqrt{p_+} & 0\\[4pt]
        \dfrac{p_T^*}{\sqrt{p_+}} & \dfrac{m}{\sqrt{p_+}}
    \end{pmatrix}.
\end{equation}
Under conjugation, the upper index $I$ becomes a lower index, reflecting the fact that the conjugated spinors transform in the dual representation of the little group $\mathrm{SU}(2)$. Little-group indices are raised and lowered with the Levi-Civita tensor,
\begin{equation}
    \tilde{\lambda}^I = \epsilon^{IJ}\tilde{\lambda}_J.
\end{equation}

The little-group index structure provides a compact representation for massive particles of arbitrary spin. For a spin-$s$ particle, the wavefunction is constructed from $2s$ symmetrized spinors \cite{Arkani-Hamed:2017jhn,Conde:2016vxs}:
\begin{equation}\begin{aligned}
    \text{spin-}s:\quad
    \lambda^{(I_{1}}_{a_{1}}\lambda^{I_{2}}_{a_{2}} \cdots \tilde{\lambda}^{I_{2s-1}}_{\dot{a}_{2s-1}} \tilde{\lambda}^{I_{2s})}_{\dot{a}_{2s}},
\end{aligned}\end{equation}
where the parentheses $(I_{1}\cdots I_{2s})$ denote full symmetrization over the little-group indices, ensuring that the wavefunction transforms irreducibly under $\mathrm{SU}(2)$. This construction is independent of the specific boost used to define the spinors, provided that little-group covariance is maintained. When the SW spinor is adopted, the same symmetric product form describes spin-$s$ particles in a basis adapted to collinear kinematics.

Having established the collinear particle states and their spinor representations, we now turn to the decomposition of massive spinors onto a fixed basis of lightlike reference vectors---the central tool for collinear power counting.

%% file: sec3_projection.tex
\section{Collinear Projection and Power Counting}
\label{sec:collinear_projection}

In the previous section, we established the SW collinear particle states and their spinor representations. To compute splitting amplitudes, we need a systematic procedure for expanding massive spinors in terms of a small number of kinematical variables, such as mass, transverse momentum, or angle variables. The key insight is that a massive momentum can be decomposed into two lightlike reference vectors, and the choice of these reference vectors determines the power-counting organization of the high-energy expansion.

For a massive particle, the momentum can be expressed as a product of its mass $m$ and the four-velocity $u^\mu$,
\begin{equation}
p_{\mu}=m u^{\mu}.
\end{equation}
This timelike unit vector $u^\mu$ can always be decomposed into two lightlike vectors $q_1^\mu$ and $q_2^\mu$:
\begin{eqnarray} \label{eq:four-velocity}
   u^\mu = q_1^\mu + q_2^\mu.
\end{eqnarray}
The normalization condition $u^2=1$ for the four-velocity then requires the two null vectors to satisfy $2 q_1\cdot q_2=1$. In spinor representation, this condition can be written as
\begin{eqnarray}
    \langle q_1 q_2\rangle=[q_2 q_1]=1,
\end{eqnarray}
which implies these two spinors are linearly independent. Thus, we can use them to define the identity operator in the spinor space:
\begin{eqnarray}
    \mathbb{I}=|q_2\rangle\langle q_1|-|q_1\rangle\langle q_2|,\\
    \mathbb{I}=-|q_2][ q_1|+|q_1][q_2|.
\end{eqnarray}
This is the spinor-helicity analogue of the completeness relation for the basis vectors $q_1$ and $q_2$. Applying the identity operator to the massive spinor $\lambda^I$ or $\tilde{\lambda}^I$ gives an expansion in this basis:
\begin{equation} \begin{aligned} \label{eq:identity_expand}
|\lambda^I\rangle &= \mathbb{I}|\lambda^I\rangle=|q_2\rangle\langle q_1\lambda^I\rangle- |q_1\rangle \langle q_2\lambda^I\rangle, \\
|\lambda^I] &=\mathbb{I}|\lambda^I]=-|q_2][q_1\lambda^I] + |q_1][ q_2\lambda^I].
\end{aligned} \end{equation}
Similarly, these two spinors together with their conjugates form four vector basis elements:
\begin{eqnarray}
    q_1 &=& |q_1\rangle[q_1|,\\
    q_2 &=& |q_2\rangle[q_2|,\\
    q_+ &=& |q_1\rangle[q_2|,\\
    q_- &=& |q_2\rangle[q_1|,
\end{eqnarray}
An arbitrary Lorentz vector can be expanded as
\begin{eqnarray}
    V_{\alpha\dot\alpha}=V_{T} |q_1\rangle[q_2|+V_{+} |q_1\rangle[q_1|+V_{\bar{T}} |q_2\rangle[q_1|+V_{-} |q_2\rangle[q_2|,
\end{eqnarray}
with the coefficients given by
\begin{eqnarray}
    V_{T}=-\langle q_2|V|q_1],\quad  V_{+}=\langle q_2|V|q_2],\quad V_{\bar{T}}=-\langle q_1|V|q_2],\quad  V_{-}=\langle q_1|V|q_1].
\end{eqnarray}
Here, $q_1$ and $q_2$ can be complex null vectors, so these four components in this case are independent. For a real vector $V^\mu$, the components satisfy $(V_T)^*=V_{\bar{T}}$ and $(V_\pm)^*=V_\pm$.

The choice of $q_1,q_2$ is directly related to the spin quantization axis. To see this, we can consider a massive particle with momentum $p$, which may point along a direction different from eq.~\eqref{eq:four-velocity}. Applying the decomposition to this massive momentum yields:
\begin{eqnarray} \label{eq:p_decom}
    p_{\alpha\dot\alpha}=p_{T} |q_1\rangle[q_2|+p_{+} |q_1\rangle[q_1|+p_{\bar{T}} |q_2\rangle[q_1|+p_{-} |q_2\rangle[q_2|.
\end{eqnarray}
For a massive particle, the spin quantization axis must be a spacelike vector orthogonal to this momentum. By extracting the two momentum components $p_+$ and $p_-$, we can define the spin axis to be
\begin{equation} \label{eq:spin_axis}
\mathbf{n}_{\alpha\dot{\alpha}}=\frac{1}{\sqrt{p_+ p_-}}(p_+|q_1\rangle[q_1| - p_- |q_2\rangle[q_2|).
\end{equation}
One can verify that this is a normalized spacelike vector, $\mathbf{n}\cdot \mathbf{n}=-1$, and orthogonal to the momentum, $p \cdot \mathbf{n}=0$. The spin state of the massive particle is described by the Pauli-Lubanski pseudovector $W^\mu$. In the spinor representation, it is given by
\begin{equation}
W_{a\dot{a}}=p^c_{\dot{a}}\lambda^I_{(c}\frac{\partial}{\partial \lambda^{Ia)}}-p^{\dot{c}}_{a}\tilde\lambda^I_{(\dot{c}} \frac{\partial}{\partial \tilde\lambda^{I\dot{a})}},
\end{equation}
Contracting it with the spin axis, we can project the spin along the chosen axis $\mathbf{n}$. This leads to the spin projection operator
\begin{equation}
S_{\mathbf{n}} =\frac{1}{m} \mathbf{n}\cdot W.
\end{equation}
where the factor of $1/m$ ensures that the eigenvalues are the half-integer spin quantum numbers. Substituting eq.~\eqref{eq:spin_axis} and doing some calculations, one can express this operator explicitly in terms of $q_1$ and $q_2$,
\begin{equation}
S_{\mathbf{n}} 
=\frac{1}{2\sqrt{ p_{+}p_{-} }}\left[ p_{+} \left([q_{1} \tilde{\lambda}^I] q^a_{1}\frac{\partial}{\partial \lambda^{Ia}} 
+\langle \lambda^I q_{1}\rangle \tilde{q}_{1}^{\dot{a}} \frac{\partial}{\partial \tilde\lambda^{I\dot{a}}}\right)
-p_{-} \left([q_{2}\tilde{\lambda}^I] q^a_{2}\frac{\partial}{\partial \lambda^{Ia}} 
+\langle \lambda^I q_{2}\rangle \tilde{q}_{2}^{\dot{a}} \frac{\partial}{\partial \tilde\lambda^{I\dot{a}}}\right) \right] .
\end{equation}
Here we used the equation of motion to convert momentum $p$ into mass $m$ in the definition of the Pauli-Lubanski pseudovector $W$. Combining with the factor $1/m$ in the spin projection operator, the resulting expression becomes independent of $m$.

Different choices of $q_1,q_2$ correspond to different spin quantization axes. In this section, we focus on the case where the two lightlike vectors $q_1^\mu = |q_1\rangle[q_1|$ and $q_2^\mu = |q_2\rangle[q_2|$ point in opposite spatial directions, so that the transformation between different bases is given by a little-group rotation. We examine two specific choices: the helicity decomposition and the collinear decomposition.

\subsection{On-shell Helicity Decomposition}

We first consider the decomposition by choosing the two null vectors as
\begin{equation}
q_1^\mu=\frac{1}{\sqrt{2-2\vec{\mathbf{e}}_1\cdot \vec{\mathbf{e}}_2}}(1,\vec{\mathbf{e}}_1),\quad 
q_2^\mu=\frac{1}{\sqrt{2-2\vec{\mathbf{e}}_1\cdot \vec{\mathbf{e}}_2}}(1,\vec{\mathbf{e}}_2).
\end{equation}
where $\vec{\mathbf{e}}_1$ and $\vec{\mathbf{e}}_2$ are two unit 3-vectors. The normalization factor $1/\sqrt{2-2\vec{\mathbf{e}}_1\cdot \vec{\mathbf{e}}_2}$ ensures that these vectors satisfy $2q_1\cdot q_2=1$. Consider a massive momentum
\begin{equation}
p^\mu=(E,\vec{p}),
\end{equation}
where $\vec{p}$ denotes the 3-momentum and $p=|\vec{p}|$ its magnitude. Using eq.~\eqref{eq:p_decom}, the massive momentum can be decomposed into four components. Focus on two components $p_+$ and $p_-$, we have 
\begin{equation} \begin{aligned} \label{eq:helicity_component}
p_+ &=\langle q_2|p|q_2] = \sqrt{\frac{2}{1-\cos\beta}} (E+p\cos\alpha_2) , \\
p_- &=\langle q_1|p|q_1] = \sqrt{\frac{2}{1-\cos\beta}} (E-p\cos\alpha_1) ,
\end{aligned} \end{equation}
where $\alpha_i$ is the angle between $\vec{\mathbf{e}}_i$ and $\vec{p}$, and $\beta$ is the angle between $\vec{\mathbf{e}}_1$ and $\vec{\mathbf{e}}_2$. These two components are subject to the constraints
\begin{equation}
E-p \;\; \le \;\; p_+,p_- \; \le \;\; E+p.
\end{equation}

\begin{figure}[htbp]
\centering
\includegraphics[width=0.49\linewidth]{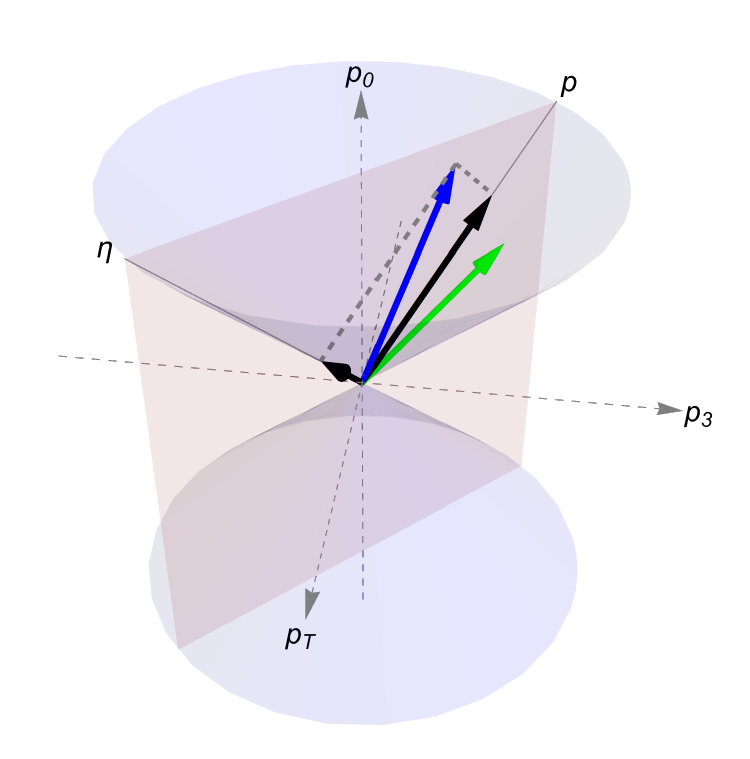}
\caption{Decomposition of a massive momentum (blue arrow) into two massless momenta (black arrows) on the light cone. In the helicity decomposition, the two massless momenta correspond to $p$ and $\eta$, spanning a pale red plane in momentum space. Both the massive momentum and the helicity spin axis (green arrow) lie in this plane.}
\label{fig:helicity_project}
\end{figure}

To define the helicity decomposition, we want it to maximally separate the large and small degrees of freedom in the high-energy regime. Therefore, we require that $p_+$ attains its maximum $(E+p)$ and $p_-$ its minimum $(E-p)$. From eq.~\eqref{eq:helicity_component}, this condition is achieved when the two lightlike vectors $q_1$ and $q_2$ align with and opposite to the momentum direction. This uniquely fixes the choice of basis vectors $q_1$ and $q_2$,
\begin{equation}
q_1^\mu=\frac{1}{2}\left(1,\frac{\vec{p}}{p}\right),\quad 
q_2^\mu=\frac{1}{2}\left(1,-\frac{\vec{p}}{p}\right).
\end{equation}
In this case, the massive momentum has a simple decomposition form 
\begin{equation}
p= (E+p) |q_1\rangle[q_1|+ (E-p) |q_2\rangle[q_2|,
\end{equation}
in which all transverse components vanish,
\begin{equation}
p_T=p_{\bar{T}}=0.
\end{equation}
This amounts to decomposing a massive momentum into two massless momenta, as illustrated in figure~\ref{fig:helicity_project}. The spin axis is then given by
\begin{equation}
\mathbf{n}_{\alpha\dot{\alpha}}=\frac{1}{m}((E+p)|q_1\rangle[q_1| - (E-p) |q_2\rangle[q_2|).
\end{equation}
It points in the same spatial direction as the momentum.
One can verify this expression by considering the rest frame, in which the spin axis reduces to $\mathbf{n}^{\mu}=(0,\vec{p}/p)$, confirming that it is a spacelike vector. In the high-energy limit $E\sim p$, the spin axis approaches the lightlike vector $q_1$.

To construct the explicit spinors, we define two spinors to absorb the factors $E+p$ and $E-p$,
\begin{equation} \begin{aligned}
|\lambda\rangle &=\sqrt{E+p} |q_1\rangle, \\
|\eta\rangle &=\sqrt{E-p} |q_2\rangle.
\end{aligned} \end{equation}
These spinors are referred to as the basis spinors of the AHH formalism. To derive their explicit form, we start from the rest-frame spinors and apply a standard Lorentz boost. In the rest frame, the massless spinors along and opposite the $z$-axis are
\begin{eqnarray}
    \lambda_0=\sqrt{m}\begin{pmatrix}
        1\\0
    \end{pmatrix},\quad \eta_0=\sqrt{m}\begin{pmatrix}
        0\\1
    \end{pmatrix}.
\end{eqnarray}
In this frame, it is convenient to reorganize the decomposition basis as
\begin{eqnarray}
    |\lambda_0\rangle[\lambda_0|+|\eta_0\rangle[\eta_0|&=&m(1,0,0,0),\\
    |\lambda_0\rangle[\lambda_0|-|\eta_0\rangle[\eta_0|&=&m(0,0,0,1),\\
    |\lambda_0\rangle[\eta_0|&=&m(0,1,i,0),\\
    |\eta_0\rangle[\lambda_0|&=&m(0,1,-i,0).
\end{eqnarray}
The first combination represents the rest-frame massive momentum, while the remaining three correspond to the three spatial polarization directions.

We then perform the AHH boost $\mathcal{L}_p$ to obtain the spinors in a general frame:
\begin{eqnarray}
    \lambda=\begin{pmatrix}
        \sqrt{E+p} c \\ \sqrt{E+p} s
    \end{pmatrix},\quad \eta=\begin{pmatrix}
        -\sqrt{E-p} s^* \\  \sqrt{E-p} c
    \end{pmatrix}.
\end{eqnarray}
Substituting these into the decomposition yields the basis in a general frame,
\begin{eqnarray}
    |\lambda\rangle[\lambda|+|\eta\rangle[\eta|&=&p,\\
    |\lambda\rangle[\lambda|-|\eta\rangle[\eta|&=&m\varepsilon_0,\\
    |\lambda\rangle[\eta|&=&m\varepsilon_+,\\
    |\eta\rangle[\lambda|&=&m\varepsilon_-.
\end{eqnarray}
where $\varepsilon_0$ represents the longitudinal polarization vector and $\varepsilon_\pm$ represent the transverse polarization vectors. Thus, in an arbitrary frame, these four bilinears provide the momentum and the three polarization vectors.

Now the massive AHH spinors $\lambda^I_{\text{AHH}}$ and $\tilde\lambda^I_{\text{AHH}}$ can be decomposed as
\begin{equation}
\begin{aligned}
|\lambda_{\text{AHH}}^-\rangle &=|\lambda\rangle,& |\lambda_{\text{AHH}}^+\rangle&=|\eta\rangle, \\
|\tilde{\lambda}_{\text{AHH}}^+] &=-|\tilde{\lambda}],& |\tilde{\lambda}_{\text{AHH}}^-] &=|\tilde{\eta}],
\end{aligned}
\end{equation}
where the index $I=\pm$ is related to the helicity of the massive particle. This is why we refer to it as the name \textit{helicity decomposition}. The two components $I=\pm$ provide a diagonal basis for the spin projection operator $S_{\mathbf{n}}$, which then takes the form
\begin{equation}
\begin{aligned}
S_{\mathbf{n}}&=\frac{1}{2m}\left( [\lambda \tilde{\lambda}^I ] \lambda^a\frac{\partial}{\partial \lambda^{Ia}} 
+\langle \lambda^I \lambda \rangle \tilde{\lambda}^{\dot{a}} \frac{\partial}{\partial \tilde\lambda^{I\dot{a}}} 
- [\eta \tilde{\lambda}^I ] \eta^a\frac{\partial}{\partial \lambda^{Ia}} 
-\langle \lambda^I \eta\rangle \tilde{\eta}^{\dot{a}} \frac{\partial}{\partial \tilde\lambda^{I\dot{a}}} \right)  \\
&=\frac{1}{2}\left(-\lambda^a\frac{\partial}{\partial \lambda^{a}} 
+ \tilde{\lambda}^{\dot{a}} \frac{\partial}{\partial \tilde\lambda^{\dot{a}}} 
+  \eta^a\frac{\partial}{\partial \eta^{a}} 
- \tilde{\eta}^{\dot{a}} \frac{\partial}{\partial \tilde\eta^{\dot{a}}} \right),
\end{aligned}
\end{equation}
where we use $\langle \lambda\eta\rangle=[\eta \lambda]=m$ to eliminate the mass dependence. Applying it to the two states gives
\begin{equation}
\begin{aligned}
S_{\mathbf{n}}\circ |\lambda_{\text{AHH}}^-\rangle &=-\frac{1}{2} |\lambda_{\text{AHH}}^+\rangle, \\
S_{\mathbf{n}}\circ |\tilde{\lambda}_{\text{AHH}}^+\rangle &=+\frac{1}{2}|\lambda_{\text{AHH}}^+\rangle. 
\end{aligned}
\end{equation}
Hence, the $I=-$ component corresponds to the spin projection eigenvalue $-1/2$, while the $I=+$ component corresponds to $+1/2$. This confirms that we can identify the component $I=\pm$ with the helicity of the massive spinor.

\subsection{On-shell Collinear Decomposition}

While the helicity decomposition is natural for single-particle states, it has a significant drawback for collider-physics applications: the reference vectors $q_1$ and $q_2$ track the momentum direction of each individual particle. In a multi-particle collinear splitting process, different particles have slightly different momentum directions, making it difficult to compare their spinor components on a common footing. The collinear decomposition resolves this by adopting \textit{fixed} reference vectors.

In the collinear decomposition, we take $q_1 = n$ to be a fixed lightlike vector along the collinear direction, and $q_2 = \bar{n}$ to be another fixed lightlike vector.
\begin{equation}
n^\mu=\frac{1}{\sqrt{2-2\vec{n}\cdot \vec{n}^\prime}}\left(1,\vec{n}\right),\quad 
\bar{n}^\mu=\frac{1}{\sqrt{2-2\vec{n}\cdot \vec{n}^\prime}}\left(1,\vec{n}^\prime\right).
\end{equation}
where $\vec{n}$ and $\vec{n}^\prime$ are two unit 3-vectors. The crucial difference from the helicity decomposition is that the basis vectors $n$ and $\bar{n}$ are independent of the momentum of any individual particle. This means they define a global reference frame for the entire splitting process.

In the general collinear decomposition, the $\bar{n}$-direction and collinear direction $n$ need not be related. Here, however, we choose the basis to maximally separate the large and small degrees of freedom. To illustrate this choice, consider the expansion of momentum. With the fixed reference spinors $|n\rangle$ and $|\bar{n}\rangle$, a real-valued momentum $p_{\alpha\dot\alpha}$ can be expanded as
\begin{eqnarray}
    p_{\alpha\dot\alpha}=p_{T} |n\rangle[\bar{n}|+p_{+} |n\rangle[n|+p_{T}^* |\bar{n}\rangle[n|+p_{-} |\bar{n}\rangle[\bar{n}|.
\end{eqnarray}
In the collinear decomposition, the momentum can have nonzero transverse momentum $p_T$. When the particle moves along the collinear direction $n^\mu$, its 3-momentum becomes $\vec{p}=p\vec{n}$. In this case, the coefficient of $n^\mu$ becomes the largest degree of freedom,
\begin{equation}
p_+=\langle \bar{n}|p|\bar{n}]= \sqrt{\frac{2}{1-\cos\alpha}}(E+p \cos \alpha) 
\end{equation}
where $\alpha$ is the angle between 3-vectors $\vec{n}$ and $\vec{n}^\prime$. Maximizing $p_+$ requries requires $\bar{n}$ to point opposite to the collinear direction $n$, so we have
\begin{equation}
n^\mu=\frac{1}{2}\left(1,\vec{n}\right),\quad 
\bar{n}^\mu=\frac{1}{2}\left(1,-\vec{n}\right).
\end{equation}

\begin{figure}[htbp]
\centering
\includegraphics[width=0.49\linewidth]{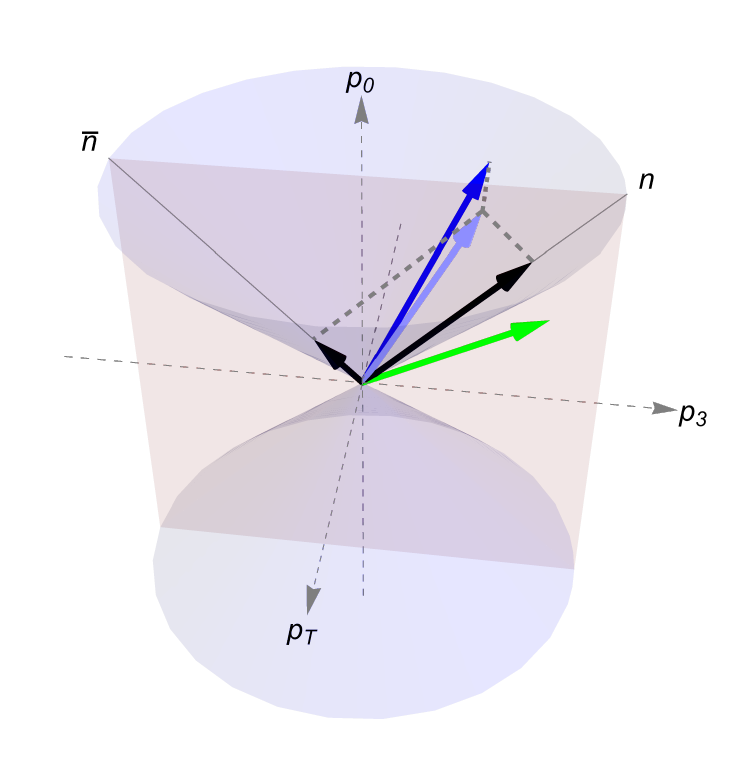}
\caption{Decomposition of a massive momentum (blue arrow) into massless momenta on the light cone. In the collinear decomposition, one first subtracts the transverse momentum; the resulting vector (light blue arrow) can then be decomposed into two massless vectors $n$ and $\bar{n}$ (black arrows). The collinear spin axis (green arrow) lies in the $n-\bar{n}$ plane, while the massive momentum does not.}
\label{fig:collinear_project}
\end{figure}

In this case, the two basis vectors $n$ and $\bar{n}$ span a plane in momentum space, which is illustrated in figure~\ref{fig:collinear_project}. The spin axis also lies in this plane and is given by
\begin{equation}
\mathbf{n}_{\alpha\dot{\alpha}}=\frac{1}{\sqrt{p_+ p_-}}(p_+|n\rangle[n| - p_- |\bar{n}\rangle[\bar{n}|).
\end{equation}
Unlike the helicity decomposition discussed in the previous subsection, the massive momentum now generally does not lie in this plane, since it can have non-zero transverse momentum $p_T$. This feature distinguishes the collinear decomposition from the helicity one.

To make the expansion explicit, we first choose a concrete representation for the reference spinors. On a convenient basis, we set
\begin{eqnarray}
    |n\rangle=\begin{pmatrix}
    1\\0
    \end{pmatrix},\quad |\bar{n}\rangle=\begin{pmatrix}
    0\\1
    \end{pmatrix}.
\end{eqnarray}

Recall that, for a massive particle with momentum $p$, the SW spinors with little group index $I=\pm$ take the form
\begin{eqnarray}\label{eq:CollinearSpinor}
    (\lambda_{\text{SW}})^-_\alpha=\begin{pmatrix}
        \sqrt{p_+} \\
        \frac{p_T}{\sqrt{p_+}} 
    \end{pmatrix},\quad 
    (\lambda_{\text{SW}})^+_\alpha=\begin{pmatrix}
        0\\
        \frac{m}{\sqrt{p_+}}
    \end{pmatrix}.
\end{eqnarray}
They are a particular solution to the massive Dirac equation in a collinear frame, and are characterized by the fact that the mass dependence is isolated in the $I=+$ component, while $p_T$ enters only through $I=-$. The contractions between SW spinors and the lightlike reference spinors $|n\rangle$, $|\bar{n}\rangle$ are 
\begin{eqnarray}
    \langle n\lambda^-_{\text{SW}}\rangle= \frac{p_T}{\sqrt{p_+}},&&\quad
    \langle \bar{n}\lambda^-_{\text{SW}}\rangle= -\sqrt{p_+},\\
    \langle n\lambda^+_{\text{SW}}\rangle=\frac{m}{\sqrt{p_+}}&,&\quad\langle \bar{n}\lambda^+_{\text{SW}}\rangle=0.
\end{eqnarray}
As shown in eq.~\eqref{eq:identity_expand}, the collinear spinors can be expanded in the fixed basis $n$, $\bar{n}$ using the identity operator, in which the coefficients are precisely the contractions. We have
\begin{eqnarray}
    |\lambda^-_{\text{SW}}\rangle= \sqrt{p_+}|n\rangle+\frac{p_T}{\sqrt{p_+}}|\bar{n}\rangle,\quad 
    |\lambda^+_{\text{SW}}\rangle=\frac{m}{\sqrt{p_+}}|\bar{n}\rangle.
\end{eqnarray}
Similarly, we can derive the decomposition for the right-handed SW spinor in the same way,
\begin{eqnarray}
    -|\lambda^+_{\text{SW}}]= \sqrt{p_+}|n]+\frac{p_T^*}{\sqrt{p_+}}|\bar{n}],\quad 
    |\lambda^-_{\text{SW}}]=\frac{m}{\sqrt{p_+}}|\bar{n}].
\end{eqnarray}

The spin projection operator in this basis is
\begin{equation}
S_{n}=\frac{1}{2\sqrt{p_{+}p_{-} }}\left[ p_{+} \left([n\tilde{\lambda}^I] n^a\frac{\partial}{\partial \lambda^{Ia}} 
+\langle \lambda^I n\rangle n^{\dot{a}} \frac{\partial}{\partial \tilde\lambda^{I\dot{a}}}\right)
-p_{-} \left([\bar{n} \tilde{\lambda}^I ] \bar{n}^a\frac{\partial}{\partial \lambda^{Ia}} 
+\langle \lambda^I \bar{n}\rangle \bar{n}^{\dot{a}} \frac{\partial}{\partial \tilde\lambda^{I\dot{a}}}\right) \right].
\end{equation}
Applying it to the state, we find
\begin{equation}
\begin{aligned}
S_{n}\circ |\lambda^+_{\text{SW}}\rangle
&=\frac{1}{2\sqrt{ p_{-} }} \left( p_T^* |n\rangle+p_{-}|\bar{n}\rangle \right)& 
&\not\propto |\lambda^+_{\text{SW}}\rangle, \\ 
S_{n}\circ |\lambda^-_{\text{SW}}\rangle
&=-\frac{1}{2\sqrt{ p_{-} }} m|n\rangle& 
&\not\propto |\lambda^-_{\text{SW}}\rangle.
\end{aligned}
\end{equation}
Therefore, the $I=\pm$ components of SW spinors are not eigenstates of the spin projection operator, and the index $\pm$ does not equal the helicity of these states. This confirms that the SW basis does not diagonalize the helicity operator, and the collinear spinors are a mixture of helicity states.

For a massless particle, the SW spinor reduces to a single Weyl spinor,
\begin{eqnarray}
    (\lambda_{\text{SW}})_\alpha=\begin{pmatrix}
        \sqrt{p_+} \\
        \frac{p_T}{\sqrt{p_+}}
    \end{pmatrix},\quad
    (\tilde{\lambda}_{\text{SW}})_\alpha=\begin{pmatrix}
    \sqrt{p_+} \\
    \frac{p_T^*}{\sqrt{p_+}}
    \end{pmatrix},
\end{eqnarray}
The spinor projection gives
\begin{eqnarray}
    \langle n\lambda_{\text{SW}}\rangle= \frac{p_T}{\sqrt{p_+}},\quad
    \langle \bar{n}\lambda_{\text{SW}}\rangle= \sqrt{p_+},\\
    \relax
    [n\lambda_{\text{SW}}]= \frac{p_T}{\sqrt{p_+}},\quad
    [\bar{n}\lambda_{\text{SW}}]= \sqrt{p_+}.
\end{eqnarray}
Then we derive the following massless-spinor expansion
\begin{eqnarray}
    |\lambda_{\text{SW}}\rangle &=& \sqrt{p_+}|n\rangle+\frac{p_T}{\sqrt{p_+}}|\bar{n}\rangle,\\
    |\lambda_{\text{SW}}] &=& \sqrt{p_+}|n]+\frac{p_T^*}{\sqrt{p_+}}|\bar{n}].
\end{eqnarray}

\subsection{Power Counting and Alignment Limit}

The helicity and collinear decompositions organize the degrees of freedom differently, and therefore yield distinct power-counting behaviors in generic kinematics. In this subsection, we characterize the high-energy scaling of each decomposition and demonstrate that they coincide in a special kinematic regime, the \textit{alignment limit}, defined by $m < p_T  \ll p_+$. In this limit, the two decompositions become interchangeable for the purpose of computing leading and subleading collinear observables.

Let us first discuss why these two decompositions give different degrees of freedom in general. As discussed in the previous subsection, the helicity decomposition moves with the spatial direction of the massive particle, while the collinear decomposition does not. This fundamental difference implies that the AHH spinors (used in the helicity decomposition) involve a single momentum scale $E$, whereas the SW spinors (used in the collinear decomposition) involve two distinct scales: $p_{T}$ and $p_{+}$. When combined with the mass scale $m$, these different kinematic variables lead to distinct power-counting behaviors for massive spinors in the high-energy expansion.

For helicity decomposition, in the high-energy limit $E\gg m$, the basis spinors exhibit the following scaling behavior:
\begin{eqnarray}
    |\lambda\rangle\sim \sqrt{E}\left(1+\mathcal{O}(\frac{m}{E})\right),\quad |\eta\rangle\sim \frac{m}{\sqrt{E}}\left(1+\mathcal{O}(\frac{m}{E})\right).
\end{eqnarray}
This means that the helicity decomposition gives a good high-energy expansion for AHH massive spinors
\begin{eqnarray}
    |\lambda^I_{\text{AHH}}\rangle=\frac{\langle \lambda^I_{\text{AHH}}\eta\rangle}{m}|\lambda\rangle+\frac{\langle\lambda \lambda^I_{\text{AHH}}\rangle}{m}|\eta\rangle
    \;\sim\; \sqrt{E}\left(1+\frac{m}{E}+\mathcal{O}(\frac{m^2}{E^2})\right),
\end{eqnarray}
It provides a straightforward power counting for massive structures: One can find the leading components survive in the high-energy regime while the flipped-helicity components will be suppressed by powers of $m/E$.

In the collinear decomposition, the basis spinor gives $|n\rangle\sim |\bar{n}\rangle\sim1$, so the power counting information is encoded in the coefficient. Nearly collinear momenta satisfy $p_+\gg p_T$, while the high-energy limit requires $p_+ \gg m$. The SW spinor therefore scales as
\begin{eqnarray}
    |\lambda_{\text{SW}}^I\rangle=\langle \lambda_{\text{SW}}^I \bar{n}\rangle|n\rangle+\langle n \lambda_{\text{SW}}^I\rangle|\bar{n}\rangle
    \;\sim\; \sqrt{p_+}+\frac{p_T}{\sqrt{p_+}}+\frac{m}{\sqrt{p_+}} ,
\end{eqnarray}

In summary, the high-energy expansion in the above two decompositions can be schematically written as
\begin{equation}\begin{aligned}
\text{helicity}:&\quad \sum_{a=0}^{\infty} \sqrt{E} \left(\frac{m}{E} \right)^a , \\
\text{collinear}:& \quad \sum_{b+c=0,1} \sqrt{p_{+}} \left(\frac{p_{T}}{p_{+}}\right)^b \left(\frac{m}{p_{+}} \right)^c .
\end{aligned}\end{equation}
A key observation is that the helicity expansion contains an infinite number of terms (in principle up to $a \to \infty$), while the collinear expansion truncates at finite order (here up to $b+c=1$), reflecting its more efficient organization of power corrections.

Since these two decompositions naturally work in different kinematic regimes, they cannot be compared directly. However, one can align them by considering the so-called alignment limit, defined by $m < p_{T} \ll p_{+}$. In this kinematic regime, the helicity decomposition yields simple angular approximations:
\begin{equation}\begin{aligned}
\cos \theta\sim 1,\quad \sin \theta \sim \theta ,
\end{aligned}\end{equation}
where we omit the high-order terms for $\theta$. Therefore, the AHH spinors take the approximate forms
\begin{equation}\begin{aligned}
\lambda \sim\begin{pmatrix}
        \sqrt{E}  \\ \sqrt{E} \theta
    \end{pmatrix}, \quad
\eta \sim\begin{pmatrix}
        -\frac{m}{\sqrt{E}} \theta \\  \frac{m}{\sqrt{E}} 
    \end{pmatrix}.
\end{aligned}\end{equation}
Plugging these into the helicity expansion formula, we obtain
\begin{equation}\begin{aligned} \label{eq:helicity_PC}
|\lambda^I_{\text{AHH}}\rangle 
&\sim \sqrt{E}\left( 1 + \theta + \frac{m}{E} + \mathcal{O}\left(\theta, \frac{m}{E}\right) \right).
\end{aligned}\end{equation}
This is an infinite expansion for both $\theta$ and $\frac{m}{E}$.

On the other hand, in the collinear decomposition, the massive spinor expansion contains only three terms:
\begin{equation}\begin{aligned} \label{eq:collinear_PC}
|\lambda_{\text{SW}}^I\rangle \sim \sqrt{p_{+}}\left( 1 + \frac{p_{T}}{p_{+}} + \frac{m}{p_{+}} \right).
\end{aligned}\end{equation}
Noting that in the alignment limit we have $\theta \sim p_{T}/p_{+}$ and $p_{+}\sim E$, it becomes evident that the two expansions in Eqs.~\eqref{eq:helicity_PC} and \eqref{eq:collinear_PC} are consistent with each other, term by term up to the first three orders. This consistency reveals an important practical advantage: the collinear decomposition automatically implements the high-energy expansion when the condition $m < p_{T} \ll p_{+}$ is satisfied. In other words, one does not need to re-express the collinear result in terms of $E$ and perform an additional expansion; the power counting is already manifest in the variables $p_{+}$ and $p_{T}$.

Finally, in the massless limit $m\to 0$, both decompositions reduce to the same set of massless spinors. 
Explicitly, the massive spinors from both formalisms become identical:
\begin{equation}\begin{aligned}
|\lambda^-_{\text{AHH}}\rangle,\;|\lambda^-_{\text{SW}}\rangle \quad&\to \quad
\begin{pmatrix}
\sqrt{ 2E }c \\
\sqrt{ 2E }s
\end{pmatrix}=
\begin{pmatrix}
\sqrt{p_+} \\
\frac{p_T}{\sqrt{p_+}}
\end{pmatrix}, \\
|\lambda^+_{\text{AHH}}\rangle,\;|\lambda^+_{\text{SW}}\rangle \quad&\to \quad 
0.
\end{aligned}\end{equation}
This convergence in the massless limit further solidifies the equivalence between the two decomposition schemes in their respective domains of validity.

With the collinear decomposition and its power-counting properties firmly established, we now have the necessary tools to construct the three-point massive amplitudes that serve as the elementary vertices for collinear splitting processes. In the next section, we apply the SW spinors and the $n$-$\bar{n}$ projection to systematically build the Standard Model three-point massive amplitudes and establish their correspondence with massless amplitudes.

%% file: sec4_3pt.tex
\section{Massive 3-point Amplitudes}
\label{sec:3pt}

The collinear decomposition and power counting established in the previous section allow us to systematically construct three-point massive amplitudes, which are the elementary building blocks from which collinear splitting functions are assembled. In this section, we first present the Standard Model massive 3-point amplitudes in a compact ``bolded" notation that incorporates the symmetrization of little-group indices. We then expand them using SW spinors, isolate the high-energy growth of longitudinal-vector components and organize its cancellation using Goldstone equivalence, and relate the resulting massive structures to massless 3-point and 4-point amplitudes.

A technical remark is in order before we proceed. The algebraic form of the 3-point amplitudes presented below is identical whether expressed in AHH or SW spinors. However, the physical interpretation of the little-group index $I$ differs between the two bases: 
\begin{itemize}
\item In the AHH (helicity) basis, $I=\pm$ labels helicity eigenstates.
\item in the SW (collinear) basis, $I=\pm$ labels the leading and subleading collinear spinor components of $\tilde{\lambda}^I$, with the assignment reversed for $\lambda^I$ ($I=\mp$).
\end{itemize} 
The two bases are related by a momentum-dependent Wigner rotation, and the amplitudes can be freely translated between them. Throughout this section, amplitudes are constructed using SW spinors, so that the power-counting properties derived in section~\ref{sec:collinear_projection} are directly applicable.

\subsection{Bolded Form and the Goldstone Equivalence Gauge}
\label{sec:bold}

We consider the massive amplitude. The on-shell scattering amplitude can be expressed in terms of the spinor products of $\lambda^I$ or $\tilde\lambda^I$ associated with different particles:
\begin{equation}
\langle 1^I 2^J \rangle\equiv \lambda_1^{Ia} \lambda_2^{J},\quad
[1^I 2^J]\equiv \tilde{\lambda}_1^{Ia} \tilde{\lambda}_2^{J}.
\end{equation}
Since the little-group indices $(I_1 \dots I_{2s})$ must be symmetrized to describe a massive particle of spin $s$, a direct expression with all indices explicit becomes cumbersome when multiple particles are involved. For convenience, we consider a \text{Bold} notation that automatically incorporates the symmetrization of little-group indices. For example, a term in the 3-point massive amplitude with spin $(s_P,s_1,s_2)=(\frac{1}{2},\frac{1}{2},1)$ can be written as
\begin{equation}
    \langle 2^{J}P^{(K_1} \rangle[P^{K_2)} 1^I] \to 
    [\mathbf{2P}]\langle \mathbf{P1} \rangle.
\end{equation}

In the Standard Model, the 3-point massive amplitude for total spin $s=s_1+s_2+s_3$ takes the compact form
\begin{equation}
\mathcal{M}_3=\frac{1}{m^{s-1}} (\mathbf{ij})^{s},
\end{equation}
where $(\mathbf{ij})$ denotes either an angle or a square bracket, namely $(\mathbf{ij})=[\mathbf{ij}]\text{ or }\langle \mathbf{ij}\rangle$ with  $\mathbf{i},\mathbf{j}=\mathbf{P},\mathbf{1},\mathbf{2}$.
The explicit forms of the 3-point massive structures in the Standard Model are summarized in the following table:
\begin{equation} \label{eq:bold_amp}
\begin{array}{c|c|c}
\hline
\text{total spin} & \text{particle} & \text{massive amplitude} \\
\hline
0 & SSS & m \\
\hline
1 & Sff & [\mathbf{12}], \langle\mathbf{12}\rangle  \\
\hline
\multirow{2}{*}{2} & Vff & \frac{1}{m} \{\langle\mathbf{2P}\rangle [\mathbf{P1}], [\mathbf{2P}] \langle\mathbf{P1}\rangle\}  \\
\cline{2-3}
 & SVV &  \frac{1}{m}\langle\mathbf{12}\rangle [\mathbf{12}] \\
\hline
3 & VVV & \makecell{ \frac{1}{m^2}\{[\mathbf{12}] [\mathbf{2P}] \langle\mathbf{P1}\rangle, [\mathbf{12}] \langle\mathbf{2P}\rangle [\mathbf{P1}], \langle\mathbf{12}\rangle [\mathbf{2P}] [\mathbf{P1}],  \\  \qquad \langle\mathbf{12}\rangle [\mathbf{2P}] \langle\mathbf{P1}\rangle,[\mathbf{12}]\langle\mathbf{2P}\rangle \langle\mathbf{P1}\rangle,\langle\mathbf{12}\rangle \langle\mathbf{2P}\rangle [\mathbf{P1}] \} } \\
\hline
\end{array}
\end{equation}
For total spin $s_1+s_2+s_3>1$, the table above omits spinor structures containing only square brackets or only angle brackets (e.g., $[\mathbf{2P}][\mathbf{P1}]$), as such terms arise only in the context of effective field theory.

We now turn to the expression of the three-point massive amplitude including coupling coefficients. The massive spinors $\lambda^I$ and $\tilde{\lambda}^I$ transform under the left- and right-handed representations of the Lorentz group. Therefore, chiral dependence must be accounted for in the coefficients. In the Standard Model, the interactions of left-handed and right-handed fermions may differ, leading to the following amplitude forms:
\begin{eqnarray}
\mathcal{M}(\mathbf{P}_{h}, {\mathbf{1}}_{f}, \mathbf{2}_{\bar{f}}) &=& {y} \langle\mathbf{12}\rangle +{y'} [\mathbf{12}], \\
\mathcal{M}(\mathbf{P}_{V}, \mathbf{1}_{f}, \mathbf{2}_{\bar{f}})
&=& X_1 \frac{\langle\mathbf{1P}\rangle [\mathbf{2P}]}{m_P} +X_2 \frac{\langle\mathbf{2P}\rangle [\mathbf{1P}]}{m_P}, \label{eq:ffV} 
\end{eqnarray}
where the subscripts $f$ and $\bar{f}$ denote fermion and anti-fermion, and $h$ denotes the Higgs boson, which is the only scalar boson in the Standard Model. 
In contrast, interactions involving only bosons do not exhibit such chiral asymmetry and therefore possess uniform coefficients:
\begin{eqnarray}
\mathcal{M}(\mathbf{P}_{V}, \mathbf{1}_{V}, \mathbf{2}_{V}) &=&  
\mathbf{f}^{\mathbf{P12}} \left(\frac{[\mathbf{12}]\langle\mathbf{2P}\rangle\langle\mathbf{P1}\rangle}{m_1 m_2}+\frac{[\mathbf{12}][\mathbf{2P}]\langle\mathbf{P1}\rangle }{m_1 m_P}+\frac{[\mathbf{12}]\langle\mathbf{2P}\rangle[\mathbf{P1}]}{m_2 m_P} \right. \nonumber \\
&&\qquad +\left.\frac{\langle\mathbf{12}\rangle[\mathbf{2P}][\mathbf{P1}]}{m_1 m_2}+\frac{\langle\mathbf{12}\rangle\langle\mathbf{2P}\rangle[\mathbf{P1}]}{m_1 m_P}+\frac{\langle\mathbf{12}\rangle[\mathbf{2P}]\langle\mathbf{P1}\rangle}{m_2 m_P} \right),
\label{eq:3V}\\
\mathcal{M}(\mathbf{P}_{h}, \mathbf{1}_{V}, \mathbf{2}_{V}) &=& 
{\mathbf{g}} \frac{\langle\mathbf{12}\rangle [\mathbf{12}]}{m_1 m_2} , 
\label{eq:VVh} \\
\mathcal{M}(\mathbf{P}_{h}, \mathbf{1}_{h}, \mathbf{2}_{h}) &=& \lambda_3, \label{eq:3h} 
\end{eqnarray}
where $\mathbf{f}^{\mathbf{P12}}$ denotes the coupling for the vector boson self-interaction, and it satisfies an antisymmetry property with respect to the superscripts $\mathbf{P},\mathbf{1},\mathbf{2}$.

We now consider the high-energy expansion of the massive amplitude described by SW spinors.
Suppose that all particles are nearly collinear, meaning that $p_{i+}$ for each particle are approximately equal, and the transverse momenta $p_{iT}$ are of the same order. In this regime, we can use $p_T$ and $m$ to characterize the scales of transverse momentum and mass. The power counting then yields
\begin{equation} \begin{aligned}
\relax
[i^+ j^+]&\sim \langle i^- j^- \rangle\sim p_T,\\
[i^+ j^-]&\sim \langle i^+ j^- \rangle\sim m,
\end{aligned} \end{equation}
with no explicit dependence on $p_{i+}$ because we set $p_{i+}/p_{j+}\sim 1$. A peculiar feature is that, when the two spinors both take the inverse sign of the bracket, the spinor contraction must vanish
\begin{equation} \begin{aligned}
\relax
[i^- j^-]=\langle i^+ j^+ \rangle=0.
\end{aligned} \end{equation}
This follows from the fact that the subleading components become collinear for each massive particle:
\begin{equation} \begin{aligned}
|1^-] &\propto |2^-] \propto |P^-] \propto |\bar{n}], \\
|1^+\rangle &\propto |2^+\rangle \propto |P^+\rangle  \propto |\bar{n}\rangle. \\
\end{aligned} \end{equation}
In the region $p_T > m$, the massive 3-point amplitude with total spin $s$ can be expanded as:
\begin{equation}
\mathcal{M}_3^s=\sum_{i=0}^s [\mathcal{M}_3]_i \sim \sum_{i=0}^s m \left(\frac{m}{p_T}\right)^{i-s},
\end{equation}
which contains $s+1$ terms. 

As an example, consider the $Sff$ case. It has total spin $1$, so its expansion should consist of two terms, 
\begin{equation}
\mathcal{M}(\mathbf{P}_h, \mathbf{1}_f, \mathbf{2}_{\bar{f}})=
[\mathcal{M}]_0+[\mathcal{M}]_1,
\end{equation}
with
\begin{equation} \begin{aligned}
{[\mathcal{M}]_0}
&={y} \langle 1^- 2^- \rangle +{y'} [1^+ 2^+], \\
[\mathcal{M}]_1
&={y} \langle 1^+ 2^- \rangle+ {y} \langle 1^- 2^+ \rangle +{y'} [1^- 2^+] +{y'} [1^+ 2^-]. \\
\end{aligned} \end{equation}
These terms yield power counting in $p_T$ and $m$, separately.

The situation becomes more subtle when massive vectors are involved. Consider the $Vff$ amplitude, for which the expansion contains three terms:
\begin{equation}
\mathcal{M}(\mathbf{P}_V, \mathbf{1}_f, \mathbf{2}_{f})=
[\mathcal{M}]_0+[\mathcal{M}]_1+[\mathcal{M}]_2.
\end{equation}
Examining its high-energy behavior reveals a critical issue: the longitudinal mode of the massive vector explicitly violates unitarity. This is apparent at leading order, where we find:
\begin{equation}
[\mathcal{M}]_0= \frac{X_1}{m_P}[2^+ P^+]\langle P^- 1^-\rangle+ \frac{X_2}{m_P} \langle 2^- P^- \rangle [P^+ 1^+] \sim \frac{p_T^2}{m}.
\end{equation}
When $p_T \gg m$, this term grows as $\frac{p_T^2}{m}$, which is faster than $p_T$ and signals apparent violation of unitarity. This indicates that the naive amplitude basis with little-group covariance is unsuitable for collinear physics.

The resolution to this issue is to reorganize the amplitude using the Goldstone boson equivalence theorem \cite{Cornwall:1974km,Chanowitz:1985hj,Yao:1988aj,Bagger:1989fc}. The longitudinal polarization vector of a massive vector can be decomposed as
\begin{equation}
\epsilon_L = \frac{\tilde{\lambda}^{+} \lambda^{-} + \tilde{\lambda}^{-} \lambda^{+}}{m}=\frac{p}{m}-2 \frac{\tilde{\lambda}^{-} \lambda^{+}}{m},
\end{equation}
where the first term corresponds to the derivative of the Goldstone boson, $\partial_\mu \pi$, and the second is a residual longitudinal mode.  By applying this decomposition at the amplitude level and using the equations of motion, we can derive an amplitude from that differs significantly from the AHH amplitude. For the $ffV$ amplitude, a key part can be rewritten as
\begin{equation} \begin{aligned}
\frac{\langle \mathbf{1} P^-\rangle [\mathbf{2} P^+]+\langle \mathbf{1} P^+\rangle [\mathbf{2} P^-]}{m_P}
&=\frac{\langle \mathbf{1} \mathbf{P} \mathbf{2} ]}{m_P}-2\frac{\langle \mathbf{1} P^+\rangle [\mathbf{2} P^-]}{m_P} \\
&=\frac{m_1[\mathbf{12}]-m_2\langle \mathbf{12}\rangle}{m_P}-2\frac{\langle \mathbf{1} P^+\rangle [\mathbf{2} P^-]}{m_P}, \\
\end{aligned} \end{equation}
where in the last step we used integration by parts (IBP) to convert the momentum of the massive vector into fermion momenta. Thus, we can keep the massive fermions in bold notation while decomposing the massive vector.

This manipulation is equivalent to working in the Goldstone equivalence gauge, introduced in Ref.~\cite{Chen:2016wkt}. In this gauge, the massive fermions carry the little-group covariance, but the massive vector does not, so the amplitude form differs from the AHH amplitude. We again consider the massive $ffV$ amplitude as an example. For a transverse mode, such as positive helicity, the expression is given by 
\begin{equation}
\mathcal{M}_{GE}(P_V^+, \mathbf{1}_f, \mathbf{2}_{f})=X_1\frac{\langle \mathbf{1} P^+\rangle[\mathbf{2} P^+]}{m_P}+X_2\frac{[\mathbf{1} P^+]\langle\mathbf{2} P^+\rangle}{m_P},
\end{equation}
where the subscript $GE$ denotes the Goldstone equivalence gauge, and the superscript $+$ denote the helicity of the massive vector. For the longitudinal mode, however, it is decomposed into a well-behaved sum,
\begin{equation}
\mathcal{M}_{GE}(P_V^0, \mathbf{1}_f, \mathbf{2}_{f})=\mathcal{M}_{\pi}+\mathcal{M}_{V},
\end{equation}
where the Goldstone amplitude $\mathcal{M}_{\pi}$ and residual vector amplitude $\mathcal{M}_V$ are given by 
\begin{equation} \begin{aligned} \label{eq:Goldstone_amp}
\mathcal{M}_{\pi}&=X_1 \frac{m_1[\mathbf{12}]-m_2\langle\mathbf{12}\rangle }{m_P} +X_2 \frac{m_1\langle\mathbf{12}\rangle-m_2[\mathbf{12}] }{m_P}, \\
\mathcal{M}_{V}&=-2X_1\frac{\langle \mathbf{1} P^+\rangle[\mathbf{2} P^-]}{m_P}-2X_2\frac{[\mathbf{1} P^+]\langle\mathbf{2} P^-\rangle}{m_P}.
\end{aligned} \end{equation}
The Goldstone amplitude $\mathcal{M}_{\pi}$ has the same power counting as the $ffS$ amplitude, with two orders $p_T$ and $m$, while $\mathcal{M}_V$ contributes only at order $m$. Neither violates unitarity.

This decomposition is not unique to the fermion-vector case. In general, when multiple vector bosons are present, the massive amplitude in the Goldstone equivalence gauge can always be expanded into only two orders,
\begin{equation}
\mathcal{M}_{GE}= [\mathcal{M}_{GE}]_0 + [\mathcal{M}_{GE}]_1 \sim p_T+m,
\end{equation}
which grows at most linearly with energy and therefore does not violate unitarity. Nevertheless, the derivation of $\mathcal{M}_{GE}$ from the AHH amplitude becomes somewhat inconvenient when more than one vector boson is involved. This motivates us to seek a more convenient object for expressing the massive amplitude in the Goldstone equivalence gauge, which is the topic of the next subsection.

\subsection{Correspondence with 3-point and 4-point Massless Amplitude}
\label{sec:correspondence}

In this subsection, we consider massless collinear amplitudes and study their relation to massive collinear amplitudes in the Goldstone equivalence gauge, as an example of what we call the \textit{massless-massive correspondence}.~\footnote{For another example of massless-massive correspondence, namely the relation between massless amplitudes and massive amplitudes with explicit little-group covariance (namely AHH amplitude), see our previous work in Ref.~\cite{Ni:2024yrr,Ni:2025xkg,Ni:2026mia,Ni:2026wiz}.} We will show that, with deformations inspired by the gauge condition and consistency with power-counting analysis, the massless amplitudes exactly match all terms in the massive ones. Consequently, the massive coefficients can be determined from the ultraviolet theory.

To reveal this massless-massive correspondence, we first examine the power counting for massless amplitudes. In the collinear context, the massless amplitude only depends on $p_+$ and $p_T$. Since $p_+$ does not appear in the Lorentz scalar object, the power counting for $n$-point massless amplitude is simple  
\begin{equation} \label{eq:massless_pc}
\mathcal{A}_n \sim (p_T)^{4-n}.
\end{equation}
Thus, the 3-point and 4-point massless amplitudes scale as $p_T$ and $p_T^0$. This scaling suggests a correspondence between the terms in the high-energy expansion of massive amplitudes $\mathcal{M}_{\text{GE}}$ and the corresponding massless amplitudes:
\begin{equation} \begin{aligned}
\mathcal{A}_3 &\to [\mathcal{M}_{GE}]_0 \sim p_T, \\
v\mathcal{A}_4 &\to [\mathcal{M}_{GE}]_1 \sim m.
\end{aligned}
\end{equation}
Here we have included the vacuum expectation value (vev) $v$ for the 4-point amplitude, since it sets the scale of spontaneous symmetry breaking and hence the masses of massive particles.

This mapping provides a powerful way to determine the coefficients of the massive theory from its massless counterpart. We will now demonstrate this matching procedure explicitly for 3-point and 4-point processes.

\paragraph{3-point matching} The matching for 3-point amplitudes involves four types of interactions: $Vff$, $Sff$, $SVS$ and $VVV$. We first consider the massless amplitudes with fermions, $Vff$ and $Sff$. They are given by
\begin{equation}
\begin{aligned}
\mathcal{A}(P^0_{S},1^{-}_{f},2^{-}_{\bar{f}})&=Y\langle 12\rangle,&
\mathcal{A}(P^0_{\bar{S}},1^{+}_{f},2_{\bar{f}})&=\tilde{Y}[12],& \\
\mathcal{A}(P^{-}_{V},1^{-}_{f},2^{+}_{\bar{f}})&=\tilde{T}_f\frac{\langle 1P \rangle^2}{\langle 12 \rangle },&
\mathcal{A}(P^{+}_{V},1^{-}_{f},2^{+}_{\bar{f}})&=\tilde{T}_f\frac{[2P]^2}{[12]},&\\
\mathcal{A}(P^{-}_{V},1^{+}_{f},2^{-}_{\bar{f}})&=T_f\frac{\langle 2P \rangle ^2}{\langle 12 \rangle },&
\mathcal{A}(P^{+}_{V},1^{+}_{f},2^{-}_{\bar{f}})&=T_f\frac{[1P]^2}{[12]},&
\end{aligned}
\end{equation}
where $Y$ and $\tilde Y$ are Yukawa couplings, while $T_f$ and $\tilde T_f$ denote fermion gauge couplings. These massless amplitudes are a functions of angle bracket $\langle ij\rangle$ and square brackets $[ij]$. No bold notation is required because massless spinors do not carry a massive $SU(2)$ little-group index.

To match these massless amplitudes to their massive counterparts, we use the dictionary
\begin{equation}\label{eq:3pt-spinor}
|i\rangle \to |i^-\rangle,\quad
|i] \to |i^+], 
\end{equation}
which maps the massless spinors to the leading components of the massive SW spinors. For example, consider the massless $ffV$ amplitude. In the light-cone gauge, it can be written as
\begin{equation} \begin{aligned}
\mathcal{A}(P^{-}_{V},1^{+}_{f},2^{-}_{\bar{f}})=T_f\frac{\langle 2P \rangle ^2}{\langle 12 \rangle }
&= T_f\frac{\langle 2P \rangle [1r]}{[Pr]}, \\
\end{aligned} \end{equation}
where $|r]$ is an arbitrary reference spinor and corresponds to light-cone gauge condition $r^\mu A_\mu=0$ for gauge boson $A_\mu$. This reference spinor can match the subleading massive state by choosing the gauge $|r]\to |P^-]$. Using this gauge choosing and the dictionary eq.~\eqref{eq:3pt-spinor}, the amplitude becomes
\begin{equation} \begin{aligned}
\mathcal{A}
\quad\to\quad T_f\frac{\langle 2^- P^- \rangle [1^+ P^-]}{[P^+ P^-]}
=T_f\frac{\langle 2^- P^- \rangle [1^+ P^-]}{m_P}. \\
\end{aligned} \end{equation}
Comparing with the massive amplitude in eq.~\eqref{eq:ffV}, we directly derive the relation between the massless and massive coefficients $T_f=X_1$.

The matching for the $Sff$ amplitude is even more direct. For instance,
\begin{equation}
\mathcal{A}(P^0_{S},1^{-}_{f},2^{-}_{\bar{f}})=Y\langle 12\rangle\quad \to\quad Y\langle 1^- 2^-\rangle,
\end{equation}
which can match to either the massive $Sff$ or $Vff$  amplitude depending on the structure.  The complete set of relations for fermionic amplitudes is summarized in Table~\ref{tab:fermion_matching}. 

\begin{table}[htbp]
\centering
\begin{tabular}{|c|c|c|}
\hline
\mydiagbox{massless}{massive} & $Sff$ & $Vff$  \\
\hline
$Sff$ & \makecell{$Y \to y$ \\ $\tilde{Y} \to y^\prime$} & \makecell{$Y \to -\frac{m_{1}}{m_{P}} X_{1}+\frac{m_{2}}{m_{P}} X_{2} $ \\ $\tilde{Y} \to \frac{m_{2}}{m_{P}} X_{1}-\frac{m_{1}}{m_{P}} X_{2} $} \\
\hline
$Vff$ & & \makecell{$T_{f}\to X_{1}$ \\ $\tilde{T}_{f}\to X_{2}$} \\
\hline
\end{tabular}
\caption{Coefficient matching from 3-point massless to 3-point massive amplitudes for processes involving fermions.}
\label{tab:fermion_matching}
\end{table}

Then we consider the amplitudes with only bosons, $SVS$ and $VVV$. They are given by 
\begin{equation}
\begin{aligned}
\mathcal{A}(P^{0}_{\bar{S}},1^{-}_{V},2^{0}_{S})&=T_s\frac{\langle12\rangle\langle P1\rangle}{\langle2P\rangle},&
\mathcal{A}(P^{0}_{\bar{S}},1^{+}_{V},2^{0}_{S})&=T_s\frac{[12][P1]}{[2P]},& \\
\mathcal{A}(P^{+}_{V},1^{-}_{V},2^{-}_{V})&=f^{P 1 2} \frac{\langle12\rangle^3}{\langle2P\rangle\langle P1\rangle},&
\mathcal{A}(P^{-}_{V},1^{+}_{V},2^{+}_{V})&=f^{P 1 2}\frac{[12]^3}{[2P][P1]},&
\end{aligned}
\end{equation}
where $T_s$ is the gauge coupling for scalars, and $f^{P12}$ is the structure constant for the gauge group. In the unbroken phase of the Standard Model, the scalar bosons can be the Higgs doublet or its conjugate, so we distinguish them by $S$ and $\bar{S}$ in the subscripts. Similarly, we can perform the matching procedure on these amplitudes. The matching result is given in Table~\ref{tab:boson_matching}.

\begin{table}[htbp]
\centering
\begin{tabular}{|c|c|c|}
\hline
\mydiagbox{massless}{massive} & $SVV$ & $VVV$ \\
\hline
$SVS$ & $T_{s}\to \mathbf{g}$ & $T_{s}\to\frac{m^2_{1}-m^2_{2}-m^2_{P}}{m_{2} m_{P}} 2\mathbf{f}^{\mathbf{12P}}$ \\
\hline
$VVV$ & & $f^{P12} \to 2\mathbf{f}^{\mathbf{P12}}$ \\
\hline
\end{tabular}
\caption{Coefficient matching from 3-point massless to 3-point massive amplitudes for processes involving only bosons.}
\label{tab:boson_matching}
\end{table}

We next formulate a four-point matching for the subleading massive amplitude. Here, an additional Higgs boson is introduced to probe the subleading SW spinor components, which are absent from the leading three-point matching. Denoting this particle as $h$, the proposed matching is
\begin{equation}
v\mathcal{A}(P,1,2,h)\to [\mathcal{M}_{\text{GE}}(P,1,2)]_1,
\end{equation}
where the vev $v$ converts the additional Higgs leg into a symmetry-breaking insertion. This relation must be understood channel by channel, and it holds only under the special kinematics specified below.

As an illustration, consider the $SffS$ amplitude. In this case, we need to consider two particle types $Sf\bar{f}\bar{S}$ and $\bar{S}f\bar{f}S$, which take the form
\begin{equation} \begin{aligned}
\mathcal{A}(P^{0}_{S},1^{-}_{f},2^{+}_{\bar{f}},h^0_{\bar{S}})
&=Y\tilde{Y} \frac{\langle 1|P|2]}{s_{1h}} + \frac{1}{2} T_f T_s\frac{\langle 1|P-p_h|2]}{s_{Ph}}, \\
\mathcal{A}(P^{0}_{\bar{S}},1^{-}_{f},2^{+}_{\bar{f}},h^0_{S})
&=\tilde{Y}Y \frac{\langle 1|P|2]}{s_{2h}} + \frac{1}{2} T_f T_s\frac{\langle 1|P-p_h|2]}{s_{Ph}},
\end{aligned}
\end{equation}
where the Mandelstam variable $s_{ij}=(p_i+p_j)^2$ denotes the massless pole structures. Notably, neither amplitude possesses all three poles $s_{Ph}$, $s_{1h}$ and $S_{2H}$. This is because the scalar $S$ and anti-scalar $\bar{S}$ are Higgs doublets carrying opposite hypercharge in the unbroken phase of the standard model. Consequently, each couples only to either $f$ or $\bar{f}$, depending on the specific particle type and chirality chosen.

\begin{figure}[htbp]
\centering
\includegraphics[width=0.5\textwidth]{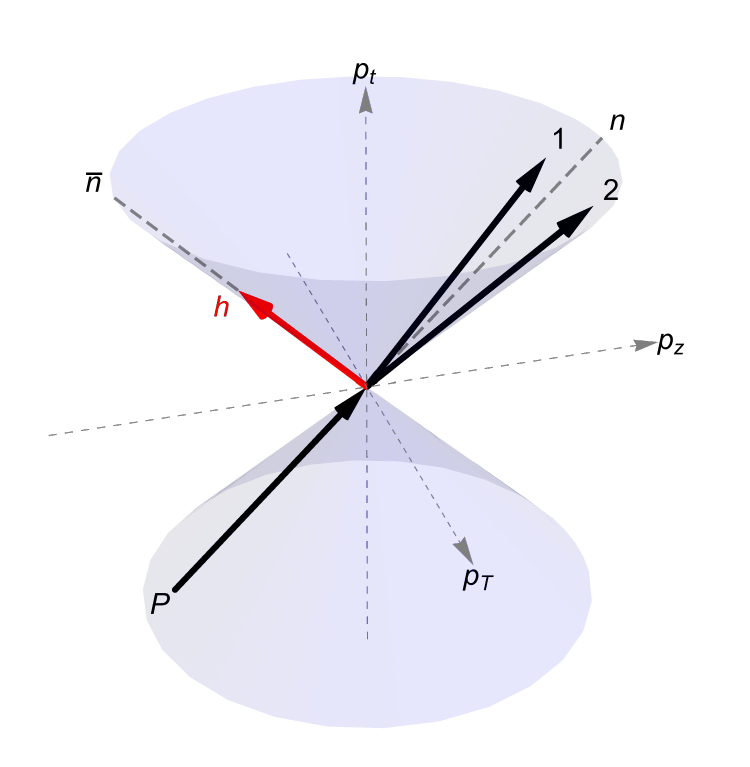}
\caption{Configuration of the four massless particles in momentum space: three particles (black arrows) are clustered along the collinear direction $n$, while the additional Higgs boson (red arrow) lies along the anti-collinear direction $\bar{n}$.}
\label{fig:momentum_space}
\end{figure}

To match this to a massive amplitude, the four massless particles must be arranged in a specific momentum-space configuration, as shown in figure~\ref{fig:momentum_space}. The first three massless particles are still matched to the leading SW spinors, following the same procedure as in the 3-point matching of eq.~\eqref{eq:3pt-spinor}. Meanwhile, the additional Higgs boson (particle $h$) is taken to point along the anti-collinear direction $\bar{n}$, allowing it to be matched to the subleading SW spinor of one of the other particles. For each channel, we apply
\begin{equation} \label{eq:higgs_match}
s_{ih}\text{-channel}:\quad |h\rangle\to|i^+\rangle,\quad
|h] \to|i^-],\quad
\end{equation}
where $i=P,1,2$. In this mapping, the massless pole will become the particle mass, 
\begin{equation} \label{eq:pole_matching}
s_{ih}=[ih]\langle hi\rangle
\quad\to \quad
[i^- i^+]\langle i^+ i^-\rangle = m_i^2.
\end{equation}
Therefore, the massless amplitude matches a local massive amplitude without pole structures.

Before we apply this mapping, we should examine the consistency of the power counting in the matching procedure. The pole matching introduces an $m^{-2}$ factor, which goes against the 4-point massless power counting in eq.~\eqref{eq:massless_pc}. To give a consistent matching, we need to find an $m^2$ contribution in the numerator after mapping. This cannot be realized by the structure $\langle 1|P|2]$ in the $SffS$ amplitude, because momentum $P$ gives the leading contribution $p_T^2$ after mapping. This inconsistency can be solved by using the momentum conservation to convert the numerator into the following form
\begin{equation} 
\langle 1|P|2]\quad \Rightarrow \quad -\langle 1|p_h|2]=\langle 1h\rangle[h2].
\end{equation}
This new structure has only $p_h$ structure, and will give a $m^2$ contribution after mapping. Each term in the $SffS$ amplitude then takes the form
\begin{equation}
\frac{\langle 1 h\rangle[h 2]}{s_{i h}}=\frac{\langle i h\rangle[h 2]}{\langle i h\rangle[h i]}.
\end{equation}
A crucial observation is that both the numerator and denominator are proportional to the Higgs momentum $p_h$, so their ratio is independent of the magnitude of $p_h$. This explains why the corresponding spinors $|h\rangle$ and $|h]$ can be rescaled to match the subleading SW spinor of different particles, as shown in  eq.~\eqref{eq:higgs_match}.

After converting the massless $SffS$ amplitude into the correct form, we can match it to obtain
\begin{equation} \begin{aligned}
\frac{\langle 1 h\rangle[h 2]}{s_{1h}} &\quad\to\quad \frac{\langle 1^- 1^+\rangle[1^- 2^+]}{m_1^2}=\frac{[1^- 2^+]}{m_1}, \\
\frac{\langle 1 h\rangle[h 2]}{s_{2h}} &\quad\to\quad \frac{\langle 1^- 2^+\rangle[2^- 2^+]}{m_2^2}=\frac{\langle 1^- 2^+\rangle}{m_2}, \\
\frac{\langle 1 h\rangle[h 2]}{s_{Ph}} &\quad\to\quad \frac{\langle 1^- P^+\rangle[P^- 2^+]}{m_P^2}.
\end{aligned}
\end{equation}
where we have used $\langle i^- i^+\rangle=[i^- i^+]=m_i$ to simplify. The two massless amplitudes thus match to
\begin{equation} \begin{aligned}
\mathcal{A}(P^{0}_{S},1^{-}_{f},2^{+}_{\bar{f}},h^0_{\bar{S}})
&\to - Y\tilde{Y} \frac{[1^- 2^+]}{m_1} - T_f T_s\frac{\langle 1^- P^+\rangle[P^- 2^+]}{m_P^2},\\
\mathcal{A}(P^{0}_{\bar{S}},1^{-}_{f},2^{+}_{\bar{f}},h^0_{S})&\to -\tilde{Y} Y \frac{\langle 1^- 2^+\rangle}{m_2} - T_f T_s\frac{\langle 1^- P^+\rangle[P^- 2^+]}{m_P^2}.
\end{aligned}
\end{equation}
Note that the coefficient of each term contains two 3-point massless coefficients. Combining the vev with one 3-point coefficient yields the particle mass:
\begin{equation}
v Y=m_1, \quad v \tilde{Y}=m_2,\quad v T_s=m_P.
\end{equation} 
The other 3-point coefficient can then match to the massive $Sff$ or $Vff$ case, as shown in table~\ref{tab:fermion_matching}. Combining the contributions of the two massless amplitudes, we derive
\begin{equation} \begin{aligned}
Sff:&\quad
\begin{cases}
- y^{\prime} m_1 \frac{[1^- 2^+]}{m_1} \\
- y m_2 \frac{\langle 1^- 2^+\rangle}{m_2}
\end{cases}, \\
Vff:&\quad
\begin{cases}
- \left( \frac{m_{2}}{m_{P}} X_{1}-\frac{m_{1}}{m_{P}} X_{2} \right)m_{1} \frac{[1^- 2^+]}{m_1} - X_1 m_P\frac{\langle 1^- P^+\rangle[P^- 2^+]}{m_P^2} \\
-\left(-\frac{m_{1}}{m_{P}} X_{1}+\frac{m_{2}}{m_{P}} X_{2} \right) m_{2} \frac{\langle 1^- 2^+\rangle}{m_2} - X_1 m_P \frac{\langle 1^- P^+\rangle[P^- 2^+]}{m_P^2}
\end{cases}.
\end{aligned}
\end{equation}
Finally, by bolding the massive spinors, we find that this result exactly matches the massive $Sff$ and $Vff$ amplitudes given in eqs.~\eqref{eq:ffV} and \eqref{eq:Goldstone_amp}.

\subsection{General 3-point Amplitude}
When we extend the discussion beyond the Standard Model, effective operators enter the analysis. In this context, the general amplitudes do not necessarily satisfy unitarity, and we are not constrained to use the Goldstone equivalence gauge. Let us return to the bolded form. In general, the 3-point massive amplitude has the following structures:
\begin{equation}
\begin{array}{c|c|c|c}
\hline
\text{total spin} & \text{particle} & \text{SM structure} & \text{EFT structure} \\
\hline
0 & SSS & 1 & \\
\hline
1 & Sff & [\mathbf{12}], \langle\mathbf{12}\rangle & \\
\hline
\multirow{2}{*}{2} & Vff & \langle\mathbf{2P}\rangle [\mathbf{P1}], [\mathbf{2P}] \langle\mathbf{P1}\rangle & \langle\mathbf{2P}\rangle \langle\mathbf{P1}\rangle, [\mathbf{2P}][\mathbf{P1}] \\
\cline{2-4}
 & SVV &  \langle\mathbf{12}\rangle [\mathbf{12}] & [\mathbf{12}]^2, \langle\mathbf{12}\rangle^2 \\
\hline
3 & VVV & \makecell{ [\mathbf{12}] [\mathbf{2P}] \langle\mathbf{P1}\rangle, [\mathbf{12}] \langle\mathbf{2P}\rangle [\mathbf{P1}], \langle\mathbf{12}\rangle [\mathbf{2P}] [\mathbf{P1}],  \\  \langle\mathbf{12}\rangle [\mathbf{2P}] \langle\mathbf{P1}\rangle,[\mathbf{12}]\langle\mathbf{2P}\rangle \langle\mathbf{P1}\rangle,\langle\mathbf{12}\rangle \langle\mathbf{2P}\rangle [\mathbf{P1}] } & 
\makecell{ {}[\mathbf{12}] [\mathbf{2P}] [\mathbf{P1}] \\ \langle\mathbf{12}\rangle \langle\mathbf{2P}\rangle \langle\mathbf{P1}\rangle } \\
\hline
\end{array}
\end{equation}
The third column corresponds to the SM structures, while the last column includes the spinor structures that appear only in effective field theory (EFT).

To derive the components of the splitting amplitudes explicitly, we parametrize the kinematics of a $1\to 2$ splitting process as follows. We take particle $P$ to be the parent particle, and particles 1 and 2 to be the daughter particles. The parent particle is taken to have zero transverse momentum, reducing the parametrization to two independent variables. The transverse momenta satisfy
\begin{equation}
P_T=0,\quad p_{1T}=-p_{2T} = p_{T}.
\end{equation}
The light-front momentum fractions are parametrized by a single variable $z$,
\begin{equation}
z=\frac{p_{1+}}{P_{+}},\quad 1-z=\frac{p_{2+}}{P_{+}}.
\end{equation}
The on-shell condition then determines the remaining momentum components via $p_{i-} = (|p_{iT}|^2 + m_i^2)/p_{i+}$. Thus the two independent variables are $x$ and $p_T$. The on-shell condition then determines the remaining momentum components via $p_{i-} = (|p_{iT}|^2 + m_i^2)/p_{i+}$. Thus the two independent variables are $z$ and $p_T$. With these parameters, the momenta of the three particles can be written as
\begin{equation} \begin{aligned}
P&=\left(P_+, \frac{m_P^2}{P_+}, 0\right),\\
p_1&=\left(z P_+, \frac{m_1^2 + |p_T|^2}{z P_+}, p_T\right),  \\
p_2&=\left((1-z) P_+, \frac{m_2^2 + |p_T|^2}{(1-z) P_+},-p_T\right).
\end{aligned}
\end{equation}

Using this parametrization, the bolded massive 3-point amplitude using SW spinors can be expressed as a matrix in the little-group indices, with entries that depend on $z$, $p_T$, and the masses $m_i$. For example, the $Sff$ amplitude can be expressed as a two-by-two matrix,
\begin{equation}
\langle 1^I 2^J \rangle_{\text{SW}}
=\begin{pmatrix}
\langle 1^- 2^- \rangle & \langle 1^+ 2^- \rangle \\
\langle 1^- 2^+ \rangle & \langle 1^+ 2^+ \rangle \\
\end{pmatrix}_{\text{SW}}
=\begin{pmatrix}
-\frac{p_T}{\sqrt{-((z-1) z)}} & -m_1 \sqrt{\frac{1}{z}-1} \\
m_2 \sqrt{-\frac{z}{z-1}} & 0 \\
\end{pmatrix}.
\end{equation}
The result is straightforward and directly gives the splitting amplitude. The four elements correspond to different spin components of the splitting process,
\begin{equation}
\begin{pmatrix}
S\to f^- f^- & S\to f^+ f^- \\
S\to f^+ f^- & S\to f^- f^+ \\
\end{pmatrix}.
\end{equation}

If we consider other amplitude forms, such as the AHH amplitude, the expression becomes more complicated,
\begin{equation}
\langle 1^I 2^J \rangle_{\text{AHH}}
=\begin{pmatrix}
\langle 1^- 2^- \rangle & \langle 1^+ 2^- \rangle \\
\langle 1^- 2^+ \rangle & \langle 1^+ 2^+ \rangle \\
\end{pmatrix}_{\text{AHH}}
=\begin{pmatrix}
\sqrt{ E_{1+} E_{2+} } (c_{1}s_{2} - c_{2}s_{1} ) &  
-\sqrt{ E_{1-} E_{2+} } (s_{1}^* s_{2}+c_{1}c_{2}) \\
\sqrt{ E_{1+} E_{2-} } (s_{1} s_{2}^*+c_{1}c_{2}) & 
\sqrt{ E_{1-} E_{2-} } (c_{1}s_{2}^* - c_{2}s_{1}^* ) \\
\end{pmatrix},
\end{equation}
where $E_{i\pm}=E_{i}\pm p_{i}$ for $i=1,2$ label the large and small components, and $c_i=\cos \frac{\theta_i}{2}$, $s_i=\sin \frac{\theta_i}{2}e^{i\phi_i}$ include the angular information. This result does not explicitly show the splitting amplitude and requires conversion from the helicity parameters $(E_i,\theta_i,\phi_i)$ to the splitting parameters $(z,p_T)$.

This conversion is necessary because the AHH amplitude employs helicity decomposition, so the spin axes for particles 1 and 2 are different and not aligned with the collinear direction. Thus, the amplitude matrix should include the two sets of angle parametrizations $(\theta_i,\phi_i)$ for particles $i=1,2$. To extract the collinear information, one must align the spin axes to a common collinear direction by hand, which is equivalent to combining one angle with the momentum to form the transverse momentum. This introduces unnecessary conversion steps in the calculation.

Recall that AHH spinors and SW spinors can be related by a little-group rotation $W$, as given in eq.~\eqref{eq:W_matrix}. This rotation can also be applied to relate the AHH and SW amplitudes as
\begin{equation}
\langle 1^I 2^J \rangle_{\text{SW}}
=W^I_K W^J_L \langle 1^K 2^L \rangle_{\text{AHH}}.
\end{equation}

Thus, by adopting the convenient SW amplitudes, we can calculate the amplitude matrix and explicitly give all possible spin components for the $1\to 2$ splitting process. In the following, we reorganize these matrix elements into tables to show the results:

\begin{itemize}

    \item $S\to ff$:  This includes two SM structures $\langle\boldsymbol{12}\rangle$ , $[\boldsymbol{12}]$:
\begin{align}
    &\begin{array}{|c|cccc|}
    \hline
    \multicolumn{5}{|c|}{\langle\boldsymbol{12}\rangle} \\
    \hline
    \to & f^- f^- & f^+ f^- & f^- f^+ & f^+ f^+  \\
    \hline
    S & -\frac{p_T}{\sqrt{-((z-1) z)}} & -m_1 \sqrt{\frac{1}{z}-1} & m_2 \sqrt{-\frac{z}{z-1}} & 0 \\ 
    \hline
    \end{array}\\
    &\begin{array}{|c|cccc|}
    \hline
    \multicolumn{5}{|c|}{[\boldsymbol{12}]} \\
    \hline
    \to & f^- f^- & f^+ f^- & f^- f^+ & f^+ f^+  \\
    \hline
    S & 0 & -m_2 \sqrt{-\frac{z}{z-1}} & m_1 \sqrt{\frac{1}{z}-1} & -\frac{p_T^*}{\sqrt{-((z-1) z)}} \\
    \hline
    \end{array}
\end{align}

\item $f\to Sf$: Similarly, we have two structures $\langle\boldsymbol{P2}\rangle$ and $[\boldsymbol{P2}]$:
\begin{align}
    &\begin{array}{|c|cc|}
    \hline
    \multicolumn{3}{|c|}{\langle\boldsymbol{P2}\rangle} \\
    \hline
    \to & S f^- &  S f^+ \\
    \hline
    f^- & m_P \sqrt{1-z} & 0  \\
    f^+ & -\frac{p_T}{\sqrt{1-z}} & \frac{m_2}{\sqrt{1-z}} \\
    \hline
    \end{array}\quad 
    \begin{array}{|c|cc|}
    \hline
    \multicolumn{3}{|c|}{[\boldsymbol{P2}]} \\
    \hline
    \to & S f^- &  S f^+ \\
    \hline
    f^- & \frac{m_2}{\sqrt{1-z}} & \frac{p_T^*}{\sqrt{1-z}} \\
    f^+ & 0 & m_P \sqrt{1-z} \\
    \hline
    \end{array}
\end{align}

\item $V\to ff$: It includes two EFT structures
$\langle\boldsymbol{1P}\rangle\langle\boldsymbol{2P}\rangle$ and $[\boldsymbol{1P}][\boldsymbol{2P}]$:
\begin{align}
        &\begin{array}{|c|cccc|}
            \hline
            \multicolumn{5}{|c|}{\langle\boldsymbol{1P}\rangle\langle\boldsymbol{2P}\rangle} \\
            \hline
            \to & f^- f^- & f^- f^+ & f^+ f^- & f^+ f^+ \\
            \hline
            V^- & m_P^2 \sqrt{(1-z) z} & 0 & 0 & 0 \\
            V^0 & \frac{m_P p_T (1-2 z)}{\sqrt{2} \sqrt{-((z-1) z)}} & \frac{m_2 m_P}{\sqrt{\frac{2}{z}-2}} & \frac{m_1 m_P \sqrt{\frac{1}{z}-1}}{\sqrt{2}} & 0 \\
            V^+ & -\frac{p_T^2}{\sqrt{-((z-1) z)}} & \frac{m_2 p_T}{\sqrt{-((z-1) z)}} & -\frac{m_1 p_T}{\sqrt{-((z-1) z)}} & \frac{m_1 m_2}{\sqrt{-((z-1) z)}} \\
            \hline
        \end{array} \\
        &\begin{array}{|c|cccc|}
            \hline
            \multicolumn{5}{|c|}{[\boldsymbol{1P}][\boldsymbol{2P}]} \\
            \hline
            \to & f^- f^- & f^- f^+ & f^+ f^- & f^+ f^+ \\
            \hline
            V^- & \frac{m_1 m_2}{\sqrt{-((z-1) z)}} & \frac{m_1 p_T^*}{\sqrt{-((z-1) z)}} & -\frac{m_2 p_T^*}{\sqrt{-((z-1) z)}} & -\frac{(p_T^*)^2}{\sqrt{-((z-1) z)}} \\
            V^0 & 0 & \frac{m_1 m_P \sqrt{\frac{1}{z}-1}}{\sqrt{2}} & \frac{m_2 m_P}{\sqrt{\frac{2}{z}-2}} & \frac{m_P p_T^* (2 z-1)}{\sqrt{2} \sqrt{-((z-1) z)}} \\
            V^+ & 0 & 0 & 0 & m_P^2 \sqrt{(1-z) z} \\
            \hline
        \end{array} 
\end{align}
and two SM structures $[\boldsymbol{1P}]\langle\boldsymbol{2P}\rangle$ and $\langle\boldsymbol{1P}\rangle[\boldsymbol{2P}]$:
\begin{align}
        &\begin{array}{|c|cccc|}
            \hline
            \multicolumn{5}{|c|}{[\boldsymbol{1P}]\langle\boldsymbol{2P}\rangle} \\
            \hline
            \to & f^- f^- & f^- f^+ & f^+ f^- & f^+ f^+ \\
            \hline
            V^- & m_1 m_P \sqrt{\frac{1}{z}-1} & 0 & -m_P p_T^* \sqrt{\frac{1}{z}-1} & 0 \\
            V^0 & -\frac{m_1 p_T}{\sqrt{2} \sqrt{-((z-1) z)}} & \frac{m_1 m_2}{\sqrt{2} \sqrt{-((z-1) z)}} & \frac{p_T p_T^*-m_P^2 (z-1) z}{\sqrt{2} \sqrt{-((z-1) z)}} & -\frac{m_2 p_T^*}{\sqrt{2} \sqrt{-((z-1) z)}} \\
            V^+ & 0 & 0 & -\frac{m_P p_T}{\sqrt{\frac{1}{z}-1}} & \frac{m_2 m_P}{\sqrt{\frac{1}{z}-1}} \\
            \hline
        \end{array} \\
        &\begin{array}{|c|cccc|}
            \hline
            \multicolumn{5}{|c|}{\langle\boldsymbol{1P}\rangle[\boldsymbol{2P}]} \\
            \hline
            \to & f^- f^- & f^- f^+ & f^+ f^- & f^+ f^+ \\
            \hline
            V^- & \frac{m_2 m_P}{\sqrt{\frac{1}{z}-1}} & \frac{m_P p_T^*}{\sqrt{\frac{1}{z}-1}} & 0 & 0 \\
            V^0 & \frac{m_2 p_T}{\sqrt{2} \sqrt{-((z-1) z)}} & \frac{p_T p_T^*-m_P^2 (z-1) z}{\sqrt{2} \sqrt{-((z-1) z)}} & \frac{m_1 m_2}{\sqrt{2} \sqrt{-((z-1) z)}} & \frac{m_1 p_T^*}{\sqrt{2} \sqrt{-((z-1) z)}} \\
            V^+ & 0 & m_P p_T \sqrt{\frac{1}{z}-1} & 0 & m_1 m_P \sqrt{\frac{1}{z}-1} \\
            \hline
        \end{array}
\end{align}

\item $f\to Vf$: Similarly, we have two EFT structures
$\langle\boldsymbol{12}\rangle\langle\boldsymbol{1P}\rangle$ and $[\boldsymbol{12}][\boldsymbol{1P}]$:
\begin{align}
        &\begin{array}{|c|cc|cc|cc|}
            \hline
            \multicolumn{7}{|c|}{\langle\boldsymbol{12}\rangle\langle\boldsymbol{1P}\rangle} \\
            \hline
            \to & V^- f^+ & V^- f^- & V^0 f^+ & V^0 f^- & V^+ f^+ & V^+ f^- \\
            \hline
            f^+ & -\frac{m_P p_T}{\sqrt{1-z}} & -\frac{p_T^2}{\sqrt{1-z} z} 
            & -\frac{m_1 m_P \sqrt{1-z}}{\sqrt{2}} & \frac{m_1 p_T (z-2)}{\sqrt{2-2 z} z} 
            & 0 & -\frac{m_1^2 \sqrt{1-z}}{z} \\
            f^- & \frac{m_2 m_P z}{\sqrt{1-z}} & \frac{m_2 p_T}{\sqrt{1-z}} 
            & 0 & \frac{m_1 m_2}{\sqrt{2-2 z}} 
            & 0 & 0 \\
            \hline
        \end{array} \\
    &\begin{array}{|c|cc|cc|cc|}
        \hline
        \multicolumn{7}{|c|}{[\boldsymbol{12}][\boldsymbol{1P}]} \\
        \hline
        \to & V^- f^+ & V^- f^- & V^0 f^+ & V^0 f^- & V^+ f^+ & V^+ f^- \\
        \hline
        f^+ & 0 & 0 & -\frac{m_1 m_2}{\sqrt{2-2 z}} & 0 & \frac{m_2 p_T^*}{\sqrt{1-z}} & -\frac{m_2 m_P z}{\sqrt{1-z}} \\
        f^- & \frac{m_1^2 \sqrt{1-z}}{z} & 0 & \frac{m_1 p_T^* (z-2)}{\sqrt{2-2 z} z} & \frac{m_1 m_P \sqrt{1-z}}{\sqrt{2}} & \frac{\left(p_T^*\right)^2}{\sqrt{1-z} z} & -\frac{m_P p_T^*}{\sqrt{1-z}} \\
        \hline
    \end{array}
\end{align}
and two SM structures $[\boldsymbol{12}]\langle\boldsymbol{1P}\rangle$, $\langle\boldsymbol{12}\rangle[\boldsymbol{1P}]$:
\begin{align}
    &\begin{array}{|c|cc|cc|cc|}
        \hline
        \multicolumn{7}{|c|}{[\boldsymbol{12}]\langle\boldsymbol{1P}\rangle} \\
        \hline
        \to & V^- f^+ & V^- f^- & V^0 f^+ & V^0 f^- & V^+ f^+ & V^+ f^- \\
        \hline
        f^+ & -\frac{m_1 p_T}{\sqrt{1-z} z} & 0 & \frac{m_1^2 (z-1)+p_T p_T^*}{\sqrt{2-2 z} z} & -\frac{m_P p_T}{\sqrt{2-2 z}} & \frac{m_1 p_T^* \sqrt{1-z}}{z} & -m_1 m_P \sqrt{1-z} \\
        f^- & \frac{m_1 m_2}{\sqrt{1-z}} & 0 & -\frac{m_2 p_T^*}{\sqrt{2-2 z}} & \frac{m_2 m_P z}{\sqrt{2-2 z}} & 0 & 0 \\
        \hline
    \end{array} \\
    &\begin{array}{|c|cc|cc|cc|}
        \hline
        \multicolumn{7}{|c|}{\langle\boldsymbol{12}\rangle[\boldsymbol{1P}]} \\
        \hline
        \to & V^- f^+ & V^- f^- & V^0 f^+ & V^0 f^- & V^+ f^+ & V^+ f^- \\
        \hline
        f^+ & 0 & 0 & -\frac{m_2 m_P z}{\sqrt{2-2 z}} & -\frac{m_2 p_T}{\sqrt{2-2 z}} & 0 & -\frac{m_1 m_2}{\sqrt{1-z}} \\
        f^- & m_1 m_P \sqrt{1-z} & \frac{m_1 p_T \sqrt{1-z}}{z} & -\frac{m_P p_T^*}{\sqrt{2-2 z}} & \frac{-m_1^2 (z-1)-p_T p_T^*}{\sqrt{2-2 z} z} & 0 & -\frac{m_1 p_T^*}{\sqrt{1-z} z} \\
        \hline
    \end{array}
\end{align}

\item $S\to VV$: It includes two EFT structures
$\langle \boldsymbol{12}\rangle^2$ and $[\boldsymbol{12}]^2$:
\begin{align}
    &\begin{array}{|c|ccc|ccc|ccc|}
        \hline
        \multicolumn{10}{|c|}{\langle \boldsymbol{12}\rangle^2} \\
        \hline
        \to & V^- V^- & V^0 V^- & V^+ V^- & V^- V^0 & V^0 V^0 & V^+ V^0 & V^- V^+ & V^0 V^+ & V^+ V^+ \\
        \hline
        S & \frac{p_T^2}{z-z^2} & \frac{\sqrt{2} m_1 p_T}{z} & m_1^2 \left(\frac{1}{z}-1\right) &
         \frac{\sqrt{2} m_2 p_T}{z-1} & -m_1 m_2 & 0 &
         -\frac{m_2^2 z}{z-1} & 0 & 0 \\
        \hline
    \end{array} \\
    &\begin{array}{|c|ccc|ccc|ccc|}
        \hline
        \multicolumn{10}{|c|}{[\boldsymbol{12}]^2} \\
        \hline
        \to & V^- V^- & V^0 V^- & V^+ V^- & V^- V^0 & V^0 V^0 & V^+ V^0 & V^- V^+ & V^0 V^+ & V^+ V^+ \\
        \hline
        S & 0 & 0 & -\frac{m_2^2 z}{z-1} 
         & 0 & -m_1 m_2 & -\frac{\sqrt{2} m_2 p_T^*}{z-1} 
         & m_1^2 \left(\frac{1}{z}-1\right) & -\frac{\sqrt{2} m_1 p_T^*}{z} & \frac{\left(p_T^*\right)^2}{z-z^2} \\
        \hline
    \end{array}
\end{align}
and one SM structure $\langle \boldsymbol{12}\rangle [\boldsymbol{12}]$
\begin{align}
    &\begin{array}{|c|ccc|ccc|ccc|}
        \hline
        \multicolumn{10}{|c|}{\langle \boldsymbol{12}\rangle [\boldsymbol{12}]} \\
        \hline
        \to & V^- V^- & V^0 V^- & V^+ V^- & V^- V^0 & V^0 V^0 & V^+ V^0 & V^- V^+ & V^0 V^+ & V^+ V^+ \\
        \hline
        S & 0 & -\frac{m_2 p_T}{\sqrt{2} (z-1)} & m_1 m_2 
         & -\frac{m_1 p_T}{\sqrt{2} z} & \frac{m_1^2 (z-1)^2+m_2^2 z^2-p_T p_T^*}{2 (z-1) z} & \frac{m_1 p_T^*}{\sqrt{2} z} 
         & m_1 m_2 & \frac{m_2 p_T^*}{\sqrt{2} (z-1)} & 0 \\
        \hline
    \end{array}
\end{align}

\item $V\to VS$: Similarly, it includes two EFT structures
$\langle \boldsymbol{1P}\rangle^2$ and $[\boldsymbol{1P}]^2$:
\begin{align}
    &\begin{array}{|c|ccc|}
        \hline
        \multicolumn{4}{|c|}{\langle \boldsymbol{1P}\rangle^2} \\
        \hline
        \to & V^- S & V^0 S & V^+ S \\
        \hline
        V^- & m_P^2 z & 0 & 0 \\
        V^0 & \sqrt{2} m_P p_T & m_1 m_P & 0 \\
        V^+ & \frac{p_T^2}{z} & \frac{\sqrt{2} m_1 p_T}{z} & \frac{m_1^2}{z} \\
        \hline
    \end{array} \quad
    \begin{array}{|c|ccc|}
        \hline
        \multicolumn{4}{|c|}{[\boldsymbol{1P}]^2} \\
        \hline
        \to & V^- S & V^0 S & V^+ S \\
        \hline
        V^- & m_1 m_P & -\frac{m_P p_T^*}{\sqrt{2}} & 0 \\
        V^0 & \frac{m_1 p_T}{\sqrt{2} z} & \frac{m_1^2+m_P^2 z^2-p_T p_T^*}{2 z} & -\frac{m_1 p_T^*}{\sqrt{2} z} \\
        V^+ & 0 & \frac{m_P p_T}{\sqrt{2}} & m_1 m_P \\
        \hline
    \end{array}
\end{align}
and one SM structure
$\langle \boldsymbol{1P}\rangle [\boldsymbol{1P}]$
\begin{align}
    &\begin{array}{|c|ccc|}
        \hline
        \multicolumn{4}{|c|}{\langle \boldsymbol{1P}\rangle [\boldsymbol{1P}]} \\
        \hline
        \to & V^- S & V^0 S & V^+ S \\
        \hline
        V^- & m_1 m_P & -\frac{m_P p_T^*}{\sqrt{2}} & 0 \\
        V^0 & \frac{m_1 p_T}{\sqrt{2} z} & \frac{m_1^2+m_P^2 z^2-p_T p_T^*}{2 z} & -\frac{m_1 p_T^*}{\sqrt{2} z} \\
        V^+ & 0 & \frac{m_P p_T}{\sqrt{2}} & m_1 m_P \\
        \hline
    \end{array}
\end{align}

\item $V\to VV$: It includes two EFT structures
$\langle\boldsymbol{1P}\rangle\langle\boldsymbol{P2}\rangle\langle\boldsymbol{12}\rangle$ and $[\boldsymbol{1P}][\boldsymbol{P2}][\boldsymbol{12}]$:
\begin{align}
    &\begin{array}{|c|ccc|ccc|}
        \hline
        \multicolumn{7}{|c|}{\langle\boldsymbol{1P}\rangle\langle\boldsymbol{P2}\rangle\langle\boldsymbol{12}\rangle} \\
        \hline
        \to & V^- V^- & V^- V^0 & V^- V^+ & V^+ V^- & V^+ V^0 & V^+ V^+ \\
        \hline
        V^- & -m_P^2 p_T & \frac{m_2 m_P^2 z}{\sqrt{2}} & 0 & 0 & 0 & 0 \\
        V^0 & \frac{m_P p_T^2 (1-2 z)}{\sqrt{2} (z-1) z} & \frac{m_2 m_P p_T z}{z-1} & -\frac{m_2^2 m_P z}{\sqrt{2} (z-1)} & \frac{m_1^2 m_P (z-1)}{\sqrt{2} z} & 0 & 0 \\
        V^+ & \frac{p_T^3}{z-z^2} & \frac{m_2 p_T^2 (z+1)}{\sqrt{2} (z-1) z} & -\frac{m_2^2 p_T}{z-1} & \frac{m_1^2 p_T}{z} & -\frac{m_1^2 m_2}{\sqrt{2} z} & 0 \\
        \hline
    \end{array} \\
    &\begin{array}{|c|ccc|}
        \hline
        \multicolumn{4}{|c|}{\langle\boldsymbol{1P}\rangle\langle\boldsymbol{P2}\rangle\langle\boldsymbol{12}\rangle} \\
        \hline
        \to & V^0 V^- & V^0 V^0 & V^0 V^+ \\
        \hline
        V^- & \frac{m_1 m_P^2 (z-1)}{\sqrt{2}} & 0 & 0 \\
        V^0 & \frac{m_1 m_P p_T (z-1)}{z} & 0 & 0 \\
        V^+ & \frac{m_1 p_T^2 (z-2)}{\sqrt{2} (z-1) z} & \frac{m_1 m_2 p_T}{(z-1) z} & -\frac{m_1 m_2^2}{\sqrt{2} (z-1)} \\
        \hline
    \end{array} 
\end{align}
\begin{align}
    &\begin{array}{|c|ccc|ccc|}
        \hline
        \multicolumn{7}{|c|}{[\boldsymbol{1P}][\boldsymbol{P2}][\boldsymbol{12}]} \\
        \hline
        \to & V^- V^- & V^- V^0 & V^- V^+ & V^+ V^- & V^+ V^0 & V^+ V^+ \\
        \hline
        V^- & 0 & \frac{m_1^2 m_2}{\sqrt{2} z} & \frac{m_1^2 p_T^*}{z} & -\frac{m_2^2 p_T^*}{z-1} & -\frac{m_2 \left(p_T^*\right)^2 (z+1)}{\sqrt{2} (z-1) z} & \frac{\left(p_T^*\right)^3}{z-z^2} \\
        V^0 & 0 & 0 & \frac{m_1^2 m_P \left(\frac{1}{z}-1\right)}{\sqrt{2}} & \frac{m_2^2 m_P z}{\sqrt{2} (z-1)} & \frac{m_2 m_P p_T^* z}{z-1} & \frac{m_P \left(p_T^*\right)^2 (2 z-1)}{\sqrt{2} (z-1) z} \\
        V^+ & 0 & 0 & 0 & 0 & -\frac{m_2 m_P^2 z}{\sqrt{2}} & -m_P^2 p_T^* \\
        \hline
    \end{array} \\
    &\begin{array}{|c|ccc|}
        \hline
        \multicolumn{4}{|c|}{[\boldsymbol{1P}][\boldsymbol{P2}][\boldsymbol{12}]} \\
        \hline
        \to & V^0 V^- & V^0 V^0 & V^0 V^+ \\
        \hline
        V^- & \frac{m_1 m_2^2}{\sqrt{2} (z-1)} & \frac{m_1 m_2 p_T^*}{(z-1) z} & -\frac{m_1 \left(p_T^*\right)^2 (z-2)}{\sqrt{2} (z-1) z} \\
        V^0 & 0 & 0 & \frac{m_1 m_P p_T^* (z-1)}{z} \\
        V^+ & 0 & 0 & -\frac{m_1 m_P^2 (z-1)}{\sqrt{2}} \\
        \hline
    \end{array}
\end{align}
and six SM structures $\langle\boldsymbol{1P}\rangle\langle\boldsymbol{P2}\rangle[\boldsymbol{12}]$, $\langle\boldsymbol{1P}\rangle[\boldsymbol{P2}]\langle\boldsymbol{12}\rangle$, $\langle\boldsymbol{1P}\rangle[\boldsymbol{P2}][\boldsymbol{12}]$, $[\boldsymbol{1P}]\langle\boldsymbol{P2}\rangle\langle\boldsymbol{12}\rangle$, $[\boldsymbol{1P}]\langle\boldsymbol{P2}\rangle[\boldsymbol{12}]$  and $[\boldsymbol{1P}][\boldsymbol{P2}]\langle\boldsymbol{12}\rangle$:
\begin{align}
    &\begin{array}{|c|ccc|ccc|}
        \hline
        \multicolumn{7}{|c|}{\langle\boldsymbol{1P}\rangle\langle\boldsymbol{P2}\rangle[\boldsymbol{12}]} \\
        \hline
        \to & V^- V^- & V^- V^0 & V^- V^+ & V^+ V^- & V^+ V^0 & V^+ V^+ \\
        \hline
        V^- & 0 & -\frac{m_1 m_P^2 (z-1)}{\sqrt{2}} & 0 & 0 & 0 & 0 \\
        V^0 & 0 & \frac{m_1 m_P p_T (1-2 z)}{2 z} & \frac{m_1 m_2 m_P}{\sqrt{2}} & -\frac{m_1 m_2 m_P}{\sqrt{2}} & -\frac{m_1 m_P p_T^*}{2 z} & 0 \\
        V^+ & 0 & -\frac{m_1 p_T^2}{\sqrt{2} z} & \frac{m_1 m_2 p_T}{z} & -\frac{m_1 m_2 p_T}{z-1} & \frac{m_1 \left(m_2^2 z-p_T p_T^*\right)}{\sqrt{2} (z-1) z} & \frac{m_1 m_2 p_T^*}{(z-1) z} \\
        \hline
    \end{array}\\
    &\begin{array}{|c|ccc|}
        \hline
        \multicolumn{4}{|c|}{\langle\boldsymbol{1P}\rangle\langle\boldsymbol{P2}\rangle[\boldsymbol{12}]} \\
        \hline
        \to & V^0 V^- & V^0 V^0 & V^0 V^+ \\
        \hline
        V^- & -\frac{m_2 m_P^2 z}{\sqrt{2}} & -\frac{1}{2} m_P^2 p_T^* & 0 \\
        V^0 & \frac{m_2 m_P p_T (1-2 z)}{2 (z-1)} & \frac{m_P \left(-m_1^2 (z-1)^2+m_2^2 z^2+p_T^* (p_T-2 p_T z)\right)}{2 \sqrt{2} (z-1) z} & \frac{m_2 m_P p_T^*}{2 (z-1)} \\
        V^+ & -\frac{m_2 p_T^2}{\sqrt{2} (z-1)} & -\frac{p_T \left(m_1^2 (z-1)-m_2^2 z+p_T p_T^*\right)}{2 (z-1) z} & \frac{m_2 \left(m_1^2 (z-1)+p_T p_T^*\right)}{\sqrt{2} (z-1) z} \\
        \hline
    \end{array}
\end{align}

\begin{align}
    &\begin{array}{|c|ccc|ccc|}
        \hline
        \multicolumn{7}{|c|}{\langle\boldsymbol{1P}\rangle[\boldsymbol{P2}]\langle\boldsymbol{12}\rangle} \\
        \hline
        \to & V^- V^- & V^- V^0 & V^- V^+ & V^+ V^- & V^+ V^0 & V^+ V^+ \\
        \hline
        V^- & \frac{m_2 m_P p_T}{z-1} & \frac{m_P \left(p_T p_T^*-m_2^2 z\right)}{\sqrt{2} (z-1)} & -\frac{m_2 m_P p_T^* z}{z-1} & 0 & 0 & 0 \\
        V^0 & \frac{m_2 p_T^2}{\sqrt{2} (z-1) z} & \frac{p_T \left(p_T p_T^*-z \left(m_2^2+m_P^2 (z-1)\right)\right)}{2 (z-1) z} & \frac{m_2 \left(m_P^2 (z-1) z-p_T p_T^*\right)}{\sqrt{2} (z-1)} & -\frac{m_1^2 m_2}{\sqrt{2} z} & -\frac{m_1^2 p_T^*}{2 z} & 0 \\
        V^+ & 0 & -\frac{m_P p_T^2}{\sqrt{2} z} & m_2 m_P p_T & 0 & \frac{m_1^2 m_P (z-1)}{\sqrt{2} z} & 0 \\
        \hline
    \end{array} \\
    &\begin{array}{|c|ccc|}
        \hline
        \multicolumn{4}{|c|}{\langle\boldsymbol{1P}\rangle[\boldsymbol{P2}]\langle\boldsymbol{12}\rangle} \\
        \hline
        \to & V^0 V^- & V^0 V^0 & V^0 V^+ \\
        \hline
        V^- & -\frac{m_1 m_2 m_P}{\sqrt{2}} & -\frac{1}{2} m_1 m_P p_T^* & 0 \\
        V^0 & -\frac{m_1 m_2 p_T (z-2)}{2 (z-1) z} & -\frac{m_1 \left(z (m_2+m_P (z-1)) (m_2-m_P z+m_P)+p_T p_T^* (z-2)\right)}{2 \sqrt{2} (z-1) z} & \frac{m_1 m_2 p_T^*}{2-2 z} \\
        V^+ & 0 & \frac{m_1 m_P p_T (z-2)}{2 z} & \frac{m_1 m_2 m_P}{\sqrt{2}} \\
        \hline
    \end{array}
\end{align}

\begin{align}
    &\begin{array}{|c|ccc|ccc|}
        \hline
        \multicolumn{7}{|c|}{\langle\boldsymbol{1P}\rangle[\boldsymbol{P2}][\boldsymbol{12}]} \\
        \hline
        \to & V^- V^- & V^- V^0 & V^- V^+ & V^+ V^- & V^+ V^0 & V^+ V^+ \\
        \hline
        V^- & 0 & \frac{m_1 m_2 m_P}{\sqrt{2}} & m_1 m_P p_T^* & 0 & 0 & 0 \\
        V^0 & 0 & \frac{m_1 m_2 p_T}{2 z} & \frac{m_1 \left(p_T p_T^*-m_P^2 (z-1) z\right)}{\sqrt{2} z} & \frac{m_1 m_2^2}{\sqrt{2} (z-1)} & \frac{m_1 m_2 p_T^* (z+1)}{2 (z-1) z} & \frac{m_1 \left(p_T^*\right)^2}{\sqrt{2} (z-1) z} \\
        V^+ & 0 & 0 & m_1 m_P p_T \left(\frac{1}{z}-1\right) & 0 & -\frac{m_1 m_2 m_P}{\sqrt{2}} & -\frac{m_1 m_P p_T^*}{z} \\
        \hline
    \end{array} \\
    &\begin{array}{|c|ccc|}
        \hline
        \multicolumn{4}{|c|}{\langle\boldsymbol{1P}\rangle[\boldsymbol{P2}][\boldsymbol{12}]} \\
        \hline
        \to & V^0 V^- & V^0 V^0 & V^0 V^+ \\
        \hline
        V^- & \frac{m_2^2 m_P z}{\sqrt{2} (z-1)} & \frac{m_2 m_P p_T^* (z+1)}{2 (z-1)} & \frac{m_P \left(p_T^*\right)^2}{\sqrt{2} (z-1)} \\
        V^0 & \frac{m_2^2 p_T}{2 (z-1)} & \frac{m_2 \left((z-1) (m_1+m_P z) (m_1-m_P z)+p_T p_T^* (z+1)\right)}{2 \sqrt{2} (z-1) z} & \frac{p_T^* \left((z-1) \left(m_1^2-m_P^2 z\right)+p_T p_T^*\right)}{2 (z-1) z} \\
        V^+ & 0 & -\frac{1}{2} m_2 m_P p_T & -\frac{m_P \left(m_1^2 (z-1)+p_T p_T^*\right)}{\sqrt{2} z} \\
        \hline
    \end{array}
\end{align}

\begin{align}
    &\begin{array}{|c|ccc|ccc|}
        \hline
        \multicolumn{7}{|c|}{[\boldsymbol{1P}]\langle\boldsymbol{P2}\rangle\langle\boldsymbol{12}\rangle} \\
        \hline
        \to & V^- V^- & V^- V^0 & V^- V^+ & V^+ V^- & V^+ V^0 & V^+ V^+ \\
        \hline
        V^- & -\frac{m_1 m_P p_T}{z} & \frac{m_1 m_2 m_P}{\sqrt{2}} & 0 & m_1 m_P p_T^* \left(\frac{1}{z}-1\right) & 0 & 0 \\
        V^0 & \frac{m_1 p_T^2}{\sqrt{2} \left(z-z^2\right)} & \frac{m_1 m_2 p_T (z+1)}{2 (z-1) z} & -\frac{m_1 m_2^2}{\sqrt{2} (z-1)} & \frac{m_1 \left(m_P^2 (z-1) z-p_T p_T^*\right)}{\sqrt{2} z} & \frac{m_1 m_2 p_T^*}{2 z} & 0 \\
        V^+ & 0 & 0 & 0 & m_1 m_P p_T & -\frac{m_1 m_2 m_P}{\sqrt{2}} & 0 \\
        \hline
    \end{array}\\
    &\begin{array}{|c|ccc|}
        \hline
        \multicolumn{4}{|c|}{[\boldsymbol{1P}]\langle\boldsymbol{P2}\rangle\langle\boldsymbol{12}\rangle} \\
        \hline
        \to & V^0 V^- & V^0 V^0 & V^0 V^+ \\
        \hline
        V^- & \frac{m_P \left(m_1^2 (z-1)+p_T p_T^*\right)}{\sqrt{2} z} & -\frac{1}{2} m_2 m_P p_T^* & 0 \\
        V^0 & \frac{p_T \left((z-1) \left(m_1^2-m_P^2 z\right)+p_T p_T^*\right)}{2 (z-1) z} & -\frac{m_2 \left((z-1) (m_1+m_P z) (m_1-m_P z)+p_T p_T^* (z+1)\right)}{2 \sqrt{2} (z-1) z} & \frac{m_2^2 p_T^*}{2 (z-1)} \\
        V^+ & \frac{m_P p_T^2}{\sqrt{2} (1-z)} & \frac{m_2 m_P p_T (z+1)}{2 (z-1)} & -\frac{m_2^2 m_P z}{\sqrt{2} (z-1)} \\
        \hline
    \end{array}
\end{align}

\begin{align}
    &\begin{array}{|c|ccc|ccc|}
        \hline
        \multicolumn{7}{|c|}{[\boldsymbol{1P}]\langle\boldsymbol{P2}\rangle[\boldsymbol{12}]} \\
        \hline
        \to & V^- V^- & V^- V^0 & V^- V^+ & V^+ V^- & V^+ V^0 & V^+ V^+ \\
        \hline
        V^- & 0 & \frac{m_1^2 m_P \left(\frac{1}{z}-1\right)}{\sqrt{2}} & 0 & m_2 m_P p_T^* & \frac{m_P \left(p_T^*\right)^2}{\sqrt{2} z} & 0 \\
        V^0 & 0 & -\frac{m_1^2 p_T}{2 z} & \frac{m_1^2 m_2}{\sqrt{2} z} & \frac{m_2 \left(p_T p_T^*-m_P^2 (z-1) z\right)}{\sqrt{2} (z-1)} & \frac{p_T^* \left(p_T p_T^*-z \left(m_2^2+m_P^2 (z-1)\right)\right)}{2 (z-1) z} & \frac{m_2 \left(p_T^*\right)^2}{\sqrt{2} \left(z-z^2\right)} \\
        V^+ & 0 & 0 & 0 & -\frac{m_2 m_P p_T z}{z-1} & \frac{m_P \left(m_2^2 z-p_T p_T^*\right)}{\sqrt{2} (z-1)} & \frac{m_2 m_P p_T^*}{z-1} \\
        \hline
    \end{array}\\
    &\begin{array}{|c|ccc|}
        \hline
        \multicolumn{4}{|c|}{[\boldsymbol{1P}]\langle\boldsymbol{P2}\rangle[\boldsymbol{12}]} \\
        \hline
        \to & V^0 V^- & V^0 V^0 & V^0 V^+ \\
        \hline
        V^- & -\frac{m_1 m_2 m_P}{\sqrt{2}} & \frac{m_1 m_P p_T^* (z-2)}{2 z} & 0 \\
        V^0 & \frac{m_1 m_2 p_T}{2-2 z} & \frac{m_1 \left(z (m_2+m_P (z-1)) (m_2-m_P z+m_P)+p_T p_T^* (z-2)\right)}{2 \sqrt{2} (z-1) z} & -\frac{m_1 m_2 p_T^* (z-2)}{2 (z-1) z} \\
        V^+ & 0 & -\frac{1}{2} m_1 m_P p_T & \frac{m_1 m_2 m_P}{\sqrt{2}} \\
        \hline
    \end{array}
\end{align}

\begin{align}
    &\begin{array}{|c|ccc|ccc|}
        \hline
        \multicolumn{7}{|c|}{[\boldsymbol{1P}][\boldsymbol{P2}]\langle\boldsymbol{12}\rangle} \\
        \hline
        \to & V^- V^- & V^- V^0 & V^- V^+ & V^+ V^- & V^+ V^0 & V^+ V^+ \\
        \hline
        V^- & \frac{m_1 m_2 p_T}{(z-1) z} & \frac{m_1 p_T p_T^*-m_1 m_2^2 z}{\sqrt{2} (z-1) z} & -\frac{m_1 m_2 p_T^*}{z-1} & \frac{m_1 m_2 p_T^*}{z} & \frac{m_1 \left(p_T^*\right)^2}{\sqrt{2} z} & 0 \\
        V^0 & 0 & -\frac{m_1 m_P p_T}{2 z} & \frac{m_1 m_2 m_P}{\sqrt{2}} & -\frac{m_1 m_2 m_P}{\sqrt{2}} & \frac{m_1 m_P p_T^* (1-2 z)}{2 z} & 0 \\
        V^+ & 0 & 0 & 0 & 0 & \frac{m_1 m_P^2 (z-1)}{\sqrt{2}} & 0 \\
        \hline
    \end{array}\\
    &\begin{array}{|c|ccc|}
        \hline
        \multicolumn{4}{|c|}{[\boldsymbol{1P}][\boldsymbol{P2}]\langle\boldsymbol{12}\rangle} \\
        \hline
        \to & V^0 V^- & V^0 V^0 & V^0 V^+ \\
        \hline
        V^- & -\frac{m_1^2 m_2 (z-1)+m_2 p_T p_T^*}{\sqrt{2} (z-1) z} & -\frac{p_T^* \left(m_1^2 (z-1)-m_2^2 z+p_T p_T^*\right)}{2 (z-1) z} & \frac{m_2 \left(p_T^*\right)^2}{\sqrt{2} (z-1)} \\
        V^0 & \frac{m_2 m_P p_T}{2 (z-1)} & \frac{m_P \left(m_1^2 (z-1)^2-m_2^2 z^2+p_T p_T^* (2 z-1)\right)}{2 \sqrt{2} (z-1) z} & \frac{m_2 m_P p_T^* (1-2 z)}{2 (z-1)} \\
        V^+ & 0 & -\frac{1}{2} \left(m_P^2 p_T\right) & \frac{m_2 m_P^2 z}{\sqrt{2}} \\
        \hline
    \end{array}
\end{align}

\end{itemize}

The above results can be converted to the form given in section~\ref{sec:bold} by considering momentum conservation,
\begin{equation}
    m^2_P= P^2=(p_1+p_2)^2= \frac{p_T p_T^*+(1-z)m_1^2+ m_2^2}{z(1-z)}.
\end{equation}
This yields a reduction rule for real-valued combinations,
\begin{equation}
    p_T p_T^* \to z(1-z)m^2_P -(1-z)m_1^2-z m_2^2.
\end{equation}
This converts the square-growth term $|p_T|^2$ into a non-growing term $m^2$, which significantly decreases the high-energy behavior. However, this reduction rule cannot be applied to the EFT structures, because they contain pure powers of $p_T$ or $p_{T}$ as complex terms, not the mixed real term $p_T p_T^*$.

The tables above collect the general 3-point splitting massive amplitudes. Together with the massless-to-massive matching relations established in this section, they provide the building blocks for the collinear splitting functions discussed next.

%% file: sec5_splitting.tex
\section{Splitting Amplitudes and Splitting Functions}
\label{sec:splitting}

In the standard DGLAP formalism \cite{Altarelli:1977zs,Dokshitzer:1977sg,Gribov:1972ri}, the collinear splitting function $P_{ba}(z)$ encodes the probability density for a parton $a$ to split into parton $b$ carrying a fraction $z$ of the longitudinal momentum. These functions govern the evolution of parton distribution functions and form the backbone of resummation programs at the LHC. In this section, we derive the complete set of leading and subleading massive splitting functions for all Standard Model particles, using the collinear amplitudes constructed in sections~\ref{sec:collinear_projection} and~\ref{sec:3pt} as input.

\begin{figure}[htbp]
\centering
\includegraphics[width=0.8\textwidth]{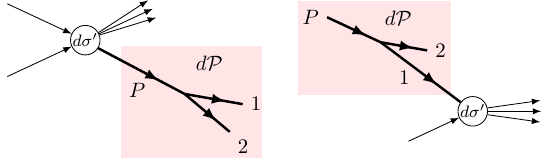}
\caption{Factorization of the cross section into the hard part $d\sigma^\prime$ and the collinear part $d\mathcal{P}$ (light red region). There are two types of factorization for the collinear splitting $P\to 1+2$: the splitting can occur in the final state (Left) or initial state (Right). }
\label{fig:ISR_FSR}
\end{figure}

In splitting problems, attention is focused on the singular parts of scattering processes, specifically divergences (for massless particles) or enhancements (for massive particles) arising from collinear behavior. In high-energy collisions, not all incoming and outgoing particles exhibit collinear behavior. The collinear part can be factorized from the total differential cross-section \cite{Chen:2016wkt}:
\begin{equation}
\begin{aligned}
d\sigma \simeq d\sigma'\times d\mathcal{P},
\end{aligned}
\end{equation}
where $d\mathcal{P}$ denotes the differential splitting function. After integration, $d\mathcal{P}$ leads to collinear divergences. This factorization may occur in the initial state or final state, corresponding to initial state radiation (ISR) or final state radiation (FSR). Figure~\ref{fig:ISR_FSR} illustrates both cases for $P\to 1+2$, with parent particle $P$ and daughters $1$ and $2$. 

Since splitting involves only a portion of the amplitude, we discuss collinear behavior between external and internal lines. This cannot be fully described by on-shell variables and necessarily mixes in off-shell degrees of freedom. For $d\mathcal{P}$, one separates on-shell and off-shell degrees of freedom:
\begin{equation}
\begin{aligned}
d\mathcal{P}\sim \underbrace{ d\Phi_{n} }_{ \text{phase space} }\times \underbrace{ \frac{1}{Q^4} }_{ \text{off-shell virtuality} } \times \underbrace{ |\mathcal{M}|^2 }_{ \text{on-shell amplitude} }.
\end{aligned}
\end{equation}
Here $Q^2$ is the virtuality of the intermediate particle and measures its departure from the mass shell. The relevant intermediate line differs between ISR and FSR, leading to the definitions:  
\begin{equation} \begin{aligned}
Q^2_{\text{initial}}&=p_1^2-m_1^2,\\
Q^2_{\text{final}}&=P^2-m_P^2.\\
\end{aligned} \end{equation}
where $P$ and $p_1$ denote the momenta of particles $P$ and $1$, and $m_P$ and $m_1$ denote their corresponding  masses.

In general, the off-shell direction may be arbitrary. However, during high-energy splitting processes, it always points in the same direction (the $p_{-}$ direction). For the remaining three on-shell degrees of freedom, since most of the momentum is concentrated along the splitting direction (the $p_{+}$ direction), the most natural kinematical decomposition is the collinear decomposition, discussed in previous sections. For the $1\to 2$ splitting process, we usually choose three independent kinematical variables as
\begin{equation}
    z=\frac{p_{1+}}{P_+},\quad p_T=p_{1x}+ip_{1y},\quad p_T^*,
\end{equation}
where the momentum fraction $z$ represents the longitudinal degrees of freedom and $p_T$ denotes the transverse one. 

After separating the longitudinal and transverse variables, the two-body FSR phase space becomes:
\begin{equation}
\begin{aligned}
d\Phi_{2}=\prod_{i=1}^2 \frac{d^3 p_{i}}{(2\pi)^3}\;\to\;\underbrace{ \frac{d p^2_{T}}{16 \pi^2}  }_{ \text{transverse} } \times \underbrace{ \frac{dz}{z \bar{z}} }_{ \text{longitudinal} },
\end{aligned}
\end{equation}
where $\bar{z}=1-z$ is the momentum fraction of particle $2$. The virtuality can also be expressed by these splitting variables with the particle masses
\begin{equation} \begin{aligned}
Q^2_{\text{final}}&=\frac{|p_{T}|^2-z\bar{z}m_{P}^2+\bar{z}m_{1}^2+ zm_{2}^2}{z\bar{z}}.\\
\end{aligned} \end{equation}
Similarly, we can also decompose the phase space and virtuality for ISR. Although their expressions differ from those for FSR, they lead to the same splitting behavior in the $d\mathcal{P}$. Thus we can focus on the discussion of FSR to extract the splitting behavior.

Then let us consider the power counting for the massive splitting problem, which can further organize the on-shell degrees of freedom. In collinear physics, we have $p_+ \gg p_T$. Mass effects can be treated as subleading contributions and become important only in the region where $p_T \gtrsim m$. Therefore, the alignment regime $p_+ \gg p_T > m$ is consistent with this power counting, and we can perform an expansion within this framework.

Since different particles have different masses, it is convenient to replace the mass parameters with the electroweak vev $v$ in the expansion. Both the virtuality and the squared amplitude can then be expanded as
\begin{equation}
\begin{aligned}
Q^2_{\text{final}} &\sim \frac{p_{T}^2}{z \bar{z}}\left(1+\mathcal{O}\left( \frac{v^2}{p_{T}^2}  \right)\right), \\
|\mathcal{M}^2| &\sim p_{T}^2 |M^{(0)}(z)|^2 + v^2 |M^{(1)}(z)|^2.
\end{aligned}
\end{equation}
where $M^{(0)}$ and $M^{(1)}$ are dimensionless functions of $z$ that also contain dimensionless couplings. There is no interference term between amplitudes at different orders, such as $M^{(0)} M^{(1)*}$, because $M^{(0)}$ and $M^{(1)*}$ correspond to different spin polarizations and are orthogonal in spin space. In the following, we also neglect the $\mathcal{O}(\frac{v^2}{p_{T}^2})$ expansion in the propagator virtuality $Q^2$. These kinematic corrections must be restored when constructing a splitting function that goes beyond the amplitude-induced mass terms. 

With these conventions, we have
\begin{equation}
\begin{aligned}
d\mathcal{P}\sim  d\mathcal{P}_{0}+d\mathcal{P}_{1}.
\end{aligned}
\end{equation}
It includes two order contributions:

\begin{enumerate}

\item \textbf{Leading order:}
\begin{equation}
\begin{aligned}
d\mathcal{P}_{0}\sim \frac{dz}{16 \pi^2}\times \underbrace{ z \bar{z}|M^{(0)}(z)|^2 }_{ \text{splitting function} }\times \frac{d p^2_{T}}{p_{T}^2}
\end{aligned}
\end{equation}
Here, integration over $dp_{T}^2 / p_{T}^2$ yields a logarithmic divergence, and the coefficient of this divergence is called the splitting function, determined by the leading-order amplitude $M^{(0)}(z)$. We see that the longitudinal and transverse degrees of freedom correspond to different physics (the divergence and its coefficient).

\item \textbf{Subleading order:}
\begin{equation}
\begin{aligned}
d\mathcal{P}_{1}\sim \frac{dz}{16 \pi^2}\times z \bar{z}|M^{(1)}(z)|^2\times \frac{v^2 d p^2_{T}}{p_{T}^4}.
\end{aligned}
\end{equation}
where $v$ is the vev of symmetry breaking.
Similarly, the dimensionless coefficient of the transverse-momentum integral defines the mass-suppressed splitting coefficient.

\end{enumerate}

For convenience, we can define these dimensionless, coupling-dressed coefficients $P^{(0)}(z)$ and $P^{(1)}(z)$ as the \textit{leading and subleading splitting functions}. They are given by
\begin{equation}
\begin{aligned}
P^{(0)}(z)&\simeq z\bar{z} |M^{(0)}(z)|^2,  \\
P^{(1)}(z)&\simeq z\bar{z} |M^{(1)}(z)|^2.
\end{aligned}
\end{equation}
These should not be confused with the conventional DGLAP splitting kernel, which is independent of the coupling constant.

These splitting functions are the truly on-shell and non-divergent quantities in the splitting problems, which describe the probability of finding one parton inside another. Therefore, the naturally required symmetry is the two-dimensional Euclidean $ISO(2)$ (the homogeneous part of the Galilean group $\text{Gal}(2)$), which leaves the longitudinal degree of freedom $z$ invariant. This implies the SW spinors deriving from $\text{Gal}(2)$ are suitable for deriving splitting functions.

The Galilean action has a geometric interpretation. As illustrated in figure~\ref{fig:momentum_space}, the pale red plane represents a hypersurface of fixed light-front momentum $p_{+}$, on which the fraction $z$ is unchanged. Under a Galilean transformation, each particle transforms along a dashed parabola,  which directly reflects the nonrelativistic form of the transverse dispersion relation. This geometry makes the Galilean organization of the transverse kinematics manifest.

\begin{figure}[htbp]
\centering
\includegraphics[width=0.49\linewidth]{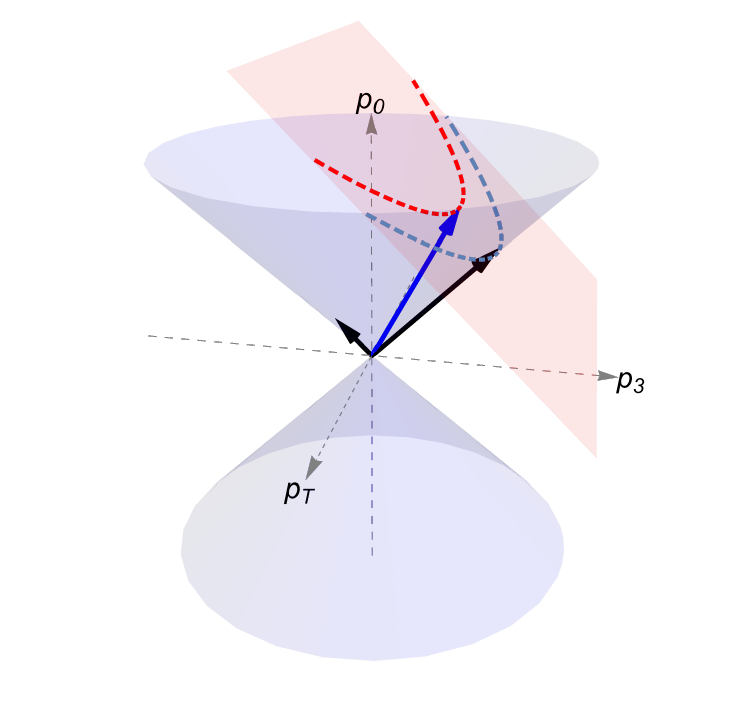}
\caption{Illustration of a Galilean boost acting on a massive particle (blue arrow) and the two massless particles (black arrows) decomposed from it. The dashed curve represents the trajectory under this continuous Galilean boost.}
\label{fig:front}
\end{figure}

Note one can decompose the massive momentum into two massless momenta lying on the light cone. This leads to two understandings of massive collinear particles:
\begin{itemize}
\item One approach is to focus on the massive particle. Under the Galilean transformation, the transverse momentum $p_T$ changed while the longitudinal fraction $p_+$ remained unchanged, so the longitudinal and transverse polarizations of the massive vector amplitude must be treated separately. This can become cumbersome when multiple vector bosons are involved.

\item The second approach decomposes each massive state into two massless spinors carrying leading and mass-suppressed components. The leading component transforms like the energetic massive state, while the suppressed component is represented by the auxiliary Higgs leg and is invariant under the relevant Galilean boost. The resulting massless amplitude is often more compact than its massive counterpart and can therefore simplify the extraction of the splitting coefficient.
\end{itemize}

In the following, let us apply these two approaches to massive splitting functions, and give a comparison.

\subsection{Massive Splitting Function}

We first derive the massive splitting functions from the massive amplitudes. Let us begin with the example of the $f\to Sf$ splitting process, which can be described by the following amplitude,
\begin{equation}
\mathcal{M}(\mathbf{P}_f, \mathbf{1}_h, \mathbf{2}_{\bar{f}})=
{y} \langle\mathbf{P2}\rangle +{y'} [\mathbf{P2}].
\end{equation}
We consider the case $f^+\to Sf$. The initial state is a fermion with a positive spin component, while the final state can have different spin components,
\begin{equation} \begin{aligned} \label{eq:fSf_amp}
\mathcal{M}(\mathbf{P}_f^-, \mathbf{1}_h^0,\mathbf{2}_{\bar{f}}^-)&=y\langle P^- 2^- \rangle=-y\frac{p_T}{\sqrt{\bar{z}}},\\
\mathcal{M}(\mathbf{P}_f^-, \mathbf{1}_h^0,\mathbf{2}_{\bar{f}}^+)&=y\langle P^- 2^+ \rangle+y^\prime[ P^- 2^+ ]=-y\frac{m_2}{\sqrt{\bar{z}}}-y^\prime \sqrt{\bar{z}}m_P.
\end{aligned} \end{equation}
For the second line, the two massive fermions in the Yukawa interaction should be the same particle, so the coupling constant and mass simplify as $m_2=m_P=m_f$ and $y=y^\prime$. Using this simplification and eliminating the scale dependence of $p_T$ and $v$, we can derive the leading and subleading splitting functions,
\begin{equation} \begin{aligned}
P^{(0)}_{Sf}&=z\bar{z} \left|-y\frac{1}{\sqrt{\bar{z}}}\right|^2=y^2 z,\\
P^{(1)}_{Sf}&=z\bar{z} \left|-y\frac{1}{\sqrt{\bar{z}}}\frac{m_f}{v}-y \sqrt{\bar{z}}\frac{m_f}{v}\right|^2=y^2(1+\bar{z})^2 z\frac{m_f}{v}.
\end{aligned} \end{equation}
Here $P^{(i)}_{Sf}$ corresponds to the probability of finding a parton $S$ inside a parton $f$, dressed with the coupling constant. 

For the case with a massive vector boson, the transverse and longitudinal polarizations must be calculated separately. Consider the splitting process $f^-\to Vf$, which is described by the $fVf$ amplitude. For a \textit{transverse} vector boson, the leading contribution includes two polarizations $\pm$ for this vector boson,
\begin{equation}\begin{aligned}
\mathcal{M}(\mathbf{P}_{f}^-, \mathbf{1}_{V}^+,\mathbf{2}_{f}^+)
&= X_1  \frac{\langle P^- 1^+\rangle [2^+ 1^+]}{m_1}
=-X_1   \frac{p_T^*}{\sqrt{\bar{z}}z},\\
\mathcal{M}(\mathbf{P}_{f}^-, \mathbf{1}_{V}^-,\mathbf{2}_{f}^+)
&= X_1  \frac{\langle P^- 1^-\rangle [2^+ 1^-]}{m_2}
= X_1  \frac{\sqrt{\bar{z}}}{z}p_T,
\end{aligned}
\end{equation}
For the subleading contribution, one spin component is included,
\begin{equation} \begin{aligned}
\mathcal{M}(\mathbf{P}_f^-, 1_V^+, \mathbf{2}_{f}^-)
&=X_1\frac{\langle P^- 1^+\rangle[2^- 1^+]}{m_1}+X_2\frac{[P^- 1^+]\langle 2^- 1^+\rangle}{m_1}
=X_1\frac{1}{\sqrt{\bar{z}}}m_2-X_2\sqrt{\bar{z}} m_P.
\end{aligned} \end{equation}
Eliminating the scale dependence of $p_T$ and $v$, we can derive the leading and subleading splitting functions,
\begin{equation} \begin{aligned}
P^{(0)}_{V_T f}&=X_1^2 \frac{1+\bar{z}^2}{z},\\
P^{(1)}_{V_T f}&=(X_1 \frac{m_2}{v}-X_2 \bar{z} \frac{m_P}{v})^2 z.\\
\end{aligned} \end{equation}
where $V_T$ denotes the transverse vector boson.

For a \textit{longitudinal} vector boson, we should work in the Goldstone equivalence gauge as described in section~\ref{sec:3pt}, decomposing the $ffV$ amplitude into a Goldstone amplitude $\mathcal{M}_\pi$ and a residual vector amplitude $\mathcal{M}_V$. At leading order, only $\mathcal{M}_\pi$ contributes:
\begin{equation} \begin{aligned}
\mathcal{M}_{\pi}(\mathbf{P}_{f}^-, \mathbf{1}_{V}^0, \mathbf{2}_{f}^-) &=-X_1 \frac{m_2\langle P^- 2^- \rangle }{m_1} +X_2 \frac{m_P\langle P^- 2^- \rangle}{m_1}
=\left(X_1 \frac{m_2}{m_1}-X_2 \frac{m_P}{m_1}\right)\frac{p_T}{\bar{z}}, \\
\mathcal{M}_{V}(\mathbf{P}_{f}^-, \mathbf{1}_{V}^0, \mathbf{2}_{f}^-) &=0 .
\end{aligned} \end{equation}
The Goldstone component $\mathcal{M}_\pi$ is structurally identical to the $fSf$ amplitude in eq.~\eqref{eq:fSf_amp}. At subleading order, both $\mathcal{M}_\pi$ and $\mathcal{M}_V$ contribute:
\begin{equation} \begin{aligned}
\mathcal{M}_{\pi}(\mathbf{P}_{f}^-, \mathbf{1}_{V}^0, \mathbf{2}_{f}^+)&=X_1 \frac{m_P[P^- 2^+]-m_2\langle P^- 2^+\rangle }{\mathbf{m}_1} +X_2 \frac{m_P\langle P^- 2^+\rangle-m_2[P^- 2^+] }{\mathbf{m}_1} \\
&=X_1 (\sqrt{\bar{z}} \frac{-m_P^2 }{m_1}+\frac{1}{\sqrt{\bar{z}}}\frac{m_2^2 }{m_1}) - X_2 \frac{m_2 m_P  }{m_1}(\frac{z}{\sqrt{\bar{z}}} ), \\
\mathcal{M}_{V}(\mathbf{P}_{f}^-, \mathbf{1}_{V}^0, \mathbf{2}_{f}^+) &=-2X_1\frac{\langle P^- 1^+\rangle[2^+ 1^-]}{m_1}
=2X_1\frac{\sqrt{\bar{z}}}{z}m_1.
\end{aligned} \end{equation}
The two fermions can now be different, so their masses $m_P,m_2$ and the two couplings $X_1,X_2$ are in general independent, which differs from the $f\to Sf$ case. Therefore, the splitting functions should be 
\begin{equation} \begin{aligned}
P^{(0)}_{V_L f}&=(X_1 \frac{m_2}{m_1}-X_2 \frac{m_P}{m_1})^2 z,\\
P^{(1)}_{V_L f}&=\left(X_1 (-\bar{z} \frac{m_P^2 }{m_1 v}+\frac{m_2^2 }{m_1 v})z -X_2 \frac{m_2 m_P  }{m_1 v}z^2 +2X_1 \bar{z} \frac{m_1}{v}\right)^2 \frac{1}{z}.
\end{aligned} \end{equation}
where $V_L$ denotes the longitudinal vector boson.

A subtle but important point should be emphasized regarding the distinction between splitting amplitudes and splitting functions. One may recall that here we use the SW amplitude, which is not equal to the AHH amplitude:
\begin{equation}
\mathcal{M}^{\mathcal{I}\mathcal{J}}_{\text{SW}} \neq \mathcal{M}^{\mathcal{I}\mathcal{J}}_{\text{AHH}}
\end{equation}
where $\mathcal{I}$ denotes the little-group index for the parent particle and $\mathcal{J}$ denotes those for the daughter particles. The two amplitudes are related by a nontrivial little-group rotation $W$. On the other hand, the splitting function is defined as the square of the splitting amplitudes summed over daughter spins,
\begin{equation}
  P^{(i)}(z)\propto \sum_{\mathcal{J}}\mathcal{M}^{\mathcal{IJ}}\mathcal{M}^*_{\mathcal{JI}},
\end{equation}
which is invariant under little-group rotations. Thus, the splitting functions are independent of the choice of spin axis.

This treatment can be applied to derive other massive splitting functions. However, when more vector bosons are included, we need to decompose each longitudinal mode into a Goldstone component and a residual vector component, making a direct calculation increasingly lengthy. In the next subsection, we will show that the massless amplitudes introduced in the previous section provide a more efficient route to the complete set of splitting functions.

\subsection{Massless Origin of the Leading and Subleading Splitting Functions}

In this subsection, we derive the massive splitting functions from massless amplitudes. To illustrate the procedure, we first review the correspondence between massless and massive splitting amplitudes. We begin by expanding the massive amplitude. In previous discussions, we did not distinguish between $p_T$ and $p_T^*$ in the power counting, as they are of the same order. Here, however, we must distinguish them to obtain an accurate description. The 3-point massive amplitude then expands as
\begin{equation}
\mathcal{M}= \mathcal{M}^{(0)} p_T + \tilde{\mathcal{M}}^{(0)} p_T^* + \mathcal{M}^{(1)} v,
\end{equation}
where the three terms are holomorphic, anti-holomorphic, and real functions, respectively, in the complex momentum space $p_T$. According to the massless-massive correspondence, they map to 3-point and 4-point amplitudes as follows,
\begin{equation} \begin{aligned}
\mathcal{M}^{(0)} &= \frac{\mathcal{A}_3}{p_T}=\frac{\mathcal{A}_3}{\sqrt{z\bar{z}}\langle 12\rangle},\quad& 
\mathcal{M}^{(1)} &= \mathcal{A}_4,& \\
\tilde{\mathcal{M}}^{(0)} &= \frac{\mathcal{A}_3}{p_T^*}= \frac{\mathcal{A}_3}{\sqrt{z\bar{z}} [12]},&\\
\end{aligned}
\end{equation}
where we have used $\langle 12\rangle=\sqrt{z\bar{z}}p_T$ and $[12]=\sqrt{z\bar{z}}p_T^*$. The contributions to the splitting function can thus be expressed in terms of massless amplitudes,
\begin{equation}
\begin{aligned}
\text{Leading}&:\quad
\left\{\begin{aligned}
z\bar{z}|\mathcal{M}^{(0)}|^2 =  \left|\frac{\mathcal{A}_3}{\langle 12 \rangle}\right|^2 \\
z\bar{z}|\tilde{\mathcal{M}}^{(0)}|^2 =  \left|\frac{\mathcal{A}_3}{[12]}\right|^2 \\
\end{aligned}\right.,\quad &  \\
\text{Subleading}&:\qquad z\bar{z}|\mathcal{M}^{(1)}|^2  =z\bar{z} |\mathcal{A}_4|^2.&
\end{aligned}
\end{equation}

We now address how to express the massless splitting amplitudes. In the on-shell formalism, these amplitudes admit multiple equivalent forms due to momentum conservation and the Schouten identity. Recall that the splitting function is a dimensionless probability, which eliminates any dependence on the scales $p_T$, $p_T^*$, or $m$. This indicates that we need to study the momentum or mass dependence of the massless spinor contractions:
\begin{equation}
\begin{aligned}
\langle ij \rangle &\sim p_{T},& [ij]&\sim p_{T}^*,& \\
\langle i 3 \rangle &\sim m_{i},& [i 3]&\sim m_{i}^*,&
\end{aligned}
\end{equation}
where $i,j=P,1,2$. Therefore, we should extract structures that are independent of momentum or mass by forming ratios of  $\lambda$ or $\tilde{\lambda}$ bracket expressions,
\begin{equation}
\frac{\langle ij\rangle }{\langle i'j' \rangle },\;
\frac{\langle i3\rangle }{\langle i'3 \rangle },\;
\frac{[ij] }{[ i'j' ] },\;
\frac{[i3] }{[ i'3 ] }.
\end{equation}
In the following, we express the massless amplitudes in terms of such building blocks.

\paragraph{Leading order:} The leading-order splitting function originates from the 3-point massless amplitude, which is determined by the helicity of each particle.  Depending on the total helicity, it can indeed be expressed solely in terms of $\lambda$ or $\tilde{\lambda}$:
\begin{equation}
\mathcal{A}_{3}(P^{\mathsf{h}_P},1^{\mathsf{h}_1},2^{\mathsf{h}_2})=\left\{\begin{aligned}
&\langle P1\rangle^{-\mathsf{h}_P-\mathsf{h}_1+\mathsf{h}_2}\langle P2\rangle^{-\mathsf{h}_P-\mathsf{h}_2+\mathsf{h}_1}\langle12\rangle^{\mathsf{h}_P-\mathsf{h}_1-\mathsf{h}_2},& \mathsf{h}_P+\mathsf{h}_{1}+\mathsf{h}_{2}&=-1,\\
&[P1]^{\mathsf{h}_P+\mathsf{h}_1-\mathsf{h}_2} [P2]^{\mathsf{h}_P+\mathsf{h}_2-\mathsf{h}_1} [12]^{-\mathsf{h}_P+\mathsf{h}_1+\mathsf{h}_2},&  \mathsf{h}_P+\mathsf{h}_{1}+\mathsf{h}_{2}&=+1.
\end{aligned}\right.
\end{equation}
Here $\mathsf{h}_i$ denotes the helicity of particles $i=P,1,2$. In a renormalizable theory, the constraint $\mathsf{h}_P+\mathsf{h}_1+\mathsf{h}_2=\pm 1$ applies.

Thus there are only two independent helicities. We choose $\mathsf{h}_P$ and $\mathsf{h}_1$ to characterize the splitting process. This leads to a universal expression for the leading-order splitting function:
\begin{equation}
P^{(0)}=
\left(\frac{1}{\sqrt{2}}\right)^{n_L}\sum_{\mathsf{h}_1} \left(\left|\frac{\mathcal{A}_{3}(P^{\mathsf{h}_P},1^{\mathsf{h}_1},2^{1-\mathsf{h}_P-\mathsf{h}_1})}{[12]}\right|^2
+ \left|\frac{\mathcal{A}_{3}(P^{\mathsf{h}_P},1^{\mathsf{h}_1},2^{-1-\mathsf{h}_P-\mathsf{h}_1})}{\langle12\rangle}\right|^2\right),
\end{equation}
where $n_L$ is the number of longitudinal modes, each normalized by $\frac{1}{\sqrt{2}}$. The massless amplitudes evaluate to
\begin{equation}
\begin{aligned}
\frac{\mathcal{A}_3}{\langle 12\rangle}=& \left(\frac{\langle P1\rangle}{\langle 12\rangle}\right)^{-2\mathsf{h}_P-2\mathsf{h}_1-1} \left(\frac{\langle P2\rangle}{\langle12\rangle}\right)^{1+2\mathsf{h}_1},&\\
\frac{\mathcal{A}_3}{[12]}=& \left(\frac{ [P1]}{[ 12]}\right)^{2\mathsf{h}_P+2\mathsf{h}_1+1} \left(\frac{ [P2] }{ [12] }\right)^{-1-2\mathsf{h}_1},&
\end{aligned}
\end{equation}
The splitting spinor contractions give
\begin{equation}
\begin{aligned}
\frac{\langle P1\rangle}{\langle 12\rangle}=-\frac{[P1]}{[ 12]}=\sqrt{\bar{z}},\qquad
\frac{\langle P2\rangle}{\langle 12\rangle}=-\frac{[P2]}{[ 12]}=-\sqrt{z}.
\end{aligned}
\end{equation}
Hence the splitting functions become functions of $z$ only.

We now turn to specific splitting processes, where the massless coefficients will be included. Using the relations in Tables~\ref{tab:fermion_matching} and \ref{tab:boson_matching}, we can match the massless coefficients to the massive ones and derive the massive leading splitting functions.

\begin{figure}[htbp]
\centering
\includegraphics[page=1,width=0.65\textwidth]{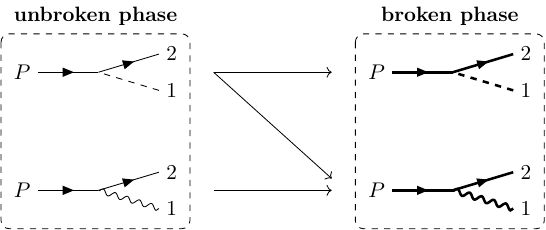}
\caption{Leading splitting function matching involving fermions. The left column corresponds to massless splittings $f\to Sf$ and $f\to Vf$, while the right column corresponds to the massive ones.}
\label{fig:leading_f_match}
\end{figure}

We first consider two typical splitting processes involving fermions:

\begin{itemize}

\item \textit{$f\to Sf$:} 
This process is described by the massless $fSf$ amplitude. When the parent fermion has positive helicity, there is only one contribution:
\begin{equation}
\mathcal{A}(P^-_{f},1^0_S,2^-_{f})=Y\langle P2\rangle.
\end{equation}
This yields the massless splitting function
\begin{equation}
P_{Sf}=\left|Y\frac{\langle P2\rangle}{\langle 12\rangle}\right|^2= Y^2 z.
\end{equation}

As illustrated in figure~\ref{fig:leading_f_match}, this massless splitting $f\to Sf$ can match to the massive splittings $f\to Sf$ or $f\to Vf$. By mapping the massless Yukawa coupling $Y$ to the massive coefficients, we obtain:
\begin{equation} \begin{aligned}
P_{S f}^{(0)} =y^2 z, \qquad
P_{V_L f}^{(0)} =\frac{1}{2}(-\frac{m_{P}}{m_{2}} X_{1}+\frac{m_{1}}{m_{P}} X_{2})^2 z.
\end{aligned} \end{equation}
The factor of $1/2$ arises from the normalization of the longitudinal mode $V_L$.

\item \textit{$f\to Vf$:} 
This process is described by the massless $fVf$ amplitude. For a positive-helicity parent fermion, there are two contributions:
\begin{equation} \begin{aligned}
\mathcal{A}(P^{-}_{f},1^{+}_{V},2^{+}_{\bar{f}})=\tilde{T}_f\frac{[12]^2}{[P2]},\qquad
\mathcal{A}(P^{-}_{f},1^{-}_{V},2^{+}_{\bar{f}})=\tilde{T}_f\frac{\langle P1 \rangle ^2}{\langle P2 \rangle }.
\end{aligned}
\end{equation}
These give the massless splitting function
\begin{equation}
P_{Vf}=\left|\tilde{T}_f\frac{[12]}{[P2]}\right|^2 + \left|\tilde{T}_f\frac{\langle P1\rangle^2}{\langle P2\rangle\langle 12\rangle}\right|^2  = \tilde{T}_f^2 (\frac{1}{z}+\frac{\bar{z}^2}{z}).
\end{equation}
It matches the massive counterpart
\begin{equation}
P^{(0)}_{V_T f} = X_2^2 \left(\frac{1}{z}+\frac{\bar{z}^2}{z}\right).
\end{equation}

\end{itemize}

\begin{figure}[htbp]
\centering
\includegraphics[page=2,width=0.65\textwidth]{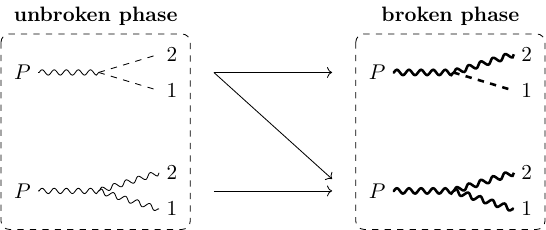}
\caption{Leading splitting function matching with bosons only. The left column corresponds to massless splittings $V\to SS$ and $V\to VV$, while the right column corresponds to the massive splittings $V\to SV$ and $V\to VV$.}
\label{fig:leading_b_match}
\end{figure}

Then we consider two typical splitting processes with only bosons:

\begin{itemize}

\item \textit{$V\to SS$:} 
This process is described by the massless $VSS$ amplitude. We again take the parent particle to have positive helicity, and there is one contribution:
\begin{equation}
\mathcal{A}(P^{-}_{V},1^{0}_{S},2^{0}_{\bar{S}})=T_s\frac{\langle P1\rangle\langle 2P\rangle}{\langle12\rangle}.
\end{equation}
It gives the massless splitting function
\begin{equation}
P_{SV}=\left|T_s \frac{\langle P1\rangle\langle P2\rangle}{\langle 12\rangle^2}\right|^2=T_s^2 z\bar{z}.
\end{equation}
As illustrated in figure~\ref{fig:leading_b_match}, this massless splitting $V\to SS$ can match to the massive splittings $V\to SV$ or $V\to VV$,
\begin{equation}
P^{(0)}_{SV_L} = \frac{1}{2}\mathbf{g}^2\, z\bar{z},\qquad
P^{(0)}_{V_L V_L} = \left(\frac{1}{2}\right)^2\left(\frac{m^2_{P}-m^2_{1}-m^2_{2}}{m_{1} m_{2}}2\mathbf{f}^{\mathbf{P12}}\right)^2\, z\bar{z}.
\end{equation}
Each factor of $\frac{1}{2}$ arise from a longitudinal mode $V_L$.

\item \textit{$V\to VV$:} This process is described by the massless $VVV$ amplitude. For a positive-helicity parent gauge boson, there are three contributions:
\begin{equation} \begin{aligned}
\mathcal{A}(P^{-}_{V},1^{-}_{V},2^{+}_{V})
&=f^{P12} \frac{\langle P1\rangle^3}{\langle 12\rangle\langle 2P\rangle},\quad&
\mathcal{A}(P^{-}_{V},1^{+}_{V},2^{-}_{V})&=f^{P12} \frac{\langle P2\rangle^3}{\langle 12\rangle\langle 2P\rangle},\\
\mathcal{A}(P^{-}_{V},1^{+}_{V},2^{+}_{V})&=f^{P12} \frac{[12]^3}{[P1][2P]}.&
\end{aligned} \end{equation}
These yield the massless splitting function
\begin{equation} \begin{aligned}
P_{VV}
&=\left|f^{P12} \frac{\langle P1\rangle^3}{\langle 12\rangle^2 \langle P2\rangle}\right|^2 
+ \left|f^{P12} \frac{\langle P2\rangle^3}{\langle 12\rangle^2 \langle P2\rangle}\right|^2
+ \left|f^{P12} \frac{[12]^2}{[P1][2P]}\right|^2 \\
&=(f^{P12})^2\left(\frac{z^3}{\bar{z}}+\frac{\bar{z}^3}{z}+\frac{1}{z\bar{z}}\right).
\end{aligned}
\end{equation}
It can match the leading massive splitting function
\begin{equation}
P^{(0)}_{V_T V_T}
=(2\mathbf{f}^{\mathbf{P12}})^2\left(\frac{z^3}{\bar{z}}+\frac{\bar{z}^3}{z}+\frac{1}{z\bar{z}}\right).
\end{equation}

\end{itemize}

Additional cases not explicitly discussed above can be derived following the same procedure from their 3-point massless amplitudes. The complete massless and leading-order massive splitting functions for the Standard Model are summarized in Tables~\ref{tab:massless_splitting} and \ref{tab:leading_splitting2}. These splitting functions are consistent with the previous result~\cite{Chen:2016wkt,Nardi:2024tce}.

\begin{table}[htbp]
\centering 
\begin{tabular}{c|ccc}
\hline
Particle P to 1 & S & V & f \\
\hline
S & \includegraphics[page=1]{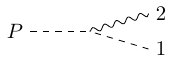} &
\includegraphics[page=2]{figure/massless_splitting.pdf} & 
\includegraphics[page=3]{figure/massless_splitting.pdf} \\
 & \rule[-10pt]{0pt}{5pt} $\displaystyle T_s^2 \left(\frac{z}{\bar{z}}\right)$ & 
$\displaystyle T_s^2 \left(\frac{\bar{z}}{z}\right)$ & $(Y^2+\tilde{Y}^2)$
 \\
\hline
V & 
\includegraphics[page=4]{figure/massless_splitting.pdf} & 
\includegraphics[page=5]{figure/massless_splitting.pdf} & 
\includegraphics[page=6]{figure/massless_splitting.pdf} \\
 & 
\rule[-12pt]{0pt}{5pt} $T_s^2(z\bar{z})$ & 
$\displaystyle f^2 \left(\frac{\bar{z}^3}{z}+\frac{z^3}{\bar{z}}+\frac{1}{z\bar{z}}\right)$ & 
$T_f^2 z^2+\tilde{T}_f^2 \bar{z}^2$ \\
\hline
f & 
\includegraphics[page=7]{figure/massless_splitting.pdf} & 
\includegraphics[page=8]{figure/massless_splitting.pdf} & \\
 & \rule[-12pt]{0pt}{5pt} $Y^2 (z)$ & $\displaystyle T_f^2 \left(\frac{1}{z}+\frac{\bar{z}^2}{z}\right)$ &  \\
\hline
\end{tabular}
\caption{Summary of massless splitting functions with Standard Model coupling constants indicated. The parent particle helicities are taken to be positive or zero. Each helicity is treated separately, and no averaging over initial-state helicities is performed.}
\label{tab:massless_splitting}
\end{table}

\begin{table}[htbp]
\centering
\adjustbox{max width=\textwidth}{
\begin{tabular}{c|c|cc|c}
\hline
Particle P to 1 & S & $V_L$ & $V_T$ & f \\
\hline
S &  &
  \multicolumn{2}{c|}{\includegraphics[page=2]{figure/massive_splitting.pdf}} &
  \includegraphics[page=3]{figure/massive_splitting.pdf} \\
 & & 
  \rule[-10pt]{0pt}{5pt} $\displaystyle \frac{1}{2}\mathbf{g}^2 \left(\frac{z}{\bar{z}}\right)$ &
  $\displaystyle \frac{1}{2}\mathbf{g}^2 \left(\frac{\bar{z}}{z}\right)$ &  
  $(y^2+y^{\prime 2})$ \\
\hline
 & 
  \includegraphics[page=4]{figure/massive_splitting.pdf} & 
  \multicolumn{2}{c|}{\includegraphics[page=5]{figure/massive_splitting.pdf}} & 
  \includegraphics[page=6]{figure/massive_splitting.pdf} \\
$V_L$ & 
  \rule[-12pt]{0pt}{5pt} $\displaystyle \frac{1}{2}\mathbf{g}^2 \left(\frac{z}{\bar{z}}\right)$ & 
  $\displaystyle \frac{1}{4}(2\mathbf{f})^2  \left(\frac{z}{\bar{z}}\right)$ & 
  $\displaystyle \frac{1}{4}(\frac{m^2_{1}-m^2_{2}-m^2_{P}}{m_{2} m_{P}}2\mathbf{f})^2 \left(\frac{\bar{z}}{z}\right)$ & 
  $(Y^2+\tilde{Y}^2)$ \\
$V_T$ & 
  \rule[-12pt]{0pt}{5pt} $\displaystyle \frac{1}{2}\mathbf{g}^2 (z\bar{z})$ & 
  $\displaystyle \frac{1}{4}(\frac{m^2_{P}-m^2_{1}-m^2_{2}}{m_{1} m_{2}}2\mathbf{f})^2 (z\bar{z})$ & 
  $\displaystyle \frac{1}{4}(2\mathbf{f})^2 \left(\frac{\bar{z}^3}{z}+\frac{z^3}{\bar{z}}+\frac{1}{z\bar{z}}\right)$ & 
  $T_f^2 z^2+\tilde{T}_f^2 \bar{z}^2$ \\
\hline
f & 
  \includegraphics[page=7]{figure/massive_splitting.pdf} & 
  \multicolumn{2}{c|}{\includegraphics[page=8]{figure/massive_splitting.pdf}} & \\
 & \rule[-12pt]{0pt}{5pt} $y^2 (z)$ & 
  $\displaystyle \frac{1}{2}(-\frac{m_{P}}{m_{2}} X_{1}+\frac{m_{1}}{m_{P}} X_{2})^2 (z)$ & 
  $\displaystyle X_1^2 \left(\frac{1}{z}+\frac{\bar{z}^2}{z}\right)$ &  \\ 
\hline
\end{tabular}
}
\caption{Summary of leading massive splitting functions in the Standard Model, with vector $V$ split into transverse polarization $V_T$ and longitudinal polarization $V_L$. The massive coefficient $\mathbf{f}$ denotes the antisymmetric structure $\mathbf{f}^{\mathbf{P12}}$. }
\label{tab:leading_splitting2}
\end{table}

\paragraph{Subleading order:} The subleading term is related to the 4-point massless amplitude, which involves an additional Higgs boson $h$ with helicity $0$. In the Standard Model, the other three particles satisfy the helicity constraint $\mathsf{h}_P+\mathsf{h}_1+\mathsf{h}_2=0$. We can therefore label the 4-point splitting amplitude by helicity $\mathsf{h}_P$ and $\mathsf{h}_1$:
\begin{equation}
\begin{aligned}
\mathcal{A}(P^{\mathsf{h}_P},1^{\mathsf{h}_1},2^{-\mathsf{h}_P-\mathsf{h}_1}, h^{0}).
\end{aligned}
\end{equation}

In this case, multiple external scalar bosons may appear. We must distinguish between UV and IR scalar bosons, which were omitted earlier for simplicity. In the Standard Model, the UV scalar bosons are complex Higgs doublet $S$ and its conjugate $\bar{S}$, while the IR scalar boson is the real Higgs boson $h$. Thus, to match the massive amplitude, we must include the contributions from the exchange of both $S$ and $\bar{S}$:
\begin{equation}
\begin{aligned}
M^{(1)}=\left(\frac{1}{\sqrt{2}}\right)^{n_L}\sum_{S,\bar{S}\to h} \mathcal{A}(P^{\mathsf{h}_P},1^{\mathsf{h}_1},2^{-\mathsf{h}_P-\mathsf{h}_1},h^{0}).
\end{aligned}
\end{equation}
As before, the normalization factor $(\frac{1}{\sqrt{2}})^{n_L}$ must be included for longitudinal vector bosons.

In the previous 3-point discussion, we considered at most two external scalar bosons. Their exchange yields the same amplitude structure. For example, exchanging particles 1 and 2 in the $VSS$ amplitude $\frac{\langle P1\rangle\langle P2\rangle}{\langle12\rangle}$ introduces a factor of $-1$. This effect is already accounted for in the coefficient matching of eq.~\eqref{eq:coeff_match}. Therefore, we only need to consider exchanges that generate different amplitude structures. After including these contributions, the splitting function is given by
\begin{equation}
\begin{aligned}
P^{(1)}=z \bar{z} \sum_{h_1}|M^{(1)}|^2.
\end{aligned}
\end{equation}

To obtain this splitting function, we must express the massless amplitude solely in terms of ratios of $\lambda$ and $\tilde{\lambda}$. In 4-point kinematics, momentum conservation relates angle and square brackets, so a given  amplitude has several equivalent representations:
\begin{equation}
\begin{aligned}
\mathcal{A}_{4}=\sum_{a+b=const.}\left( \frac{\langle . \rangle }{\langle . \rangle} \right)^a \left( \frac{[ . ] }{[ . ]} \right)^b.
\end{aligned}
\end{equation}
When both angle and square brackets appear, the amplitude expression may generate terms proportional to $m/p_T$ via momentum conservation. To avoid this, the amplitude should be expressed using only $\lambda$ or only $\tilde{\lambda}$. In this form, the 4-point massless amplitude takes the same form as in $\mathcal{N}=4$ SYM (except for the 4-scalar amplitude), because 4-point $\mathcal{N}=4$ SYM amplitudes are either MHV or anti-MHV. This property follows from the conformal symmetry in the 2D transverse plane, as discussed in appendix~\ref{app:conformal}.

Before presenting examples of subleading splittings, we must specify our treatment of the additional Higgs boson $h$. As stated in section~\ref{sec:correspondence}, its momentum magnitude does not affect the splitting behavior, so we may set it to the unit anti-collinear vector $\bar{n}$. The corresponding spinors are
\begin{equation}
\begin{aligned}
|h]\to|\bar{n}],\quad |h\rangle\to|\bar{n}\rangle.
\end{aligned}    
\end{equation}
For the massless coefficient, the 4-point amplitude typically contains products of two 3-point couplings. We label the coupling involving particle $i$ and the additional Higgs boson with a superscript $(ih)$: 
\begin{equation}
\text{labeling}:\quad
Y^{(ih)},\; 
T_s^{(ih)},
\end{equation}
where $i=P,1,2$. When matching to the massive coefficients, these structures yield ratios of particle masses $m_i$ to the vev $v$: 
\begin{equation} \begin{aligned}
\text{fermion mass}:&\quad Y^{(ih)}\to \frac{m_i}{v}, \\
\text{vector mass}:&\quad 
T_s^{(ih)}\to\left\{\begin{aligned}
  &\frac{m_i}{v},& &i\neq h \\
  &0,& & i=h
\end{aligned}\right. . \\
\end{aligned}
\end{equation}

\begin{figure}[htbp]
\centering
\includegraphics[page=1, width=0.95\textwidth]{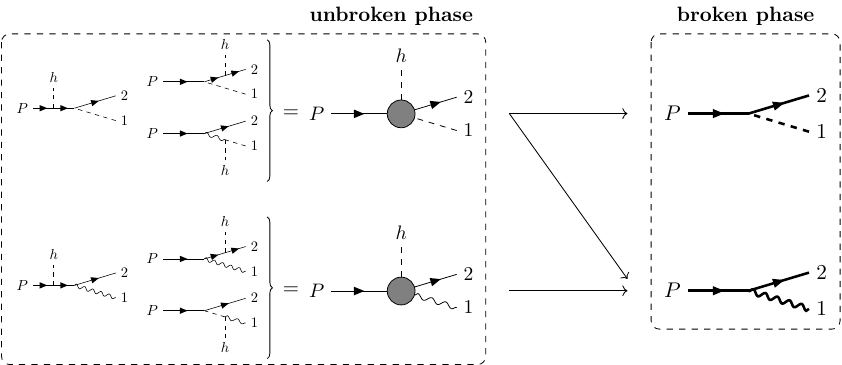}
\caption{Subleading splitting function matching involving fermions. The left column corresponds to massless splitting $f\to SfS$ and $f\to VfS$, while the right column corresponds to the massive splitting $f\to Sf$ and $f\to Vf$.}
\label{fig:subleading_f_match}
\end{figure}

We first consider two typical splitting processes involving fermions:

\begin{itemize}
\item $f\to SfS$: We consider the 4-point massless $fSfS$ amplitude written in terms of only $\lambda$ spinors:
\begin{equation} \begin{aligned}
\mathcal{A}(P^-_{f},1^0_{S},2^+_{\bar{f}},h^0_{\bar{S}})
&=Y \tilde{Y}^{(Ph)}\frac{\langle P1\rangle}{\langle 12\rangle}+T_f T_s^{(1h)}\frac{\langle P1\rangle\langle hP\rangle}{\langle 2P\rangle\langle h1\rangle}, \\
\mathcal{A}(P^-_{f},1^0_{\bar{S}},2^+_{\bar{f}},h^0_{S})
&=\tilde{Y} Y^{(2h)}\frac{\langle Ph\rangle}{\langle 2h\rangle}+T_f T_s^{(1h)}\frac{\langle P1\rangle\langle hP\rangle}{\langle 2P\rangle\langle h1\rangle}.
\end{aligned}
\end{equation}
As illustrated in figure~\ref{fig:subleading_f_match}, the three pole structures correspond to three diagrams with different channels, in which the additional Higgs boson $h$ is inserted into the external lines $P$, $1$, and $2$, respectively. This insertion is also reflected in the pole structure of each term. Expressed in terms of splitting parameters, the amplitude becomes
\begin{equation} \begin{aligned}
\mathcal{A}(P^-_{f},1^0_{S},2^+_{\bar{f}},h^0_{\bar{S}})
&= -Y\tilde{Y}^{(Ph)}\sqrt{\bar{z}}-T_f T_s^{(1h)}\frac{\sqrt{\bar{z}}}{z}. \\
\mathcal{A}(P^-_{f},1^0_{S},2^+_{\bar{f}},h^0_{\bar{S}})
&= -\tilde{Y} Y^{(2h)}\frac{1}{\sqrt{\bar{z}}}-T_f T_s^{(1h)}\frac{\sqrt{\bar{z}}}{z}. \\
\end{aligned}
\end{equation}
Combining these two contributions, this massless splitting can match two massive splitting processes, $f\to Sf$ and $f\to Vf$. In the former case, particles 1 and $h$ are identical, so the $T^{(ih)}_s$ term vanishes. In the latter case, particles 1 and $h$ are different, so the $T^{(ih)}_s$ term contributes. So we have
\begin{equation} \begin{aligned}
\mathcal{M}^{(1)}(P^-_{f},1^0_{h},2^+_{\bar{f}})&=-y \frac{m_2}{v}\frac{1}{\sqrt{\bar{z}}} -y \frac{m_P}{v} \sqrt{\bar{z}}, \\
\mathcal{M}^{(1)}(P^-_{f},1^0_{V},2^+_{\bar{f}})&=\frac{1}{\sqrt{2}}\left(-\left(-\frac{m_P}{m_1}X_1+\frac{m_2}{m_1}X_2 \right) \frac{m_{2}}{v}\frac{1}{\sqrt{\bar{z}}}-\left( \frac{m_2}{m_1}X_1 -\frac{m_P}{m_1}X_2 \right)\frac{m_{P}}{v}\bar{z}-2X_{2} \frac{m_{1}}{v}\frac{\sqrt{\bar{z}}}{z}\right). \\
\end{aligned}
\end{equation}
Thus we derive the subleading splitting functions,
\begin{equation} \begin{aligned}
P^{(1)}_{Sf}&=(-y \frac{m_2}{v} -y \frac{m_P}{v} \bar{z})^2 z, \\
P^{(1)}_{V_L f}&=\frac{1}{2}(-\left(-\frac{m_P}{m_2}X_1+\frac{m_1}{m_2}X_2 \right) \frac{m_{1}}{v}-\left( \frac{m_1}{m_2}X_1 -\frac{m_P}{m_2}X_2 \right)\frac{m_{P}}{v}\bar{z}-X_{2} \frac{m_{2}}{v}\frac{\bar{z}}{z})^2 z. 
\end{aligned}
\end{equation}

\item $f\to VfS$: This corresponds to the 4-point massless $fVfS$ amplitude. As shown in figure~\ref{fig:subleading_f_match}, this amplitude also receives contributions from three massless diagrams, where the additional Higgs boson $h$ is inserted into different external lines. However, these diagrams are gauge-dependent individually. In the on-shell formalism, one must combine contributions from different diagrams to obtain a gauge-invariant amplitude. This amplitude can be expressed using only $\lambda$ spinors as:
\begin{equation} \label{eq:amp_fVfS}
\begin{aligned}
\mathcal{A}(P^{-}_{f},1^{+}_{V},2^{-}_{\bar{f}},h^0_{S})
&=\tilde{T}_f Y^{(2h)}\frac{\langle P 2 \rangle \langle P h \rangle }{\langle 1 P \rangle\langle 1h \rangle}
+T_f Y^{(P h)} \frac{\langle 2 P \rangle\langle h2 \rangle }{\langle 12 \rangle\langle 1h \rangle}\\
&=\tilde{T}_f Y^{(2h)}\sqrt{\bar{z}}
+T_f Y^{(P h)} \frac{1}{\sqrt{\bar{z}}}.
\end{aligned}
\end{equation}
Here each term combines two diagram contributions: the first involves insertions of $h$ into lines $P$ and $1$, while the second involves insertions into lines $1$ and $2$.~\footnote{The amplitude form is not unique. Here we choose the form with coefficients involving only $T_{f}$ and $Y$. One could alternatively choose a form involving the $T_{s}$ coupling of the scalar boson:
\begin{equation}
\mathcal{A}(P^{-}_{f},1^{+}_{V},2^{-}_{\bar{f}},h^0_{S})
=\tilde{T}_f Y^{(1h)}\frac{\langle P 2 \rangle^2 }{\langle 1 P \rangle\langle 12 \rangle}
-Y T_{s}^{(2h)} \frac{\langle 2 P \rangle\langle 32 \rangle }{\langle 12 \rangle\langle 13 \rangle}
\end{equation} 
This yields the same final result as eq.~\eqref{eq:amp_fVfS}, since the splitting function is independent of the chosen form of the massless amplitude.} It matches the massive counterpart
\begin{equation}
\begin{aligned}
\mathcal{M}^{(1)}(P^{-}_{f},1^{+}_{V},2^{-}_{\bar{f}})&= X_{2} \frac{m_{2}}{v} \sqrt{\bar{z}}
+X_{1} \frac{m_{P}}{v} \frac{1}{\sqrt{\bar{z}}}. 
\end{aligned}
\end{equation}
Thus the massive subleading splitting function is
\begin{equation}
P^{(1)}_{V_T f}=(X_{2} \frac{m_{2}}{v} \bar{z}
+X_{1} \frac{m_{P}}{v} )^2 z.
\end{equation}

\begin{figure}[htbp]
\centering
\includegraphics[page=2, width=0.65\textwidth]{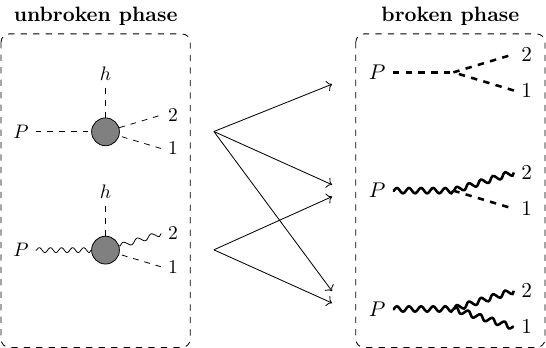}
\caption{Subleading splitting function matching with boson only. The left column corresponds to massless splitting $S\to SSS$ and $V\to SVS$, while the right column corresponds to the massive splitting $S\to SS$, $V\to SV$ and $V\to VV$.}
\label{fig:subleading_b_match}
\end{figure}

\end{itemize}

We now consider splitting processes involving only bosons, which include two gauge couplings $T_s$. In this case, we label the coupling not involving the additional Higgs as $T_s^{(ij)}$ with $i,j=P,1,2$. The coefficient matching can then be written more explicitly as
\begin{equation} \label{eq:coeff_match}
\begin{aligned}
\begin{array}{cc|c|c}
\text{particle } i & \text{particle } j & \text{match } VVV & \text{match } VSV \\ 
\hline 
V & S & {\displaystyle T_{s}^{(ij)} \to \frac{m_{i}^2-m_{j}^2-m_{k}^2}{m_{j}m_{k}} 2\mathbf{f}^{\mathbf{ijk}}} & \mbox{-} \\
V & \bar{S} & {\displaystyle T_{s}^{(ij)} \to \frac{m_{i}^2-m_{k}^2-m_{j}^2}{m_{k}m_{j}} 2\mathbf{f}^{\mathbf{ikj}}} & \mbox{-} \\
S & \bar{S} & {\displaystyle T_{s}^{(ij)} \to  \frac{m_{k}^2-m_{i}^2-m_{j}^2}{m_{i}m_{j}} 2\mathbf{f}^{\mathbf{kij}}} & {\displaystyle T_{s}^{(ij)} \to  \mathbf{g}} \\
\bar{S} & S & {\displaystyle T_{s}^{(ij)} \to  \frac{m_{k}^2-m_{j}^2-m_{i}^2}{m_{j}m_{i}} 2\mathbf{f}^{\mathbf{kji}}} & {\displaystyle T_{s}^{(ij)} \to  -\mathbf{g} }  \\
\end{array}
\end{aligned}
\end{equation}
Besides this, we also need to label the coupling of the four-scalar interaction as $\lambda^{(ij|kh)}$. When we sum over all contributions of scalar and anti-scalar permutations, they will match the scalar mass:
\begin{equation}
\begin{array}{c|ccc}
\to & SSS & VSV & VVV \\ 
\hline
\sum_{\sigma} \lambda^{(\sigma(i)\sigma(j)|\sigma(k)h)}
& {\displaystyle -3\frac{m_h^2}{v^2}}
& {\displaystyle 2\frac{m_h^2}{v^2}}
& 0
\end{array}
\end{equation}
where $\sigma$ denotes a permutation of the scalar and anti-scalar bosons.

To illustrate the matching with bosons only, we consider two typical splitting processes:

\begin{itemize}
\item $V\to SVS$: We consider the 4-point massless $VSVS$ amplitude written in terms of angle brackets only:

\begin{equation}
\begin{aligned}
\mathcal{A}(P_{V}^-,1_{S}^0,2_{V}^+,h_{\bar{S}}^0)&=T_s^{(P1)} T_s^{(2h)} \frac{\langle P h \rangle^2\langle P 1 \rangle}{\langle P 2\rangle\langle1h\rangle\langle h2\rangle}
+T_s^{(21)} T_s^{(P h)}\frac{\langle P 1\rangle^2\langle P h\rangle}{\langle P 2\rangle\langle 12\rangle\langle 1h\rangle} \\
&=T_s^{(P1)} T_s^{(2h)} \frac{1}{z}
+T_s^{(21)} T_s^{(P h)} \frac{\bar{z}}{z}.
\end{aligned}
\end{equation}
One can verify that exchanging the scalar bosons $1$ and $h$ merely maps the two terms into each other. This operation does not generate new amplitude structures, so a single massless amplitude is sufficient to derive the massive counterpart. As illustrated in figure~\ref{fig:subleading_b_match}, the massless splitting $V\to SVS$ can match two massive splitting processes, $V\to SV$ and $V\to VV$:
\begin{equation}
\begin{aligned}
M^{(1)}(P^-_{V},1^0_{h},2^+_{V}) 
&= \frac{1}{\sqrt{2}}\left(\mathbf{g} \frac{m_{2}}{v} \frac{1}{z} - \mathbf{g} \frac{m_{P}}{v} \frac{\bar{z}}{z}\right)
= \frac{1}{\sqrt{2}}\mathbf{g} \frac{m_{V}}{v}, \\
M^{(1)}(P^-_{V},1^0_{V},2^+_{V}) 
&= \frac{m_{P}^2-m_{1}^2-m_{2}^2}{m_{1}m_{2}} 2\mathbf{f}^{\mathbf{P12}} \frac{m_{2}}{v} \frac{1}{z} - \frac{m_{2}^2-m_{1}^2-m_{P}^2}{m_{1}m_{P}} 2\mathbf{f}^{\mathbf{21P}} \frac{m_{P}}{v} \frac{\bar{z}}{z}\\ 
&= 2\mathbf{f}^{\mathbf{P12}}\left( \frac{m_{P}^2-m_{2}^2}{m_{1} v} - \frac{m_{1}}{v} \left( \frac{1+\bar{z}}{z}\right) \right). \\
\end{aligned}
\end{equation}
In the first line, the expression simplifies using $m_V=m_2=m_P$. The subleading splitting functions are then
\begin{equation} \begin{aligned}
P^{(1)}_{S V_T} &= \frac{1}{2} \mathbf{g}^2 \left(\frac{m_{V}}{v}\right)^2 z\bar{z}, \\
P^{(1)}_{V_L V_T} &= (2\mathbf{f}^{\mathbf{P12}})^2 \left( \frac{m_{P}^2-m_{2}^2}{m_{1} v} z - \frac{m_{1}}{v} (1+\bar{z}) \right)^2 \frac{\bar{z}}{z}.
\end{aligned}
\end{equation}

\item $S\to SSS$: This corresponds to four-scalar amplitudes. In this case, we can write three massless amplitudes $SS\bar{S}\bar{S}$, $S\bar{S}S\bar{S}$ and $\bar{S}SS\bar{S}$ with different amplitude structures:
\begin{equation}
\begin{aligned}
\mathcal{A}(P_{S}^0,1_{S}^0,2_{\bar{S}}^0,3_{\bar{S}}^0)  
&=T_s^{(P 2)} T_s^{(13)} \left(\frac{\langle 23 \rangle\langle 1P \rangle}{\langle 13 \rangle\langle 2P\rangle}-\frac{1}{2}\right)
+T_s^{(12)} T_s^{(P3)} \left(\frac{\langle 32\rangle\langle 1P \rangle}{\langle 12\rangle\langle 3P \rangle}-\frac{1}{2}\right)-\lambda^{(P1|23)}, \\
\mathcal{A}(P_{S}^0,1_{\bar{S}}^0,2_{S}^0,3_{\bar{S}}^0)  
&=T_s^{(P1)} T_s^{(23)} \left(\frac{\langle 13 \rangle\langle 2P \rangle}{\langle 23 \rangle\langle 1P\rangle}-\frac{1}{2}\right)
+T_s^{(21)} T_s^{(P3)}\left(\frac{\langle 31\rangle\langle 2P \rangle}{\langle 21\rangle\langle 3P \rangle}-\frac{1}{2}\right)-\lambda^{(P2|13)}, \\ 
\mathcal{A}(P_{\bar{S}}^0,1_{S}^0,2_{S}^0,3_{\bar{S}}^0)  
&=T_s^{(1P)} T_s^{(23)} \left(\frac{\langle P3 \rangle\langle 21 \rangle}{\langle 23 \rangle\langle P 1\rangle}-\frac{1}{2}\right)
+T_s^{(2P)} T_s^{(1 3)}\left(\frac{\langle 31\rangle\langle 2P \rangle}{\langle 21\rangle\langle 3P \rangle}-\frac{1}{2}\right)-\lambda^{(12|P3)}.  
\end{aligned}
\end{equation}
Expressed in terms of splitting parameters, these become
\begin{equation}
\begin{aligned}
\mathcal{A}(P_{S}^0,1_{S}^0,2_{\bar{S}}^0,3_{\bar{S}}^0) 
&= T_s^{(P 2)} T_s^{(13)} \left( -\frac{1+\bar{z}}{2z} \right) + T_s^{(12)} T_s^{(P3)} \left(\frac{\bar{z}-z}{2} \right)-\lambda^{(P1|23)}, \\
\mathcal{A}(P_{S}^0,1_{\bar{S}}^0,2_{S}^0,3_{\bar{S}}^0)  
&= T_s^{(P1)} T_s^{(23)}  \left(-\frac{1+z}{2\bar{z}} \right) + T_s^{(21)} T_s^{(P3)} \left( -\frac{\bar{z}-z}{2} \right)-\lambda^{(P2|13)}, \\
\mathcal{A}(P_{\bar{S}}^0,1_{S}^0,2_{S}^0,3_{\bar{S}}^0)
&= T_s^{(1P)} T_s^{(23)} \left(\frac{1+z}{2\bar{z}} \right) + T_s^{(2P)} T_s^{(1 3)} \left( \frac{1+\bar{z}}{2z} \right)-\lambda^{(12|P3)}. \\
\end{aligned}
\end{equation}
As illustrated in figure~\ref{fig:subleading_b_match}, it can match three kinds of massive splitting amplitudes:
\begin{equation}
\begin{aligned}
M^{(1)}(P^0_{h},1^0_{h},2^0_{h}) 
=& -3 \frac{m_h^2}{v^2}=\frac{\lambda_3}{v}, \\
M^{(1)}(P^0_{V},1^0_{h},2^0_{V}) 
=& \left(\frac{1}{\sqrt{2}}\right)^2 \left(-\mathbf{g}\frac{m_{V}}{v} \frac{2(1-z\bar{z})}{\bar{z}} -2 \frac{m_h^2}{v^2}\right), \\
M^{(1)}(P^0_{V},1^0_{V},2^0_{V}) 
=& \left(\frac{1}{\sqrt{2}}\right)^3 2\mathbf{f}^{\mathbf{P12}} \left(  \frac{m_{1}(m_{1}^2-m_{P}^2-m_{2}^2)}{m_{P}m_{2} v} \frac{1+\bar{z}}{z} 
+ \frac{m_{P} (m_{P}^2-m_{1}^2-m_{2}^2)}{m_{1}m_{2} v} (\bar{z}-z)\right. \\
&\qquad\quad \left. -\frac{m_{2} (m_{2}^2-m_{P}^2-m_{1}^2)}{m_{P}m_{1} v} \frac{1+z}{\bar{z}} \right). \\ 
\end{aligned}
\end{equation}
The corresponding splitting functions are given by
\begin{equation} \begin{aligned}
P^{(1)}_{SS} =& 9 (\frac{m_h^2}{v^2})^2 z\bar{z} , \\
P^{(1)}_{SV_L} =&  \left(\frac{1}{2}\right)^2 \mathbf{g}^2(-\frac{m_V}{v}(1-z\bar{z}) -\frac{m_h^2}{v^2} \bar{z} )^2 \frac{z}{\bar{z}}, \\
P^{(1)}_{V_L V_L} =& \left(\frac{1}{2}\right)^3 (2\mathbf{f}^{\mathbf{P12}})^2 \left(  \frac{m_{1}(m_{1}^2-m_{P}^2-m_{2}^2)}{m_{P}m_{2} v} \frac{1+\bar{z}}{z} 
+ \frac{m_{P} (m_{P}^2-m_{1}^2-m_{2}^2)}{m_{1}m_{2} v} (\bar{z}-z)\right. \\
&\qquad\qquad \left. -\frac{m_{2} (m_{2}^2-m_{P}^2-m_{1}^2)}{m_{P}m_{1} v} \frac{1+z}{\bar{z}} \right)^2. \\ 
\end{aligned}
\end{equation}

\end{itemize}

These results agree with the corresponding massive-amplitude calculations available in~\cite{Chen:2016wkt,Nardi:2024tce}. They constitute the central $1\to 2$ results of this work. In the next section, we show how these elementary $1\to 2$ splitting amplitudes can be composed into sequential higher-multiplicity amplitudes by transforming the splitting variables between successive collinear frames.


%% file: sec6_recursive.tex
\section{Recursive Construction for Higher-Point Splitting}
\label{sec:recursive}

The $1\to 2$ splitting functions derived in section~\ref{sec:splitting} suffice for the leading-logarithmic evolution of parton distributions. However, many collider observables such as jet substructure and multi-parton final states require splitting functions for $1\to 3$, $1\to 4$, and higher multiplicities. Computing these directly from Feynman diagrams is cumbersome~\cite{Catani:1998nv, Campbell:1997hg}, particularly when mass effects are included~\cite{Dhani:2023uxu,Craft:2023aew}. In this section, we show that the on-shell framework admits a recursive construction: any $1\to n$ splitting amplitude can be built by composing elementary $1\to 2$ splittings, with a universal substitution rule for the splitting variables derived from Galilean symmetry.

For a general $1\to n$ process, we use following parameterization for the momenta 
\begin{equation} \begin{aligned}
P^{\mu} =(P_+,0,0), \qquad
p^{\mu}_i =\left( z_i P_{+}, \frac{|p_{iT}|^2}{z_i P_+},p_{iT} \right),
\end{aligned} \end{equation}
where the longitudinal momentum fractions $z_i$ and transverse momenta $p_{iT}$ are constrained by momentum conservation in these directions 
\begin{equation} \label{eq:momentum_conservation}
\begin{aligned}
\sum_i z_{i}=1, \qquad
\sum_i p_{iT}=0.
\end{aligned}
\end{equation}

Thus, the general splitting parameters are $(z_i, p_{iT})$.
We will show how to generate them from the $1\to 2$ splitting parameters $(z, p_{T})$. The key physical insight is that the intermediate particle in a sequential splitting carries non-zero transverse momentum, but this can be absorbed into a redefinition of the splitting variables for the subsequent $1\to 2$ process via a standard boost from Galilean symmetry. This yields a universal \textit{substitution rule} that replaces the kinematic parameters $(z, p_T)$ of the elementary splitting function with appropriate functions of the multi-particle kinematics.

\subsection{Recursive Construction: Leading Order}

\begin{figure}[htbp]
\centering
\includegraphics[width=0.4\textwidth]{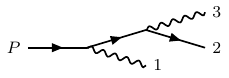}
\caption{Typical decomposition of a $1 \to 3$ process. The parent particle $P$ first undergoes a splitting into particle $1$ and an intermediate state $q$, which subsequently decays into particles $2$ and $3$.}
\label{fig:1to3}
\end{figure}

Consider a $1\to 3$ process, illustrated in figure~\ref{fig:1to3}. It can be decomposed into two sequential $1\to 2$ processes,
\begin{equation}
\begin{aligned}
\mathcal{A}(P;1,2,3)\to \mathcal{A}(P;1,q)\times \mathcal{A}(q;2,3),
\end{aligned}
\end{equation}
where the intermediate particle carries momentum $q=P-p_1$. The on-shell part of $q$ is
\begin{equation}
\begin{aligned}
q^\mu =\left( (1-z_{1})P_{+}, \frac{|p_{1T}^2|}{(1-z_{1})P_{+}} ,-p_{1T}   \right).
\end{aligned}
\end{equation}
While the first splitting is directly given by a 3-point splitting amplitude from the previous section, the second splitting does not immediately reduce to a standard $1\to 2$ splitting amplitude, because the intermediate particle possesses non-zero transverse momentum $q_T=-p_{1T} \neq 0$.

To address this, we perform a standard boost $L_p$ that brings the second $1 \to 2$ process into the standard splitting parametrization, where the transverse momentum of the intermediate particle vanishes:
\begin{equation}
\begin{aligned}
\mathcal{A}(q;2,3) \xrightarrow{L_p} \mathcal{A}(P;k_{1},k_{2}),
\end{aligned}
\end{equation}
with $P=(P_+,0,0)$ still denoting the momentum of the parent particle, and $k_i$ given by
\begin{equation} \begin{aligned} \label{eq:target_form}
k^{\mu}_1 &=\left( w P_{+}, \frac{|k_T|^2}{w P_+},k_{T} \right), \\
k^{\mu}_2 &=\left( (1-w) P_{+}, \frac{|k_T|^2}{(1-w) P_+},-k_{T} \right),
\end{aligned} \end{equation}
where $w$ and $k_T$ are new splitting parameters. The second splitting amplitude should be described by $w$ and $k_T$, whose values are not yet known at this stage.

To determine the second splitting amplitude, we need the explicit form of the standard boost $L_p$. This boost was discussed in section~\ref{sec:particle_state}, let us review its definition
\begin{equation}
    L_p =B_z\times \Lambda= \exp(ivT+iv^*\Bar{T})\times\exp(i\zeta K_3),
\end{equation}
where $B_z$ is a light-front scale transformation that changes $p_+$ and $p_-$, and $\Lambda$ is a Galilean standard boost that changes $p_T$. The boost parameters $v$ and $\zeta$ are not arbitrary. Only when we set
\begin{equation}
\begin{aligned}
v e^{i \zeta}=\frac{p_{1T}}{P_{+}},
\end{aligned}
\end{equation}
does this standard boost map the intermediate particle's momentum to the parent momentum:
\begin{equation}
\begin{aligned}
q^\mu =\left( (1-z_{1})P_{+}, \frac{|p_{1T}^2|}{(1-z_{1})P_{+}} ,-p_{1T}   \right)
\xrightarrow{B_{z}} \left( P_{+}, \frac{|p_{1T}^2|}{P_{+}} ,-p_{1T} \right) 
\xrightarrow{\Lambda} (P_{+},0,0).
\end{aligned}
\end{equation}

The momenta of particles $2$ and $3$ undergo the same sequence of transformations, mapping to $k_1$ and $k_2$. Consider particle 2 explicitly:
\begin{equation}
\begin{aligned}
p_2 &= \left( z_{2} P_+, \frac{|p_{2T}|^2}{z_{2} P_+}, p_{2T} \right)  \\
&\xrightarrow{B_{z}} \left( \frac{z_{2}}{1-z_{1}} P_+, \frac{|p_{2T}|^2}{\frac{z_{2}}{1-z_{1}} P_+}, p_{2T} \right)  \\
&\xrightarrow{\Lambda} \left( \frac{z_{2}}{1-z_{1}} P_+, \frac{|p_{2T} + \frac{z_{2}}{1-z_{1}} p_{1T}|^2}{\frac{z_{2}}{1-z_{1}} P_+}, p_{2T} + \frac{z_{2}}{1-z_{1}} p_{1T} \right).
\end{aligned}
\end{equation}
The resulting momentum $k_1$ should take the desired form in eq.~\eqref{eq:target_form}, so we derive the splitting parameters
\begin{equation} \label{eq:sub_parameter}
\begin{aligned}
w &=\frac{z_2}{1-z_1}=\frac{z_{2}}{z_{2}+z_{3}}, \\
q_{T} &=p_{2T}+\frac{z_2}{1-z_1}p_{1T}=\frac{ z_{3} p_{2T} - z_{2} p_{3T} }{z_{2}+z_{3}},
\end{aligned}
\end{equation}
where we have used momentum conservation eq.~\eqref{eq:momentum_conservation} to relate the parameters of particle $1$ to those of the other two particles.

The second splitting process can now be described by the new splitting parameters $w$, $q_T$. These parameters can be obtained via a universal replacement, without requiring case-by-case calculations. For the leading-order massive $1 \to 2$ splitting amplitude, the substitution directly yields
\begin{equation} \label{eq:substitution}
\begin{aligned}
\mathcal{A}_3(z,p_{T})\Rightarrow \mathcal{A}_3\left( \frac{z_2}{z_2+z_3} ,\frac{ z_{3} p_{2T} - z_{2} p_{3T} }{z_{2}+z_{3}}  \right).
\end{aligned}
\end{equation}
As a verification, let us return to the splitting process $f\to VfV$ in figure~\ref{fig:1to3}. Consider the intermediate fermion $f^-$ mediates the splitting $f^+ \to f^+ V^+$. We begin with a $1\to 2$ splitting from the parent particle:
\begin{equation}
\begin{aligned}
\mathcal{A}(z,P_T) &= \frac{1}{(1 - z) \sqrt{z}} p_T,  \\
\end{aligned}
\end{equation}
and then apply the substitution eq.~\eqref{eq:substitution} to derive
\begin{equation}
\begin{aligned}
\mathcal{A}(w,q_T) &\Rightarrow \frac{1}{\frac{z_3}{z_{2}+z_{3}} \sqrt{\frac{z_2}{z_{2}+z_{3}}}} \frac{z_3 p_{2T} - z_2 p_{3T}}{z_2+z_{3}} \\
&= \frac{\sqrt{z_2+z_{3}}}{z_{3} \sqrt{z_2}} (z_3 p_{2T} - z_2 p_{3T}).
\end{aligned}
\end{equation}
Thus the corresponding splitting amplitude $f^-\to V^+ f^+ V^+$ takes the form
\begin{equation}
\begin{aligned}
\mathcal{A}(P_f^-, 1_V^+, 2_{\bar{f}}^+, 3_V^+) 
&\simeq \mathcal{A}(P_f^-, 1_V^+, q_{\bar{f}}^+, )\times\frac{1}{\tilde{Q}^2}\times \mathcal{A}(q_f^-, 2_{\bar{f}}^+, 3_V^+)\\
&=\frac{1}{z_1 \sqrt{1-z_1}} p_{1T} \times \frac{1}{\tilde{Q}^2} \times \frac{\sqrt{z_2+z_{3}}}{z_{3} \sqrt{z_2}} (z_3 p_{2T} - z_2 p_{3T}),
\end{aligned}
\end{equation}
where the virtuality $\tilde{Q}^2$ is given by
\begin{equation}
\tilde{Q}^2=\sum_{i=2}^3\frac{m_{i}^2+|p_{iT}|^2}{z_{i}}-\frac{m_{q}^2+|p_{1T}^2|}{1-z_{1}}.
\end{equation}

We have thus shown that the substitution rule maps an elementary $1\to 2$ splitting amplitude to the second step of a $1\to 3$ process. This construction generalizes immediately to $1\to n$ processes. For an $n$-particle final state, the splitting amplitude factorizes as
\begin{equation}
\label{eq:recursive_bootstrap}
\mathcal{A}_{1\to n}(z_1,\dots,z_n; p_{1T},\dots,p_{nT}) = \sum_{\text{int}} \frac{1}{\tilde{Q}^2_{n-3}}\mathcal{A}_{1\to 2}(w^\prime, q_T^\prime)\, \mathcal{A}_{1\to (n-1)}\bigl(z_1,\dots,z_{n-1}; p_{1T},\dots,p_{(n-1)T}\bigr),
\end{equation}
where the sum runs over allowed intermediate particle types and $w^\prime,q_T^\prime$ correspond to the standard boost $L_p$ that maps the intermediate particle in the last step to the parent momentum. The initial condition for the recursion is the elementary $1\to 2$ splitting amplitude $\mathcal{A}_{1\to 2}$, which is derived in the previous section. This provides a recursive construction for splitting amplitudes in both the massless and leading massive cases.

\subsection{Recursive Construction: Subleading Order}

We now consider the subleading contributions. Recall that the subleading $1\to 2$ splitting function is obtained from 4-point massless amplitudes with an additional Higgs boson, whose momentum is taken along the $\bar{n}$ direction. Note that  $\bar{n}$ is invariant under the standard boost, and the scale of the Higgs momentum does not affect the subleading contribution. Therefore, the same substitution rule applies, giving
\begin{equation}
\begin{aligned}
\mathcal{A}_4(z,p_{T},\bar{n})\to \mathcal{A}_4\left( \frac{z_2}{z_2+z_3} ,\frac{ z_{3} p_{2T} - z_{2} p_{3T} }{z_{2}+z_{3}},\bar{n}  \right).
\end{aligned}
\end{equation}

\begin{figure}[htbp]
\centering
\includegraphics[width=0.8\textwidth]{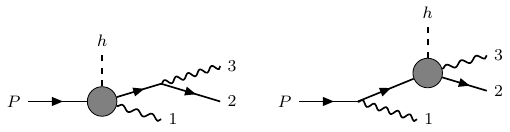}
\caption{Typical massless amplitude corresponding to the subleading contributions in a sequential $1 \to 3$ splitting process. The additional Higgs boson $h$ can be inserted into the first step (left) or the second step (right).}
\label{fig:1to3_Higgs}
\end{figure}

We again use the example of the splitting process $f\to VfV$ to illustrate the substitution. As shown in figure~\ref{fig:1to3_Higgs}, there are two contributions: the additional Higgs boson can be inserted into the first or the second step. Inserting it into the first step gives a calculation similar to that in the previous subsection.  Let us focus on the case where it is inserted into the second step, where the intermediate fermion $f^-$ mediates the splitting $f^+ \to f^+ V^+ S$. The original and substituted 4-point massless splitting amplitudes are
\begin{equation}
\begin{aligned}
\mathcal{A}(z,p_T) &= \tilde{T}_f Y^{(2h)}\sqrt{ z }
+T_f Y^{(P h)} \frac{1}{\sqrt{ z }}, \\
\mathcal{A}(w,q_T) &\Rightarrow \tilde{T}_f Y^{(2h)}\sqrt{ \frac{z_{2}}{z_{2}+z_{3}} }
+T_f Y^{(P h)} \sqrt{ \frac{z_{2}+z_{3}}{z_{2}} }.
\end{aligned}
\end{equation}
The corresponding massive splitting amplitudes are
\begin{equation}
\begin{aligned}
\mathcal{M}^{\text{(sub)}}(z,P_T)
=& X_{2}m_{1} \sqrt{ z } + X_{1}m_{P} \frac{1}{\sqrt{ z }}, \\
\mathcal{M}^{\text{(sub)}}(w,q_T)
\Rightarrow& X_{2}m_{2} \sqrt{ \frac{z_{2}}{z_{2}+z_{3}} } + X_{1}m_{Q} \sqrt{ \frac{z_{2}+z_{3}}{z_{2}} }.
\end{aligned}
\end{equation}

Therefore, the recursive construction extends naturally to the subleading order. The same substitution rule applies to the subleading amplitudes without modification. The general subleading recursive formula takes the form
\begin{equation}
\label{eq:recursive_subleading}
\mathcal{M}_{1\to n}^{\text{(sub)}}(z_i, p_{iT}) = \sum_{\text{int}} \Bigl[
\mathcal{M}_{1\to 2}^{\text{(lead)}}(w^\prime, q_{T}^\prime)\, \mathcal{M}_{1\to (n-1)}^{\text{(sub)}}(z_i, p_{iT})
+ \mathcal{M}_{1\to 2}^{\text{(sub)}}(w^\prime, q_{T}^\prime)\, \mathcal{M}_{1\to (n-1)}^{\text{(lead)}}(z_i, p_{iT})
\Bigr],
\end{equation}
where the first term combines a leading-order first splitting with a subleading subsequent splitting, and the second term accounts for the subleading correction to the first splitting followed by a leading-order subsequent splitting. Both channels contribute at the same order in the $m/p_T$ expansion.

This recursive structure completes the on-shell program for massive splitting functions: starting from the collinear decomposition of massive spinors and the elementary three-point amplitudes, we have derived the complete set of $1\to 2$ splitting functions, and shown how they can be composed to construct splitting functions of arbitrary multiplicity through the universal substitution rule. Each step of this discussion fully exploits the nature of Galilean symmetry, illustrating how the appropriate symmetry simplifies the calculation in the collinear context.

%% file: sec7_con.tex
\section{Summary and Outlook}
\label{sec:summary}

In this paper, we have presented a complete on-shell formalism for constructing collinear splitting functions of massive particles using the spinor-helicity method. Our framework is built on three pillars: the Soper-Weinberg collinear spinors derived from the light-front Galilean subgroup of the Poincar\'e group (section~2), the systematic decomposition of massive momenta and spinors onto fixed lightlike reference vectors $n$ and $\bar{n}$ with manifest power counting (section~3), and the matching between massive and massless amplitudes at the three- and four-point levels (section~4). Based on these, the following results of this work are in order:
\begin{itemize}
    \item \textbf{Complete massive splitting functions.} We have derived the full set of leading and subleading massive splitting functions for all Standard Model particles,  expressed in closed analytic form in terms of the longitudinal momentum fraction $z$ and the transverse momentum $P_T$ (section~5). The leading-order splitting functions reproduce the standard Altarelli-Parisi kernels in the massless limit, while the subleading contributions capture mass corrections that are essential for electroweak precision physics.

    \item \textbf{Massless-to-massive amplitude matching.} We have established a systematic dictionary between massless and massive coupling coefficients. At leading order, massive three-point amplitudes are matched to massless three-point amplitudes via the identification of the leading SW spinor components with massless spinors. At subleading order, the Higgs insertion technique, where an additional Higgs boson probes the subleading ($I=+$) spinor component, maps massless four-point amplitudes to subleading massive splitting functions (section~4.2). This matching is tabulated for all SM interactions and provides a constructive alternative to the diagrammatic extraction of collinear limits.

    \item \textbf{Recursive bootstrap for higher-point splitting.} We have shown that higher-point splitting functions can be constructed from elementary $1\to 2$ splittings via a universal substitution rule (section~6). This rule, a consequence of the Galilean symmetry of the collinear subgroup, replaces the kinematic parameters $(z, P_T)$ with appropriate functions of the multi-particle kinematics and applies identically to both leading and subleading orders. Two explicit worked examples ($f\to fV$ and $V\to VV$) confirm the validity of the construction.
\end{itemize}

Taken together, these results provide a self-contained on-shell framework that constructs massive splitting functions from first principles, without recourse to the collinear limit of full matrix elements.
Beyond the specific results enumerated above, this work introduces several methodological advances that may be of independent interest.

\begin{itemize}
    \item First, the collinear decomposition with \textit{fixed} reference vectors $n$ and $\bar{n}$, as opposed to the momentum-dependent basis used in the helicity decomposition, provides a global reference frame for multi-particle collinear processes. This is essential for the recursive bootstrap, where the intermediate particle's spinor components must be compared across different splitting steps. The fixed basis also makes the power counting manifest: the large, intermediate, and small scales ($P_+$, $P_T$, $m$) are cleanly separated into distinct spinor components.

\item Second, the identification of the alignment limit $m < P_T   \ll P_+$, where the helicity and collinear decompositions yield coincident power counting, clarifies the regime of applicability of each decomposition. The collinear decomposition is shown to be the natural choice for splitting problems because its expansion truncates at finite order, in contrast to the infinite series generated by the helicity expansion.

\item Third, the massless-to-massive matching via Higgs insertion, converting massless poles into physical masses ($s_{i4} \to m_i^2$), provides a systematic probe of subleading spinor components. To the best of our knowledge, this is the first time that the complete set of subleading massive splitting functions has been obtained from on-shell four-point amplitudes in a unified framework.

\end{itemize}
Beside, the on-shell framework makes the group-theoretic structure of collinear factorization manifest. The appearance of the 2D Galilean group as the symmetry of the collinear regime clarifies why the splitting functions depend only on $z$ and not on the absolute scale $P_+$, and the ISO(2) invariance of the splitting kernel follows directly from the Galilean algebra.

The results of this paper have direct relevance to several active areas of collider physics. First, modern parton shower Monte Carlo programs, such as Pythia~\cite{Brooks:2021kji}, Herwig~\cite{Masouminia:2021kne}, and Sherpa~\cite{Bothmann:2020sxm}, rely on splitting functions as their fundamental branching kernels. The massive splitting functions derived here can be directly incorporated into shower algorithms to improve the treatment of heavy particle thresholds, where finite-mass effects modify both the branching probability and the kinematic distributions of the decay products. The recursive bootstrap further enables the inclusion of higher-point splittings for jet substructure observables. Second, at the energies accessible to the high-luminosity LHC and future colliders, electroweak Sudakov logarithms $\sim \alpha_W \log^2(Q^2/M_W^2)$ become numerically significant. The massive splitting functions derived here, which include the full mass dependence of the $W$, $Z$, and Higgs bosons, provide the necessary ingredients for next-to-leading-logarithmic electroweak resummation, complementing existing calculations \cite{Chen:2016wkt,Ciafaloni:2000df}. Finally, the collinear decomposition with fixed $n$ and $\bar{n}$ vectors is closely related to the soft-collinear effective theory (SCET)~\cite{Bauer:2000yr,Bauer:2001yt} formulation of collinear physics, where the $n$-collinear and $\bar{n}$-anti-collinear fields play an analogous role. Matching the on-shell results of this paper to SCET operator bases, particularly for the subleading mass corrections, would connect the amplitude-level construction to the effective field theory language more commonly used in factorization proofs, and could facilitate the incorporation of interplay between the collinear and soft modes.


In summary, the on-shell spinor-helicity framework developed in this paper provides a unified, constructive approach to massive collinear splitting functions. By making the underlying group-theoretic and kinematical structure manifest, it offers a versatile foundation and a constructive starting point for precision collider physics in the era of high-luminosity LHC and future colliders.

%% file: app_conformal.tex
\begin{appendix}

\section{Splitting Function and Conformal Symmetry}
\label{app:conformal}

In the main text, we have shown that the splitting function is directly related to amplitudes written solely in terms of angle or square brackets. In fact, these amplitude forms can be explained by enhancing the Galilean symmetry $\text{Gal}(2)$ to a larger symmetry.

Let us begin by discussing the property of deriving the subleading massive splitting function from 4-point massless amplitudes. We found that these massless amplitudes involve a Higgs boson carrying momentum along a fixed $\bar{n}$-direction. More importantly, the resulting splitting function is independent of the magnitude of the Higgs momentum. Since the $\bar{n}$-direction corresponds to the Hamiltonian $P_-$ of the Galilean group, this independence  equivalent implies that the value of $P_-$ does not need to be conserved in the splitting process, signaling the presence of scale invariance. Therefore, the symmetry should be enhanced from the Galilean group to the conformal group of the two-dimensional transverse space, $SO(3,1)$ (isomorphic to $SL(2,\mathbb{C})$), which acts on the 2-dimensional transverse space.

Since we want to compute the splitting function using Lorentz-covariant spinor variables, this $SO(3,1)$ symmetry can be viewed as the restriction of the full 4-dimensional conformal symmetry $SO(4,2)$ to the 2-dimensional transverse space. The conformal generators are given by
\begin{equation} \begin{aligned}
J_{ab}&=\frac{1}{2}\sum_i \left(\lambda_{i a}\frac{\partial}{\partial \lambda_{i}^b}-\lambda_{i b}\frac{\partial}{\partial \lambda_{i}^a}\right),&
\tilde{J}_{\dot{a} \dot{b}}&=\frac{1}{2}\sum_i \left(\tilde{\lambda}_{i \dot{a}}\frac{\partial}{\partial \tilde{\lambda}_{i}^{\dot{b}}}-\tilde{\lambda}_{i \dot{b}}\frac{\partial}{\partial \tilde{\lambda}_{i}^{\dot{a}}}\right), \\
\mathcal{P}_{a \dot{b}}&=\sum_i \lambda_{i a} \tilde{\lambda}_{i \dot{b}}, \\
D&=\sum_i \left(\frac{1}{2}\lambda_{i a} \frac{\partial}{\partial \lambda_{i}^a}+\frac{1}{2}\tilde{\lambda}_{i \dot{a}} \frac{\partial}{\partial \tilde{\lambda}_{i}^{\dot{a}}} +1\right), \\
K_{a \dot{b}} &=\sum_i \frac{\partial}{\partial \lambda_{i}^{a}} \frac{\partial}{\partial \tilde{\lambda}_{i}^{\dot{b}}},
\end{aligned} \end{equation}
where $J$ and $\tilde J$ are the Lorentz generators, $\mathcal{P}$ is the translation generator, $D$ is the dilatation generator and $K$ is the special conformal genrator. Applying these conformal generators to a scattering amplitude $\mathcal{A}$ must therefore yield zero. This explains why we can write the amplitude using only angle or square brackets: such forms satisfy the condition $K_{a\dot{b}} \mathcal{A}=0$.

Although conformal symmetry constrains the form of the amplitude, it does not uniquely determine it. To find the desired amplitude, we can consider the maximal symmetry for 4-dimensional theories with fields of spin $\le 1$ field theory, namely $\mathcal{N}=4$ super-Yang-Mills (SYM) theory. In this theory, conformal symmetry is combined with supersymmetry to yield superconformal symmetry. Since this superconformal symmetry include an $\mathcal{N}=4$ supersymmetry, the theory contains $4^2=16$ particle states, organized by the $SU(4)$ R-symmetry of the $\mathcal{N}=4$ algebra. These 16 helicity states are
\begin{itemize}
    \item one positive-helicity gluon: $g^+$, with helicity $+1$;
    \item one negative-helicity gluon: $g^-$, with helicity $-1$;
    \item four positive-helicity gluinos (fermions): $\bar{f}^+$ , with helicity $+\frac12$;
    \item four negative-helicity gluinos: $f^-$,  with helicity $-\frac12$;
    \item six real scalars: $S$ and $\bar{S}$, which can be organized into three complex pairs and have helicity $0$.
\end{itemize}
All these states are related by the action of the supercharges and can be organized into a single massless supermultiplet.

The on-shell superspace formalism provides a powerful way to express scattering amplitudes in a compact form. In the on-shell superspace, the supermultiplet is represented by a superfield
\begin{equation}
\Phi = g^+ + \eta_A \bar{f}^A + \frac{1}{2!}\eta_A\eta_B S^{AB} + \frac{1}{3!}\eta_A\eta_B\eta_C f^{ABC} + \frac{1}{4!}\eta_A\eta_B\eta_C\eta_D \epsilon^{ABCD}\, g^-,
\end{equation}
where $\eta^A$ are Grassmann variables labeled by the $SU(4)$ index $A=1,\dots,4$. A particular particle state can be obtained as a component of this superfield by taking derivatives with respect to the Grassmann variables,
\begin{equation} \begin{aligned} \label{eq:derivative}
\left.\Phi\right|_{\eta=0} &=g^+,\quad&
\left.\left(\prod_{A=1}^4\frac{\partial}{\partial \eta_{A}}\right)\Phi\right|_{\eta=0} &=g^-,&\\
\left.\frac{\partial}{\partial \eta_{1}}\Phi\right|_{\eta=0} &=\bar{f}^1,\quad&
\left.\left(\prod_{A=1}^3\frac{\partial}{\partial \eta_{A}}\right)\Phi\right|_{\eta=0} &=f^{123},&\\
&& \left.\left(\prod_{A=1}^2\frac{\partial}{\partial \eta_{A}}\right)\Phi\right|_{\eta=0} &=S^{12}.&
\end{aligned}
\end{equation}

In $\mathcal{N}=4$ SYM, scattering amplitudes can be decomposed into color-ordered partial amplitudes, as in pure gluon scattering:
\begin{equation}
\mathcal{A}_n=\sum_{\text{perms }\sigma} \text{tr}(T^{a_1}T^{\sigma(a_2}\dots T^{a_n)})\; \mathcal{A}_n[\Phi_1,\sigma(\Phi_2,\dots,\Phi_n)],
\end{equation}
where the traces $\text{tr}(T^{a_1} T^{a_2} \dots T^{a_n})$ form a color basis for the $SU(N)$ gauge theory. Color-orderded amplitudes involving only angle brackets are components of the maximally helicity-violating (MHV) superamplitude. At tree-level, the $n$-point MHV superamplitude is
\begin{equation}
\mathcal{A}_n^{\text{MHV}}[\Phi_1,\Phi_2,\dots,\Phi_n] = \frac{\delta^{(4)}(\mathcal{P}) \, \delta^{(8)}(\tilde{\mathcal{Q}})}{\langle 12 \rangle \langle 23 \rangle \cdots \langle n1 \rangle},
\end{equation}
where $\tilde{\mathcal{Q}}_A = \sum_{i=1}^n \lambda_i \eta_{iA}$ is the supercharge, and the pole structure $\langle 12 \rangle \langle 23 \rangle \cdots \langle n1 \rangle$ depends on the color ordering. The delta functions $\delta^{(4)}(P)$ and $\delta^{(8)}(\tilde{\mathcal{Q}})$ impose  momentum and supermomentum conservation. The numerator can be written in terms of anglue brackets and Grassmann variables as
\begin{equation}
\delta^{(8)}\left(\tilde{Q}\right) = \frac{1}{2^4} \prod_{A=1}^{4} \tilde{Q}_{A a} \tilde{Q}_A^{ a} = \frac{1}{2^4} \prod_{A=1}^{4} \sum_{i,j=1}^{n} \langle ij \rangle \eta_{iA} \eta_{jA}.
\end{equation}
Consequently, each R-symmetry index $A$ appears twice in every component term, through a pair $\eta_{iA}$ and $\eta_{jA}$.

Similarly, color-ordered amplitudes involving only square brakces are unified into anti-MHV superamplitudes. These  are obtained by exchanging angle and square brackets:
\begin{equation}
\mathcal{A}_n^{\overline{\text{MHV}}}[1,2,\dots,n] = \frac{\delta^{(4)}(\mathcal{P}) \, \delta^{(8)}(\mathcal{Q})}{[12][23]\cdots[n1]},
\end{equation}
with $\mathcal{Q}^A = \sum_{i=1}^n \tilde{\lambda}_i \frac{\partial}{\partial \eta_{iA}}$.

In the main text, we use the amplitude with only angle brackets, so we focus on the MHV superamplitudes. Although these superamplitudes contain various helicity states, only a few types are relevant to the subleading splitting function: $ffSS$, $ffVS$, $VVSS$ and $SSSS$ amplitude.

\paragraph{ffSS}
We first consider the $ffSS$ case with helicity $(h_1,h_2,h_3,h_4)=(-\frac{1}{2},+\frac{1}{2},0,0)$. The two scalar bosons must be Hermitian conjugates, so they carry opposite R-symmetry indices, such as $S^{12}$ and $S^{34}$. The other two particles must then also be Hermitian conjugates, namely the gluinos $f^{123}$ and $f^{4}$. To derive this helicity amplitude, one applies Grassmann derivatives as in eq.~\eqref{eq:derivative} to the superamplitude:
\begin{equation} \begin{aligned}
\mathcal{A}[1_{f^{123}},3_{S^{12}},2_{f^{4}},4_{S^{34}}]
&=\left.\left(\prod_{A=1}^3\frac{\partial}{\partial \eta_{1 A}}\right) \frac{\partial}{\partial \eta_{2 4}} \left(\prod_{B=1}^2\frac{\partial}{\partial \eta_{3 B}}\right)  \left(\prod_{C=3}^4\frac{\partial}{\partial \eta_{4 C}}\right)
\mathcal{A}[\Phi_1,\Phi_3,\Phi_2,\Phi_4]\right|_{\eta=0} \\
&=\frac{\langle 13\rangle^2 \langle 14\rangle \langle 24\rangle}{\langle13\rangle\langle32\rangle\langle24\rangle\langle41\rangle}=\frac{\langle 13\rangle}{\langle23\rangle}.\\
\end{aligned}
\end{equation}
This is not the only result; a different color ordering yields a different amplitude structure:
\begin{equation} \begin{aligned}
\mathcal{A}[1_{f^{123}},3_{S^{12}},4_{S^{34}},2_{f^{4}}]
&=\left.\left(\prod_{A=1}^3\frac{\partial}{\partial \eta_{1 A}}\right) \frac{\partial}{\partial \eta_{2 4}} \left(\prod_{B=1}^2\frac{\partial}{\partial \eta_{3 B}}\right) \left(\prod_{C=3}^4\frac{\partial}{\partial \eta_{4 C}}\right) 
\mathcal{A}[\Phi_1,\Phi_3,\Phi_4,\Phi_2]\right|_{\eta=0} \\
&=\frac{\langle 13\rangle^2 \langle 14\rangle \langle 24\rangle}{\langle13\rangle\langle 34\rangle\langle42\rangle\langle21\rangle}
=\frac{\langle 13\rangle\langle 41\rangle}{\langle 43\rangle\langle 21\rangle}.
\end{aligned}
\end{equation}
Since particles 3 and 4 are scalar bosons, we may exchange their R-symmetry indices to obtain
\begin{equation} \begin{aligned}
\mathcal{A}[1_{f^{123}},3_{S^{34}},2_{f^{4}},4_{S^{12}}]
&=\frac{\langle 14\rangle^2 \langle 13\rangle \langle 23\rangle}{\langle13\rangle\langle32\rangle\langle24\rangle\langle41\rangle}
=\frac{\langle14\rangle}{\langle24\rangle}.
\end{aligned}
\end{equation}
Although other color-ordered amplitudes exist, only the above results are relevant to the massive splitting function. We now return to a theory without supersymmetry, meaning the coefficients of the above structures can differ. Using the coefficients from the Standard Model, we obtain
\begin{equation} \begin{aligned}
\mathcal{A}(1^-_{f},2^+_{\bar{f}},3^0_{S},4^0_{\bar{S}})
&=Y \tilde{Y}^{(14)} \frac{\langle 13\rangle}{\langle 23\rangle}+T_f T_s^{(34)}\frac{\langle 13\rangle\langle 41\rangle}{\langle 21\rangle\langle 43\rangle}, \\
\mathcal{A}(1^-_{f},2^+_{\bar{f}},3^0_{\bar{S}},4^0_{S})
&=\tilde{Y} Y^{(24)}\frac{\langle 14\rangle}{\langle 24\rangle}+T_f T_s^{(34)}\frac{\langle 13\rangle\langle 41\rangle}{\langle 21\rangle\langle 43\rangle}.
\end{aligned}
\end{equation}

\paragraph{ffVS} Nest, we consider the $ffVS$ case with helicity $(h_1,h_2,h_3,h_4)=(-\frac{1}{2},-\frac{1}{2},+1,0)$. Since the gluon carries positive helicity, it has no R-symmetry index. Therefore, the two gluinos must share two identical indices, yielding
\begin{equation} \begin{aligned}
\mathcal{A}[1_{f^{123}},2_{f^{124}},4_{S^{34}},3_{g^+}]
&=\frac{\langle 12\rangle^2 \langle 14\rangle \langle 24\rangle }{\langle12\rangle\langle24\rangle\langle43\rangle\langle31\rangle}, \\
\mathcal{A}[1_{f^{123}},2_{f^{124}},3_{g^+},4_{S^{34}}]
&=\frac{\langle 12\rangle^2 \langle 14\rangle \langle 24\rangle }{\langle12\rangle\langle23\rangle\langle34\rangle\langle41\rangle}.
\end{aligned}
\end{equation}
Again, we return to a theory with only conformal symmetry. After simplifying  and assigning different coefficients, these two structures yield the amplitude
\begin{equation}
\begin{aligned}
\mathcal{A}(1^{-}_{f},2^{-}_{\bar{f}},3^{+}_{V},4^0_{S})
&=\tilde{T}_f Y^{(24)}\frac{\langle 1 2 \rangle \langle 1 4 \rangle }{\langle 3 1 \rangle\langle 3 4 \rangle}
+T_f Y^{(1 4)} \frac{\langle 2 1 \rangle\langle 4 2 \rangle }{\langle 3 2 \rangle\langle 3 4 \rangle}.
\end{aligned}
\end{equation}

\paragraph{VVSS} For the $VVSS$ case, we consider helicity $(h_1,h_2,h_3,h_4)=(-1,+1,0,0)$. The two scalar bosons are again Hermitian conjugates, $S^{12}$ and $S^{34}$. Taking derivatives of the superamplitude gives
\begin{equation} \begin{aligned}
\mathcal{A}[1_{g^{-}},2_{g^{+}},4_{S^{12}},3_{S^{34}}]
&=\frac{\langle 14\rangle^2 \langle 13\rangle^2 }{\langle12\rangle\langle24\rangle\langle43\rangle\langle31\rangle}, \\
\mathcal{A}[1_{g^{-}},2_{g^{+}},3_{S^{34}},4_{S^{12}}]
&=\frac{\langle 14\rangle^2 \langle 13\rangle^2 }{\langle12\rangle\langle23\rangle\langle34\rangle\langle41\rangle}.
\end{aligned}
\end{equation}
Upon dropping supersymmetry, these two structures determine the form of the $VVSS$ amplitude in the Standard model,
\begin{equation}
\begin{aligned}
\mathcal{A}(1_{V}^-,2_{V}^+,3_{S}^0,4_{\bar{S}}^0)&=T_s^{(13)} T_s^{(24)} \frac{\langle 1 4 \rangle^2\langle 1 3 \rangle}{\langle 1 2\rangle\langle34\rangle\langle42\rangle}
+T_s^{(23)} T_s^{(1 4)}\frac{\langle 1 3 \rangle^2\langle 1 4\rangle}{\langle 1 2\rangle\langle 23\rangle\langle34\rangle}.
\end{aligned}
\end{equation}

\paragraph{SSSS}
The only exception is the $SSSS$ amplitude. We require particles 1 and 2 to be identical, while 3 and 4 are the  corresponding antiparticles. The superamplitude then gives
\begin{equation} \begin{aligned}
\mathcal{A}[1_{S^{12}},3_{S^{12}},4_{S^{34}},2_{S^{34}}]
&=\frac{\langle 13\rangle^2 \langle 24\rangle^2 }{\langle13\rangle\langle34\rangle\langle42\rangle\langle21\rangle}, \\
\mathcal{A}[1_{S^{12}},3_{S^{12}},2_{S^{34}},4_{S^{34}}]
&=\frac{\langle 13\rangle^2 \langle 24\rangle^2 }{\langle13\rangle\langle32\rangle\langle24\rangle\langle41\rangle}.
\end{aligned}
\end{equation}
Superconformal symmetry excludes the momentum-independent four-scalar contact structure. In a nonsupersymmetric theory, this structure must be restored with its own coefficient
\begin{equation}
\begin{aligned}
\mathcal{A}(1_{S}^0,2_{\bar{S}}^0,3_{S}^0,4_{\bar{S}}^0)
=&T_s^{(12)} T_s^{(34)} \frac{\langle 24 \rangle\langle 3 1 \rangle}{\langle 34 \rangle\langle 21\rangle}
+T_s^{(32)} T_s^{(1 4)}\frac{\langle 42\rangle\langle 3 1 \rangle}{\langle 32\rangle\langle 4 1 \rangle}\\
&-\left(\frac{1}{2}T_s^{(12)} T_s^{(34)}+\frac{1}{2} T_s^{(32)} T_s^{(1 4)}+\lambda^{(12|34)}\right).
\end{aligned}
\end{equation}

\end{appendix}